\documentclass[traditabstract,longauth,a4paper]{aa}

\def\setsymbol#1#2{\expandafter\def\csname #1\endcsname{#2}}
\def\getsymbol#1{\csname #1\endcsname}

\def\Planck{\textit{Planck}}

\def\all2013resultspapers{\nocite{planck2013-p01, planck2013-p02, planck2013-p02a, planck2013-p02d, planck2013-p02b, planck2013-p03, planck2013-p03c, planck2013-p03f, planck2013-p03d, planck2013-p03e, planck2013-p01a, planck2013-p06, planck2013-p03a, planck2013-pip88, planck2013-p08, planck2013-p11, planck2013-p12, planck2013-p13, planck2013-p14, planck2013-p15, planck2013-p05b, planck2013-p17, planck2013-p09, planck2013-p09a, planck2013-p20, planck2013-p19, planck2013-pipaberration, planck2013-p05, planck2013-p05a, planck2013-pip56, planck2013-p06b}}

\newbox\tablebox    \newdimen\tablewidth
\def\leaderfil{\leaders\hbox to 5pt{\hss.\hss}\hfil}
\def\endPlancktable{\tablewidth=\columnwidth 
    $$\hss\copy\tablebox\hss$$
    \vskip-\lastskip\vskip -2pt}
\def\endPlancktablewide{\tablewidth=\textwidth 
    $$\hss\copy\tablebox\hss$$
    \vskip-\lastskip\vskip -2pt}
\def\tablenote#1 #2\par{\begingroup \parindent=0.8em
    \abovedisplayshortskip=0pt\belowdisplayshortskip=0pt
    \noindent
    $$\hss\vbox{\hsize\tablewidth \hangindent=\parindent \hangafter=1 \noindent
    \hbox to \parindent{$^#1$\hss}\strut#2\strut\par}\hss$$
    \endgroup}
\def\doubleline{\vskip 3pt\hrule \vskip 1.5pt \hrule \vskip 5pt}

\def\L2{\ifmmode L_2\else $L_2$\fi}

\def\DeltaT{\ifmmode \Delta T\else $\Delta T$\fi}
\def\deltat{\ifmmode \Delta t\else $\Delta t$\fi}
\def\fknee{\ifmmode f_{\rm knee}\else $f_{\rm knee}$\fi}
\def\Fmax{\ifmmode F_{\rm max}\else $F_{\rm max}$\fi}
\def\solar{\ifmmode{\rm M}_{\mathord\odot}\else${\rm M}_{\mathord\odot}$\fi}
\def\Msolar{\ifmmode{\rm M}_{\mathord\odot}\else${\rm M}_{\mathord\odot}$\fi}
\def\Lsolar{\ifmmode{\rm L}_{\mathord\odot}\else${\rm L}_{\mathord\odot}$\fi}

\def\inv{\ifmmode^{-1}\else$^{-1}$\fi}
\def\mo{\ifmmode^{-1}\else$^{-1}$\fi}
\def\sup#1{\ifmmode ^{\rm #1}\else $^{\rm #1}$\fi}
\def\expo#1{\ifmmode \times 10^{#1}\else $\times 10^{#1}$\fi}
\def\,{\thinspace}
\def\lsim{\mathrel{\raise .4ex\hbox{\rlap{$<$}\lower 1.2ex\hbox{$\sim$}}}}
\def\gsim{\mathrel{\raise .4ex\hbox{\rlap{$>$}\lower 1.2ex\hbox{$\sim$}}}}

\def\simprop{\mathrel{\raise .4ex\hbox{\rlap{$\propto$}\lower 1.2ex\hbox{$\sim$}}}}
\def\deg{\ifmmode^\circ\else$^\circ$\fi}
\def\pdeg{\ifmmode $\setbox0=\hbox{$^{\circ}$}\rlap{\hskip.11\wd0 .}$^{\circ}
          \else \setbox0=\hbox{$^{\circ}$}\rlap{\hskip.11\wd0 .}$^{\circ}$\fi}
\def\arcs{\ifmmode {^{\scriptstyle\prime\prime}}
          \else $^{\scriptstyle\prime\prime}$\fi}
\def\arcm{\ifmmode {^{\scriptstyle\prime}}
          \else $^{\scriptstyle\prime}$\fi}
\newdimen\sa  \newdimen\sb
\def\parcs{\sa=.07em \sb=.03em
     \ifmmode \hbox{\rlap{.}}^{\scriptstyle\prime\kern -\sb\prime}\hbox{\kern -\sa}
     \else \rlap{.}$^{\scriptstyle\prime\kern -\sb\prime}$\kern -\sa\fi}
\def\parcm{\sa=.08em \sb=.03em
     \ifmmode \hbox{\rlap{.}\kern\sa}^{\scriptstyle\prime}\hbox{\kern-\sb}
     \else \rlap{.}\kern\sa$^{\scriptstyle\prime}$\kern-\sb\fi}
\def\ra[#1 #2 #3.#4]{#1\sup{h}#2\sup{m}#3\sup{s}\llap.#4}
\def\dec[#1 #2 #3.#4]{#1\deg#2\arcm#3\arcs\llap.#4}
\def\deco[#1 #2 #3]{#1\deg#2\arcm#3\arcs}
\def\rra[#1 #2]{#1\sup{h}#2\sup{m}}

\def\dots{\relax\ifmmode \ldots\else $\ldots$\fi}
\def\WHzsr{\ifmmode $W\,Hz\mo\,sr\mo$\else W\,Hz\mo\,sr\mo\fi}
\def\mHz{\ifmmode $\,mHz$\else \,mHz\fi}
\def\GHz{\ifmmode $\,GHz$\else \,GHz\fi}
\def\mKs{\ifmmode $\,mK\,s$^{1/2}\else \,mK\,s$^{1/2}$\fi}
\def\muKs{\ifmmode \,\mu$K\,s$^{1/2}\else \,$\mu$K\,s$^{1/2}$\fi}
\def\muKRJs{\ifmmode \,\mu$K$_{\rm RJ}$\,s$^{1/2}\else \,$\mu$K$_{\rm RJ}$\,s$^{1/2}$\fi}
\def\muKHz{\ifmmode \,\mu$K\,Hz$^{-1/2}\else \,$\mu$K\,Hz$^{-1/2}$\fi}
\def\MJysr{\ifmmode \,$MJy\,sr\mo$\else \,MJy\,sr\mo\fi}
\def\MJysrmK{\ifmmode \,$MJy\,sr\mo$\,mK$_{\rm CMB}\mo\else \,MJy\,sr\mo\,mK$_{\rm CMB}\mo$\fi}
\def\microns{\ifmmode \,\mu$m$\else \,$\mu$m\fi}

\def\muK{\ifmmode \,\mu$K$\else \,$\mu$\hbox{K}\fi}
\def\microK{\ifmmode \,\mu$K$\else \,$\mu$\hbox{K}\fi}
\def\muW{\ifmmode \,\mu$W$\else \,$\mu$\hbox{W}\fi}
\def\kms{\ifmmode $\,km\,s$^{-1}\else \,km\,s$^{-1}$\fi}
\def\kmsMpc{\ifmmode $\,\kms\,Mpc\mo$\else \,\kms\,Mpc\mo\fi}
\setsymbol{LFI:center:frequency:70GHz:units}{70.3\,GHz}
\setsymbol{LFI:center:frequency:44GHz:units}{44.1\,GHz}
\setsymbol{LFI:center:frequency:30GHz:units}{28.5\,GHz}

\setsymbol{LFI:center:frequency:70GHz}{70.3}
\setsymbol{LFI:center:frequency:44GHz}{44.1}
\setsymbol{LFI:center:frequency:30GHz}{28.5}

\setsymbol{LFI:center:frequency:LFI18:Rad:M:units}{71.7\GHz}
\setsymbol{LFI:center:frequency:LFI19:Rad:M:units}{67.5\GHz}
\setsymbol{LFI:center:frequency:LFI20:Rad:M:units}{69.2\GHz}
\setsymbol{LFI:center:frequency:LFI21:Rad:M:units}{70.4\GHz}
\setsymbol{LFI:center:frequency:LFI22:Rad:M:units}{71.5\GHz}
\setsymbol{LFI:center:frequency:LFI23:Rad:M:units}{70.8\GHz}
\setsymbol{LFI:center:frequency:LFI24:Rad:M:units}{44.4\GHz}
\setsymbol{LFI:center:frequency:LFI25:Rad:M:units}{44.0\GHz}
\setsymbol{LFI:center:frequency:LFI26:Rad:M:units}{43.9\GHz}
\setsymbol{LFI:center:frequency:LFI27:Rad:M:units}{28.3\GHz}
\setsymbol{LFI:center:frequency:LFI28:Rad:M:units}{28.8\GHz}
\setsymbol{LFI:center:frequency:LFI18:Rad:S:units}{70.1\GHz}
\setsymbol{LFI:center:frequency:LFI19:Rad:S:units}{69.6\GHz}
\setsymbol{LFI:center:frequency:LFI20:Rad:S:units}{69.5\GHz}
\setsymbol{LFI:center:frequency:LFI21:Rad:S:units}{69.5\GHz}
\setsymbol{LFI:center:frequency:LFI22:Rad:S:units}{72.8\GHz}
\setsymbol{LFI:center:frequency:LFI23:Rad:S:units}{71.3\GHz}
\setsymbol{LFI:center:frequency:LFI24:Rad:S:units}{44.1\GHz}
\setsymbol{LFI:center:frequency:LFI25:Rad:S:units}{44.1\GHz}
\setsymbol{LFI:center:frequency:LFI26:Rad:S:units}{44.1\GHz}
\setsymbol{LFI:center:frequency:LFI27:Rad:S:units}{28.5\GHz}
\setsymbol{LFI:center:frequency:LFI28:Rad:S:units}{28.2\GHz}

\setsymbol{LFI:center:frequency:LFI18:Rad:M}{71.7}
\setsymbol{LFI:center:frequency:LFI19:Rad:M}{67.5}
\setsymbol{LFI:center:frequency:LFI20:Rad:M}{69.2}
\setsymbol{LFI:center:frequency:LFI21:Rad:M}{70.4}
\setsymbol{LFI:center:frequency:LFI22:Rad:M}{71.5}
\setsymbol{LFI:center:frequency:LFI23:Rad:M}{70.8}
\setsymbol{LFI:center:frequency:LFI24:Rad:M}{44.4}
\setsymbol{LFI:center:frequency:LFI25:Rad:M}{44.0}
\setsymbol{LFI:center:frequency:LFI26:Rad:M}{43.9}
\setsymbol{LFI:center:frequency:LFI27:Rad:M}{28.3}
\setsymbol{LFI:center:frequency:LFI28:Rad:M}{28.8}
\setsymbol{LFI:center:frequency:LFI18:Rad:S}{70.1}
\setsymbol{LFI:center:frequency:LFI19:Rad:S}{69.6}
\setsymbol{LFI:center:frequency:LFI20:Rad:S}{69.5}
\setsymbol{LFI:center:frequency:LFI21:Rad:S}{69.5}
\setsymbol{LFI:center:frequency:LFI22:Rad:S}{72.8}
\setsymbol{LFI:center:frequency:LFI23:Rad:S}{71.3}
\setsymbol{LFI:center:frequency:LFI24:Rad:S}{44.1}
\setsymbol{LFI:center:frequency:LFI25:Rad:S}{44.1}
\setsymbol{LFI:center:frequency:LFI26:Rad:S}{44.1}
\setsymbol{LFI:center:frequency:LFI27:Rad:S}{28.5}
\setsymbol{LFI:center:frequency:LFI28:Rad:S}{28.2}

\setsymbol{LFI:white:noise:sensitivity:70GHz:units}{134.7\muKs}
\setsymbol{LFI:white:noise:sensitivity:44GHz:units}{164.7\muKs}
\setsymbol{LFI:white:noise:sensitivity:30GHz:units}{143.4\muKs}

\setsymbol{LFI:white:noise:sensitivity:70GHz}{134.7}
\setsymbol{LFI:white:noise:sensitivity:44GHz}{164.7}
\setsymbol{LFI:white:noise:sensitivity:30GHz}{143.4}

\setsymbol{LFI:white:noise:sensitivity:LFI18:Rad:M:units}{512.0\muKs}
\setsymbol{LFI:white:noise:sensitivity:LFI19:Rad:M:units}{581.4\muKs}
\setsymbol{LFI:white:noise:sensitivity:LFI20:Rad:M:units}{590.8\muKs}
\setsymbol{LFI:white:noise:sensitivity:LFI21:Rad:M:units}{455.2\muKs}
\setsymbol{LFI:white:noise:sensitivity:LFI22:Rad:M:units}{492.0\muKs}
\setsymbol{LFI:white:noise:sensitivity:LFI23:Rad:M:units}{507.7\muKs}
\setsymbol{LFI:white:noise:sensitivity:LFI24:Rad:M:units}{462.2\muKs}
\setsymbol{LFI:white:noise:sensitivity:LFI25:Rad:M:units}{413.6\muKs}
\setsymbol{LFI:white:noise:sensitivity:LFI26:Rad:M:units}{478.6\muKs}
\setsymbol{LFI:white:noise:sensitivity:LFI27:Rad:M:units}{277.7\muKs}
\setsymbol{LFI:white:noise:sensitivity:LFI28:Rad:M:units}{312.3\muKs}
\setsymbol{LFI:white:noise:sensitivity:LFI18:Rad:S:units}{465.7\muKs}
\setsymbol{LFI:white:noise:sensitivity:LFI19:Rad:S:units}{555.6\muKs}
\setsymbol{LFI:white:noise:sensitivity:LFI20:Rad:S:units}{623.2\muKs}
\setsymbol{LFI:white:noise:sensitivity:LFI21:Rad:S:units}{564.1\muKs}
\setsymbol{LFI:white:noise:sensitivity:LFI22:Rad:S:units}{534.4\muKs}
\setsymbol{LFI:white:noise:sensitivity:LFI23:Rad:S:units}{542.4\muKs}
\setsymbol{LFI:white:noise:sensitivity:LFI24:Rad:S:units}{399.2\muKs}
\setsymbol{LFI:white:noise:sensitivity:LFI25:Rad:S:units}{392.6\muKs}
\setsymbol{LFI:white:noise:sensitivity:LFI26:Rad:S:units}{418.6\muKs}
\setsymbol{LFI:white:noise:sensitivity:LFI27:Rad:S:units}{302.9\muKs}
\setsymbol{LFI:white:noise:sensitivity:LFI28:Rad:S:units}{285.3\muKs}

\setsymbol{LFI:white:noise:sensitivity:LFI18:Rad:M}{512.0}
\setsymbol{LFI:white:noise:sensitivity:LFI19:Rad:M}{581.4}
\setsymbol{LFI:white:noise:sensitivity:LFI20:Rad:M}{590.8}
\setsymbol{LFI:white:noise:sensitivity:LFI21:Rad:M}{455.2}
\setsymbol{LFI:white:noise:sensitivity:LFI22:Rad:M}{492.0}
\setsymbol{LFI:white:noise:sensitivity:LFI23:Rad:M}{507.7}
\setsymbol{LFI:white:noise:sensitivity:LFI24:Rad:M}{462.2}
\setsymbol{LFI:white:noise:sensitivity:LFI25:Rad:M}{413.6}
\setsymbol{LFI:white:noise:sensitivity:LFI26:Rad:M}{478.6}
\setsymbol{LFI:white:noise:sensitivity:LFI27:Rad:M}{277.7}
\setsymbol{LFI:white:noise:sensitivity:LFI28:Rad:M}{312.3}
\setsymbol{LFI:white:noise:sensitivity:LFI18:Rad:S}{465.7}
\setsymbol{LFI:white:noise:sensitivity:LFI19:Rad:S}{555.6}
\setsymbol{LFI:white:noise:sensitivity:LFI20:Rad:S}{623.2}
\setsymbol{LFI:white:noise:sensitivity:LFI21:Rad:S}{564.1}
\setsymbol{LFI:white:noise:sensitivity:LFI22:Rad:S}{534.4}
\setsymbol{LFI:white:noise:sensitivity:LFI23:Rad:S}{542.4}
\setsymbol{LFI:white:noise:sensitivity:LFI24:Rad:S}{399.2}
\setsymbol{LFI:white:noise:sensitivity:LFI25:Rad:S}{392.6}
\setsymbol{LFI:white:noise:sensitivity:LFI26:Rad:S}{418.6}
\setsymbol{LFI:white:noise:sensitivity:LFI27:Rad:S}{302.9}
\setsymbol{LFI:white:noise:sensitivity:LFI28:Rad:S}{285.3}

\setsymbol{LFI:knee:frequency:70GHz:units}{29.5\mHz}
\setsymbol{LFI:knee:frequency:44GHz:units}{56.2\mHz}
\setsymbol{LFI:knee:frequency:30GHz:units}{113.7\mHz}

\setsymbol{LFI:knee:frequency:70GHz}{29.5}
\setsymbol{LFI:knee:frequency:44GHz}{56.2}
\setsymbol{LFI:knee:frequency:30GHz}{113.7}

\setsymbol{LFI:knee:frequency:LFI18:Rad:M:units}{16.3\mHz}
\setsymbol{LFI:knee:frequency:LFI19:Rad:M:units}{15.1\mHz}
\setsymbol{LFI:knee:frequency:LFI20:Rad:M:units}{18.7\mHz}
\setsymbol{LFI:knee:frequency:LFI21:Rad:M:units}{37.2\mHz}
\setsymbol{LFI:knee:frequency:LFI22:Rad:M:units}{12.7\mHz}
\setsymbol{LFI:knee:frequency:LFI23:Rad:M:units}{34.6\mHz}
\setsymbol{LFI:knee:frequency:LFI24:Rad:M:units}{46.2\mHz}
\setsymbol{LFI:knee:frequency:LFI25:Rad:M:units}{24.9\mHz}
\setsymbol{LFI:knee:frequency:LFI26:Rad:M:units}{67.6\mHz}
\setsymbol{LFI:knee:frequency:LFI27:Rad:M:units}{187.4\mHz}
\setsymbol{LFI:knee:frequency:LFI28:Rad:M:units}{122.2\mHz}
\setsymbol{LFI:knee:frequency:LFI18:Rad:S:units}{17.7\mHz}
\setsymbol{LFI:knee:frequency:LFI19:Rad:S:units}{22.0\mHz}
\setsymbol{LFI:knee:frequency:LFI20:Rad:S:units}{8.7\mHz}
\setsymbol{LFI:knee:frequency:LFI21:Rad:S:units}{25.9\mHz}
\setsymbol{LFI:knee:frequency:LFI22:Rad:S:units}{15.8\mHz}
\setsymbol{LFI:knee:frequency:LFI23:Rad:S:units}{129.8\mHz}
\setsymbol{LFI:knee:frequency:LFI24:Rad:S:units}{100.9\mHz}
\setsymbol{LFI:knee:frequency:LFI25:Rad:S:units}{38.9\mHz}
\setsymbol{LFI:knee:frequency:LFI26:Rad:S:units}{58.9\mHz}
\setsymbol{LFI:knee:frequency:LFI27:Rad:S:units}{104.4\mHz}
\setsymbol{LFI:knee:frequency:LFI28:Rad:S:units}{40.7\mHz}

\setsymbol{LFI:knee:frequency:LFI18:Rad:M}{16.3}
\setsymbol{LFI:knee:frequency:LFI19:Rad:M}{15.1}
\setsymbol{LFI:knee:frequency:LFI20:Rad:M}{18.7}
\setsymbol{LFI:knee:frequency:LFI21:Rad:M}{37.2}
\setsymbol{LFI:knee:frequency:LFI22:Rad:M}{12.7}
\setsymbol{LFI:knee:frequency:LFI23:Rad:M}{34.6}
\setsymbol{LFI:knee:frequency:LFI24:Rad:M}{46.2}
\setsymbol{LFI:knee:frequency:LFI25:Rad:M}{24.9}
\setsymbol{LFI:knee:frequency:LFI26:Rad:M}{67.6}
\setsymbol{LFI:knee:frequency:LFI27:Rad:M}{187.4}
\setsymbol{LFI:knee:frequency:LFI28:Rad:M}{122.2}
\setsymbol{LFI:knee:frequency:LFI18:Rad:S}{17.7}
\setsymbol{LFI:knee:frequency:LFI19:Rad:S}{22.0}
\setsymbol{LFI:knee:frequency:LFI20:Rad:S}{8.7}
\setsymbol{LFI:knee:frequency:LFI21:Rad:S}{25.9}
\setsymbol{LFI:knee:frequency:LFI22:Rad:S}{15.8}
\setsymbol{LFI:knee:frequency:LFI23:Rad:S}{129.8}
\setsymbol{LFI:knee:frequency:LFI24:Rad:S}{100.9}
\setsymbol{LFI:knee:frequency:LFI25:Rad:S}{38.9}
\setsymbol{LFI:knee:frequency:LFI26:Rad:S}{58.9}
\setsymbol{LFI:knee:frequency:LFI27:Rad:S}{104.4}
\setsymbol{LFI:knee:frequency:LFI28:Rad:S}{40.7}

\setsymbol{LFI:slope:70GHz:units}{$-1.03$\mHz}
\setsymbol{LFI:slope:44GHz:units}{$-0.89$\mHz}
\setsymbol{LFI:slope:30GHz:units}{$-0.87$\mHz}

\setsymbol{LFI:slope:70GHz}{$-1.03$}
\setsymbol{LFI:slope:44GHz}{$-0.89$}
\setsymbol{LFI:slope:30GHz}{$-0.87$}

\setsymbol{LFI:slope:LFI18:Rad:M:units}{$-1.04$\mHz}
\setsymbol{LFI:slope:LFI19:Rad:M:units}{$-1.09$\mHz}
\setsymbol{LFI:slope:LFI20:Rad:M:units}{$-0.69$\mHz}
\setsymbol{LFI:slope:LFI21:Rad:M:units}{$-1.56$\mHz}
\setsymbol{LFI:slope:LFI22:Rad:M:units}{$-1.01$\mHz}
\setsymbol{LFI:slope:LFI23:Rad:M:units}{$-0.96$\mHz}
\setsymbol{LFI:slope:LFI24:Rad:M:units}{$-0.83$\mHz}
\setsymbol{LFI:slope:LFI25:Rad:M:units}{$-0.91$\mHz}
\setsymbol{LFI:slope:LFI26:Rad:M:units}{$-0.95$\mHz}
\setsymbol{LFI:slope:LFI27:Rad:M:units}{$-0.87$\mHz}
\setsymbol{LFI:slope:LFI28:Rad:M:units}{$-0.88$\mHz}
\setsymbol{LFI:slope:LFI18:Rad:S:units}{$-1.15$\mHz}
\setsymbol{LFI:slope:LFI19:Rad:S:units}{$-1.00$\mHz}
\setsymbol{LFI:slope:LFI20:Rad:S:units}{$-0.95$\mHz}
\setsymbol{LFI:slope:LFI21:Rad:S:units}{$-0.92$\mHz}
\setsymbol{LFI:slope:LFI22:Rad:S:units}{$-1.01$\mHz}
\setsymbol{LFI:slope:LFI23:Rad:S:units}{$-0.95$\mHz}
\setsymbol{LFI:slope:LFI24:Rad:S:units}{$-0.73$\mHz}
\setsymbol{LFI:slope:LFI25:Rad:S:units}{$-1.16$\mHz}
\setsymbol{LFI:slope:LFI26:Rad:S:units}{$-0.79$\mHz}
\setsymbol{LFI:slope:LFI27:Rad:S:units}{$-0.82$\mHz}
\setsymbol{LFI:slope:LFI28:Rad:S:units}{$-0.91$\mHz}

\setsymbol{LFI:slope:LFI18:Rad:M}{$-1.04$}
\setsymbol{LFI:slope:LFI19:Rad:M}{$-1.09$}
\setsymbol{LFI:slope:LFI20:Rad:M}{$-0.69$}
\setsymbol{LFI:slope:LFI21:Rad:M}{$-1.56$}
\setsymbol{LFI:slope:LFI22:Rad:M}{$-1.01$}
\setsymbol{LFI:slope:LFI23:Rad:M}{$-0.96$}
\setsymbol{LFI:slope:LFI24:Rad:M}{$-0.83$}
\setsymbol{LFI:slope:LFI25:Rad:M}{$-0.91$}
\setsymbol{LFI:slope:LFI26:Rad:M}{$-0.95$}
\setsymbol{LFI:slope:LFI27:Rad:M}{$-0.87$}
\setsymbol{LFI:slope:LFI28:Rad:M}{$-0.88$}
\setsymbol{LFI:slope:LFI18:Rad:S}{$-1.15$}
\setsymbol{LFI:slope:LFI19:Rad:S}{$-1.00$}
\setsymbol{LFI:slope:LFI20:Rad:S}{$-0.95$}
\setsymbol{LFI:slope:LFI21:Rad:S}{$-0.92$}
\setsymbol{LFI:slope:LFI22:Rad:S}{$-1.01$}
\setsymbol{LFI:slope:LFI23:Rad:S}{$-0.95$}
\setsymbol{LFI:slope:LFI24:Rad:S}{$-0.73$}
\setsymbol{LFI:slope:LFI25:Rad:S}{$-1.16$}
\setsymbol{LFI:slope:LFI26:Rad:S}{$-0.79$}
\setsymbol{LFI:slope:LFI27:Rad:S}{$-0.82$}
\setsymbol{LFI:slope:LFI28:Rad:S}{$-0.91$}

\setsymbol{LFI:FWHM:70GHz:units}{13\parcm01}
\setsymbol{LFI:FWHM:44GHz:units}{27\parcm92}
\setsymbol{LFI:FWHM:30GHz:units}{32\parcm65}

\setsymbol{LFI:FWHM:70GHz}{13.01}
\setsymbol{LFI:FWHM:44GHz}{27.92}
\setsymbol{LFI:FWHM:30GHz}{32.65}

\setsymbol{LFI:FWHM:LFI18:units}{13\parcm39}
\setsymbol{LFI:FWHM:LFI19:units}{13\parcm01}
\setsymbol{LFI:FWHM:LFI20:units}{12\parcm75}
\setsymbol{LFI:FWHM:LFI21:units}{12\parcm74}
\setsymbol{LFI:FWHM:LFI22:units}{12\parcm87}
\setsymbol{LFI:FWHM:LFI23:units}{13\parcm27}
\setsymbol{LFI:FWHM:LFI24:units}{22\parcm98}
\setsymbol{LFI:FWHM:LFI25:units}{30\parcm46}
\setsymbol{LFI:FWHM:LFI26:units}{30\parcm31}
\setsymbol{LFI:FWHM:LFI27:units}{32\parcm65}
\setsymbol{LFI:FWHM:LFI28:units}{32\parcm66}

\setsymbol{LFI:FWHM:LFI18}{13.39}
\setsymbol{LFI:FWHM:LFI19}{13.01}
\setsymbol{LFI:FWHM:LFI20}{12.75}
\setsymbol{LFI:FWHM:LFI21}{12.74}
\setsymbol{LFI:FWHM:LFI22}{12.87}
\setsymbol{LFI:FWHM:LFI23}{13.27}
\setsymbol{LFI:FWHM:LFI24}{22.98}
\setsymbol{LFI:FWHM:LFI25}{30.46}
\setsymbol{LFI:FWHM:LFI26}{30.31}
\setsymbol{LFI:FWHM:LFI27}{32.65}
\setsymbol{LFI:FWHM:LFI28}{32.66}

\setsymbol{LFI:FWHM:uncertainty:LFI18:units}{0.170\arcm}
\setsymbol{LFI:FWHM:uncertainty:LFI19:units}{0.174\arcm}
\setsymbol{LFI:FWHM:uncertainty:LFI20:units}{0.170\arcm}
\setsymbol{LFI:FWHM:uncertainty:LFI21:units}{0.156\arcm}
\setsymbol{LFI:FWHM:uncertainty:LFI22:units}{0.164\arcm}
\setsymbol{LFI:FWHM:uncertainty:LFI23:units}{0.171\arcm}
\setsymbol{LFI:FWHM:uncertainty:LFI24:units}{0.652\arcm}
\setsymbol{LFI:FWHM:uncertainty:LFI25:units}{1.075\arcm}
\setsymbol{LFI:FWHM:uncertainty:LFI26:units}{1.131\arcm}
\setsymbol{LFI:FWHM:uncertainty:LFI27:units}{1.266\arcm}
\setsymbol{LFI:FWHM:uncertainty:LFI28:units}{1.287\arcm}

\setsymbol{LFI:FWHM:uncertainty:LFI18}{0.170}
\setsymbol{LFI:FWHM:uncertainty:LFI19}{0.174}
\setsymbol{LFI:FWHM:uncertainty:LFI20}{0.170}
\setsymbol{LFI:FWHM:uncertainty:LFI21}{0.156}
\setsymbol{LFI:FWHM:uncertainty:LFI22}{0.164}
\setsymbol{LFI:FWHM:uncertainty:LFI23}{0.171}
\setsymbol{LFI:FWHM:uncertainty:LFI24}{0.652}
\setsymbol{LFI:FWHM:uncertainty:LFI25}{1.075}
\setsymbol{LFI:FWHM:uncertainty:LFI26}{1.131}
\setsymbol{LFI:FWHM:uncertainty:LFI27}{1.266}
\setsymbol{LFI:FWHM:uncertainty:LFI28}{1.287}

\setsymbol{HFI:center:frequency:100GHz:units}{100\,GHz}
\setsymbol{HFI:center:frequency:143GHz:units}{143\,GHz}
\setsymbol{HFI:center:frequency:217GHz:units}{217\,GHz}
\setsymbol{HFI:center:frequency:353GHz:units}{353\,GHz}
\setsymbol{HFI:center:frequency:545GHz:units}{545\,GHz}
\setsymbol{HFI:center:frequency:857GHz:units}{857\,GHz}

\setsymbol{HFI:center:frequency:100GHz}{100}
\setsymbol{HFI:center:frequency:143GHz}{143}
\setsymbol{HFI:center:frequency:217GHz}{217}
\setsymbol{HFI:center:frequency:353GHz}{353}
\setsymbol{HFI:center:frequency:545GHz}{545}
\setsymbol{HFI:center:frequency:857GHz}{857}

\setsymbol{HFI:Ndetectors:100GHz}{8}
\setsymbol{HFI:Ndetectors:143GHz}{11}
\setsymbol{HFI:Ndetectors:217GHz}{12}
\setsymbol{HFI:Ndetectors:353GHz}{12}
\setsymbol{HFI:Ndetectors:545GHz}{3}
\setsymbol{HFI:Ndetectors:857GHz}{4}

\setsymbol{HFI:FWHM:Maps:100GHz:units}{9\parcm88}
\setsymbol{HFI:FWHM:Maps:143GHz:units}{7\parcm18}
\setsymbol{HFI:FWHM:Maps:217GHz:units}{4\parcm87}
\setsymbol{HFI:FWHM:Maps:353GHz:units}{4\parcm65}
\setsymbol{HFI:FWHM:Maps:545GHz:units}{4\parcm72}
\setsymbol{HFI:FWHM:Maps:857GHz:units}{4\parcm39}
\setsymbol{HFI:FWHM:Maps:100GHz}{9.88}
\setsymbol{HFI:FWHM:Maps:143GHz}{7.18}
\setsymbol{HFI:FWHM:Maps:217GHz}{4.87}
\setsymbol{HFI:FWHM:Maps:353GHz}{4.65}
\setsymbol{HFI:FWHM:Maps:545GHz}{4.72}
\setsymbol{HFI:FWHM:Maps:857GHz}{4.39}

\setsymbol{HFI:beam:ellipticity:Maps:100GHz}{1.15}
\setsymbol{HFI:beam:ellipticity:Maps:143GHz}{1.01}
\setsymbol{HFI:beam:ellipticity:Maps:217GHz}{1.06}
\setsymbol{HFI:beam:ellipticity:Maps:353GHz}{1.05}
\setsymbol{HFI:beam:ellipticity:Maps:545GHz}{1.14}
\setsymbol{HFI:beam:ellipticity:Maps:857GHz}{1.19}

\setsymbol{HFI:FWHM:Mars:100GHz:units}{9\parcm37}
\setsymbol{HFI:FWHM:Mars:143GHz:units}{7\parcm04}
\setsymbol{HFI:FWHM:Mars:217GHz:units}{4\parcm68}
\setsymbol{HFI:FWHM:Mars:353GHz:units}{4\parcm43}
\setsymbol{HFI:FWHM:Mars:545GHz:units}{3\parcm80}
\setsymbol{HFI:FWHM:Mars:857GHz:units}{3\parcm67}

\setsymbol{HFI:FWHM:Mars:100GHz}{9.37}
\setsymbol{HFI:FWHM:Mars:143GHz}{7.04}
\setsymbol{HFI:FWHM:Mars:217GHz}{4.68}
\setsymbol{HFI:FWHM:Mars:353GHz}{4.43}
\setsymbol{HFI:FWHM:Mars:545GHz}{3.80}
\setsymbol{HFI:FWHM:Mars:857GHz}{3.67}

\setsymbol{HFI:beam:ellipticity:Mars:100GHz}{1.18}
\setsymbol{HFI:beam:ellipticity:Mars:143GHz}{1.03}
\setsymbol{HFI:beam:ellipticity:Mars:217GHz}{1.14}
\setsymbol{HFI:beam:ellipticity:Mars:353GHz}{1.09}
\setsymbol{HFI:beam:ellipticity:Mars:545GHz}{1.25}
\setsymbol{HFI:beam:ellipticity:Mars:857GHz}{1.03}

\setsymbol{HFI:CMB:relative:calibration:100GHz}{$\lsim 1\%$}
\setsymbol{HFI:CMB:relative:calibration:143GHz}{$\lsim 1\%$}
\setsymbol{HFI:CMB:relative:calibration:217GHz}{$\lsim 1\%$}
\setsymbol{HFI:CMB:relative:calibration:353GHz}{$\lsim 1\%$}
\setsymbol{HFI:CMB:relative:calibration:545GHz}{}
\setsymbol{HFI:CMB:relative:calibration:857GHz}{}

\setsymbol{HFI:CMB:absolute:calibration:100GHz}{$\lsim 2\%$}
\setsymbol{HFI:CMB:absolute:calibration:143GHz}{$\lsim 2\%$}
\setsymbol{HFI:CMB:absolute:calibration:217GHz}{$\lsim 2\%$}
\setsymbol{HFI:CMB:absolute:calibration:353GHz}{$\lsim 2\%$}
\setsymbol{HFI:CMB:absolute:calibration:545GHz}{}
\setsymbol{HFI:CMB:absolute:calibration:857GHz}{}

\setsymbol{HFI:FIRAS:gain:calibration:accuracy:statistical:100GHz}{}
\setsymbol{HFI:FIRAS:gain:calibration:accuracy:statistical:143GHz}{}
\setsymbol{HFI:FIRAS:gain:calibration:accuracy:statistical:217GHz}{}
\setsymbol{HFI:FIRAS:gain:calibration:accuracy:statistical:353GHz}{2.5\%}
\setsymbol{HFI:FIRAS:gain:calibration:accuracy:statistical:545GHz}{1\%}
\setsymbol{HFI:FIRAS:gain:calibration:accuracy:statistical:857GHz}{0.5\%}

\setsymbol{HFI:FIRAS:gain:calibration:accuracy:systematic:100GHz}{}
\setsymbol{HFI:FIRAS:gain:calibration:accuracy:systematic:143GHz}{}
\setsymbol{HFI:FIRAS:gain:calibration:accuracy:systematic:217GHz}{}
\setsymbol{HFI:FIRAS:gain:calibration:accuracy:systematic:353GHz}{}
\setsymbol{HFI:FIRAS:gain:calibration:accuracy:systematic:545GHz}{7\%}
\setsymbol{HFI:FIRAS:gain:calibration:accuracy:systematic:857GHz}{7\%}

\setsymbol{HFI:FIRAS:zero:point:accuracy:100GHz:units}{0.8\MJysr}
\setsymbol{HFI:FIRAS:zero:point:accuracy:143GHz:units}{}
\setsymbol{HFI:FIRAS:zero:point:accuracy:217GHz:units}{}
\setsymbol{HFI:FIRAS:zero:point:accuracy:353GHz:units}{1.4\MJysr}
\setsymbol{HFI:FIRAS:zero:point:accuracy:545GHz:units}{2.2\MJysr}
\setsymbol{HFI:FIRAS:zero:point:accuracy:857GHz:units}{1.7\MJysr}

\setsymbol{HFI:FIRAS:zero:point:accuracy:100GHz}{0.8}
\setsymbol{HFI:FIRAS:zero:point:accuracy:143GHz}{}
\setsymbol{HFI:FIRAS:zero:point:accuracy:217GHz}{}
\setsymbol{HFI:FIRAS:zero:point:accuracy:353GHz}{1.4}
\setsymbol{HFI:FIRAS:zero:point:accuracy:545GHz}{2.2}
\setsymbol{HFI:FIRAS:zero:point:accuracy:857GHz}{1.7}

\setsymbol{HFI:unit:conversion:100GHz:units}{0.2415\MJysrmK}
\setsymbol{HFI:unit:conversion:143GHz:units}{0.3694\MJysrmK}
\setsymbol{HFI:unit:conversion:217GHz:units}{0.4811\MJysrmK}
\setsymbol{HFI:unit:conversion:353GHz:units}{0.2883\MJysrmK}
\setsymbol{HFI:unit:conversion:545GHz:units}{0.05826\MJysrmK}
\setsymbol{HFI:unit:conversion:857GHz:units}{0.002238\MJysrmK}

\setsymbol{HFI:unit:conversion:100GHz}{0.2415}
\setsymbol{HFI:unit:conversion:143GHz}{0.3694}
\setsymbol{HFI:unit:conversion:217GHz}{0.4811}
\setsymbol{HFI:unit:conversion:353GHz}{0.2883}
\setsymbol{HFI:unit:conversion:545GHz}{0.05826}
\setsymbol{HFI:unit:conversion:857GHz}{0.002238}

\setsymbol{HFI:colour:correction:alpha=-2:V1.01:100GHz}{0.9893}
\setsymbol{HFI:colour:correction:alpha=-2:V1.01:143GHz}{0.9759}
\setsymbol{HFI:colour:correction:alpha=-2:V1.01:217GHz}{1.0007}
\setsymbol{HFI:colour:correction:alpha=-2:V1.01:353GHz}{1.0028}
\setsymbol{HFI:colour:correction:alpha=-2:V1.01:545GHz}{1.0019}
\setsymbol{HFI:colour:correction:alpha=-2:V1.01:857GHz}{0.9889}

\setsymbol{HFI:colour:correction:alpha=0:V1.01:100GHz}{1.0008}
\setsymbol{HFI:colour:correction:alpha=0:V1.01:143GHz}{1.0148}
\setsymbol{HFI:colour:correction:alpha=0:V1.01:217GHz}{0.9909}
\setsymbol{HFI:colour:correction:alpha=0:V1.01:353GHz}{0.9888}
\setsymbol{HFI:colour:correction:alpha=0:V1.01:545GHz}{0.9878}
\setsymbol{HFI:colour:correction:alpha=0:V1.01:857GHz}{1.0014}

\providecommand{\sorthelp}[1]{}

\usepackage{graphicx,epsfig}
\usepackage{enumerate}
\usepackage{natbib,pjournal}
\bibpunct{(}{)}{;}{a}{}{,} % to follow the A&A style
\usepackage{color}
\usepackage{float}
\usepackage[english]{babel}
\usepackage{txfonts}
\usepackage{url}
\usepackage{multibib}

\begin{document}

%This author list corresponds to \title{Author list for SVN PIP\_102\_Dole: Sources discovered by Planck and confirmed by Herschel-SPIRE}
%Prepared by M. Lopez-Caniego (Marcos.Lopez.Caniego@sciops.esa.int), ESAC/ESA
%This version is from Wed Jan 28 10:27:25 2015 CET
%\subtitle{There are 185 co-authors in this list}
\author{\small
Planck Collaboration: N.~Aghanim\inst{54}
\and
B.~Altieri\inst{35}
\and
M.~Arnaud\inst{66}
\and
M.~Ashdown\inst{63, 6}
\and
J.~Aumont\inst{54}
\and
C.~Baccigalupi\inst{79}
\and
A.~J.~Banday\inst{88, 10}
\and
R.~B.~Barreiro\inst{59}
\and
N.~Bartolo\inst{27, 60}
\and
E.~Battaner\inst{90, 91}
% ...
}
\institute{\small
APC, AstroParticule et Cosmologie, Universit\'{e} Paris Diderot, CNRS/IN2P3, CEA/lrfu, Observatoire de Paris, Sorbonne Paris Cit\'{e}, 10, rue Alice Domon et L\'{e}onie Duquet, 75205 Paris Cedex 13, France\goodbreak
\and
African Institute for Mathematical Sciences, 6-8 Melrose Road, Muizenberg, Cape Town, South Africa\goodbreak
\and
Agenzia Spaziale Italiana Science Data Center, Via del Politecnico snc, 00133, Roma, Italy\goodbreak
\and
Agenzia Spaziale Italiana, Viale Liegi 26, Roma, Italy\goodbreak
\and
Aix Marseille Universit\'{e}, CNRS, LAM (Laboratoire d'Astrophysique de Marseille) UMR 7326, 13388, Marseille, France\goodbreak
\and
Astrophysics Group, Cavendish Laboratory, University of Cambridge, J J Thomson Avenue, Cambridge CB3 0HE, U.K.\goodbreak
\and
Astrophysics \& Cosmology Research Unit, School of Mathematics, Statistics \& Computer Science, University of KwaZulu-Natal, Westville Campus, Private Bag X54001, Durban 4000, South Africa\goodbreak
\and
Atacama Large Millimeter/submillimeter Array, ALMA Santiago Central Offices, Alonso de Cordova 3107, Vitacura, Casilla 763 0355, Santiago, Chile\goodbreak
\and
CITA, University of Toronto, 60 St. George St., Toronto, ON M5S 3H8, Canada\goodbreak
\and
CNRS, IRAP, 9 Av. colonel Roche, BP 44346, F-31028 Toulouse cedex 4, France\goodbreak
\and
California Institute of Technology, Pasadena, California, U.S.A.\goodbreak
\and
Centro de Estudios de F\'{i}sica del Cosmos de Arag\'{o}n (CEFCA), Plaza San Juan, 1, planta 2, E-44001, Teruel, Spain\goodbreak
\and
Computational Cosmology Center, Lawrence Berkeley National Laboratory, Berkeley, California, U.S.A.\goodbreak
\and
DSM/Irfu/SPP, CEA-Saclay, F-91191 Gif-sur-Yvette Cedex, France\goodbreak
\and
DTU Space, National Space Institute, Technical University of Denmark, Elektrovej 327, DK-2800 Kgs. Lyngby, Denmark\goodbreak
\and
D\'{e}partement de Physique Th\'{e}orique, Universit\'{e} de Gen\`{e}ve, 24, Quai E. Ansermet,1211 Gen\`{e}ve 4, Switzerland\goodbreak
\and
Departamento de F\'{\i}sica, Universidad de Oviedo, Avda. Calvo Sotelo s/n, Oviedo, Spain\goodbreak
\and
Department of Astronomy and Department of Physics, University of Illinois at Urbana-Champaign, 1002 West Green Street, Urbana, Illinois, U.S.A.\goodbreak
\and
Department of Astrophysics/IMAPP, Radboud University Nijmegen, P.O. Box 9010, 6500 GL Nijmegen, The Netherlands\goodbreak
\and
Department of Physics \& Astronomy, University of British Columbia, 6224 Agricultural Road, Vancouver, British Columbia, Canada\goodbreak
\and
Department of Physics and Astronomy, University College London, London WC1E 6BT, U.K.\goodbreak
\and
Department of Physics, Florida State University, Keen Physics Building, 77 Chieftan Way, Tallahassee, Florida, U.S.A.\goodbreak
\and
Department of Physics, Gustaf H\"{a}llstr\"{o}min katu 2a, University of Helsinki, Helsinki, Finland\goodbreak
\and
Department of Physics, Princeton University, Princeton, New Jersey, U.S.A.\goodbreak
\and
Department of Physics, University of California, Santa Barbara, California, U.S.A.\goodbreak
\and
Department of Physics, University of Illinois at Urbana-Champaign, 1110 West Green Street, Urbana, Illinois, U.S.A.\goodbreak
\and
Dipartimento di Fisica e Astronomia G. Galilei, Universit\`{a} degli Studi di Padova, via Marzolo 8, 35131 Padova, Italy\goodbreak
\and
Dipartimento di Fisica e Scienze della Terra, Universit\`{a} di Ferrara, Via Saragat 1, 44122 Ferrara, Italy\goodbreak
\and
Dipartimento di Fisica, Universit\`{a} La Sapienza, P. le A. Moro 2, Roma, Italy\goodbreak
\and
Dipartimento di Fisica, Universit\`{a} degli Studi di Milano, Via Celoria, 16, Milano, Italy\goodbreak
\and
Dipartimento di Fisica, Universit\`{a} degli Studi di Trieste, via A. Valerio 2, Trieste, Italy\goodbreak
\and
Dipartimento di Fisica, Universit\`{a} di Roma Tor Vergata, Via della Ricerca Scientifica, 1, Roma, Italy\goodbreak
\and
Discovery Center, Niels Bohr Institute, Blegdamsvej 17, Copenhagen, Denmark\goodbreak
\and
European Southern Observatory, ESO Vitacura, Alonso de Cordova 3107, Vitacura, Casilla 19001, Santiago, Chile\goodbreak
\and
European Space Agency, ESAC, Camino bajo del Castillo, s/n, Urbanizaci\'{o}n Villafranca del Castillo, Villanueva de la Ca\~{n}ada, Madrid, Spain\goodbreak
\and
European Space Agency, ESAC, Planck Science Office, Camino bajo del Castillo, s/n, Urbanizaci\'{o}n Villafranca del Castillo, Villanueva de la Ca\~{n}ada, Madrid, Spain\goodbreak
\and
European Space Agency, ESTEC, Keplerlaan 1, 2201 AZ Noordwijk, The Netherlands\goodbreak
\and
Facolt\`{a} di Ingegneria, Universit\`{a} degli Studi e-Campus, Via Isimbardi 10, Novedrate (CO), 22060, Italy\goodbreak
\and
Gran Sasso Science Institute, INFN, viale F. Crispi 7, 67100 L'Aquila, Italy\goodbreak
\and
HGSFP and University of Heidelberg, Theoretical Physics Department, Philosophenweg 16, 69120, Heidelberg, Germany\goodbreak
\and
Helsinki Institute of Physics, Gustaf H\"{a}llstr\"{o}min katu 2, University of Helsinki, Helsinki, Finland\goodbreak
\and
INAF - Osservatorio Astrofisico di Catania, Via S. Sofia 78, Catania, Italy\goodbreak
\and
INAF - Osservatorio Astronomico di Padova, Vicolo dell'Osservatorio 5, Padova, Italy\goodbreak
\and
INAF - Osservatorio Astronomico di Roma, via di Frascati 33, Monte Porzio Catone, Italy\goodbreak
\and
INAF - Osservatorio Astronomico di Trieste, Via G.B. Tiepolo 11, Trieste, Italy\goodbreak
\and
INAF/IASF Bologna, Via Gobetti 101, Bologna, Italy\goodbreak
\and
INAF/IASF Milano, Via E. Bassini 15, Milano, Italy\goodbreak
\and
INFN, Sezione di Bologna, Via Irnerio 46, I-40126, Bologna, Italy\goodbreak
\and
INFN, Sezione di Roma 1, Universit\`{a} di Roma Sapienza, Piazzale Aldo Moro 2, 00185, Roma, Italy\goodbreak
\and
INFN/National Institute for Nuclear Physics, Via Valerio 2, I-34127 Trieste, Italy\goodbreak
\and
Imperial College London, Astrophysics group, Blackett Laboratory, Prince Consort Road, London, SW7 2AZ, U.K.\goodbreak
\and
Infrared Processing and Analysis Center, California Institute of Technology, Pasadena, CA 91125, U.S.A.\goodbreak
\and
Institut Universitaire de France, 103, bd Saint-Michel, 75005, Paris, France\goodbreak
\and
Institut d'Astrophysique Spatiale, CNRS (UMR8617) Universit\'{e} Paris-Sud 11, B\^{a}timent 121, Orsay, France\goodbreak
\and
Institut d'Astrophysique de Paris, CNRS (UMR7095), 98 bis Boulevard Arago, F-75014, Paris, France\goodbreak
\and
Institute for Space Sciences, Bucharest-Magurale, Romania\goodbreak
\and
Institute of Astronomy, University of Cambridge, Madingley Road, Cambridge CB3 0HA, U.K.\goodbreak
\and
Institute of Theoretical Astrophysics, University of Oslo, Blindern, Oslo, Norway\goodbreak
\and
Instituto de F\'{\i}sica de Cantabria (CSIC-Universidad de Cantabria), Avda. de los Castros s/n, Santander, Spain\goodbreak
\and
Istituto Nazionale di Fisica Nucleare, Sezione di Padova, via Marzolo 8, I-35131 Padova, Italy\goodbreak
\and
Jet Propulsion Laboratory, California Institute of Technology, 4800 Oak Grove Drive, Pasadena, California, U.S.A.\goodbreak
\and
Jodrell Bank Centre for Astrophysics, Alan Turing Building, School of Physics and Astronomy, The University of Manchester, Oxford Road, Manchester, M13 9PL, U.K.\goodbreak
\and
Kavli Institute for Cosmology Cambridge, Madingley Road, Cambridge, CB3 0HA, U.K.\goodbreak
\and
LAL, Universit\'{e} Paris-Sud, CNRS/IN2P3, Orsay, France\goodbreak
\and
LERMA, CNRS, Observatoire de Paris, 61 Avenue de l'Observatoire, Paris, France\goodbreak
\and
Laboratoire AIM, IRFU/Service d'Astrophysique - CEA/DSM - CNRS - Universit\'{e} Paris Diderot, B\^{a}t. 709, CEA-Saclay, F-91191 Gif-sur-Yvette Cedex, France\goodbreak
\and
Laboratoire Traitement et Communication de l'Information, CNRS (UMR 5141) and T\'{e}l\'{e}com ParisTech, 46 rue Barrault F-75634 Paris Cedex 13, France\goodbreak
\and
Laboratoire de Physique Subatomique et Cosmologie, Universit\'{e} Grenoble-Alpes, CNRS/IN2P3, 53, rue des Martyrs, 38026 Grenoble Cedex, France\goodbreak
\and
Laboratoire de Physique Th\'{e}orique, Universit\'{e} Paris-Sud 11 \& CNRS, B\^{a}timent 210, 91405 Orsay, France\goodbreak
\and
Lawrence Berkeley National Laboratory, Berkeley, California, U.S.A.\goodbreak
\and
Lebedev Physical Institute of the Russian Academy of Sciences, Astro Space Centre, 84/32 Profsoyuznaya st., Moscow, GSP-7, 117997, Russia\goodbreak
\and
Max-Planck-Institut f\"{u}r Astrophysik, Karl-Schwarzschild-Str. 1, 85741 Garching, Germany\goodbreak
\and
McGill Physics, Ernest Rutherford Physics Building, McGill University, 3600 rue University, Montr\'{e}al, QC, H3A 2T8, Canada\goodbreak
\and
National University of Ireland, Department of Experimental Physics, Maynooth, Co. Kildare, Ireland\goodbreak
\and
Niels Bohr Institute, Blegdamsvej 17, Copenhagen, Denmark\goodbreak
\and
Observational Cosmology, Mail Stop 367-17, California Institute of Technology, Pasadena, CA, 91125, U.S.A.\goodbreak
\and
Optical Science Laboratory, University College London, Gower Street, London, U.K.\goodbreak
\and
Paris, France\goodbreak
\and
SISSA, Astrophysics Sector, via Bonomea 265, 34136, Trieste, Italy\goodbreak
\and
School of Physics and Astronomy, Cardiff University, Queens Buildings, The Parade, Cardiff, CF24 3AA, U.K.\goodbreak
\and
Sorbonne Universit\'{e}-UPMC, UMR7095, Institut d'Astrophysique de Paris, 98 bis Boulevard Arago, F-75014, Paris, France\goodbreak
\and
Space Research Institute (IKI), Russian Academy of Sciences, Profsoyuznaya Str, 84/32, Moscow, 117997, Russia\goodbreak
\and
Space Sciences Laboratory, University of California, Berkeley, California, U.S.A.\goodbreak
\and
Special Astrophysical Observatory, Russian Academy of Sciences, Nizhnij Arkhyz, Zelenchukskiy region, Karachai-Cherkessian Republic, 369167, Russia\goodbreak
\and
Steward Observatory, University of Arizona, Tucson, AZ, U.S.A.\goodbreak
\and
Sub-Department of Astrophysics, University of Oxford, Keble Road, Oxford OX1 3RH, U.K.\goodbreak
\and
UPMC Univ Paris 06, UMR7095, 98 bis Boulevard Arago, F-75014, Paris, France\goodbreak
\and
Universit\'{e} de Toulouse, UPS-OMP, IRAP, F-31028 Toulouse cedex 4, France\goodbreak
\and
Universities Space Research Association, Stratospheric Observatory for Infrared Astronomy, MS 232-11, Moffett Field, CA 94035, U.S.A.\goodbreak
\and
University of Granada, Departamento de F\'{\i}sica Te\'{o}rica y del Cosmos, Facultad de Ciencias, Granada, Spain\goodbreak
\and
University of Granada, Instituto Carlos I de F\'{\i}sica Te\'{o}rica y Computacional, Granada, Spain\goodbreak
% ...
}

\title{\Planck\ intermediate results. XXVII. High-redshift infrared galaxy
 overdensity candidates and lensed sources discovered by \Planck\
 and confirmed by \textit{Herschel}-SPIRE\thanks{corresponding
    author \email{herve.dole@ias.u-psud.fr}}}

\titlerunning{high-$z$ sources
  discovered by \Planck\ and confirmed by \textit{Herschel}} 

\authorrunning{Planck Collaboration}

\date{Submitted 11 Aug 2014 / Accepted 13 March 2015}
% {.} { .} {.} {.}
\abstract{We have used the \Planck\ all-sky submillimetre and millimetre
  maps to search for rare sources distinguished by extreme brightness,
  a few hundred millijanskies, and their potential for being
  situated at high redshift.  These ``cold'' \Planck\ sources,
  selected using the High Frequency Instrument (HFI) directly from the
  maps and from the Planck Catalogue of Compact Sources (PCCS), all
  satisfy the criterion of having their rest-frame far-infrared peak
  redshifted to the frequency range 353 -- 857\,GHz.  This
  colour-selection favours galaxies in the redshift range $z=2$--4,
  which we consider as cold peaks in the cosmic infrared background.
  With a $4\farcm5$ beam at the four highest frequencies, our sample
  is expected to include overdensities of galaxies in groups or
  clusters, lensed galaxies, and chance line-of-sight projections.  We
  perform a dedicated {\it Herschel}-SPIRE follow-up of 234 such
  \Planck\ targets, finding a significant excess of red 350 and
  500$\,\mu$m sources, in comparison to reference SPIRE fields.  About
  94\,\% of the SPIRE sources in the \Planck\ fields are consistent
  with being overdensities of galaxies peaking at 350$\,\mu$m, with
  3\,\% peaking at 500$\,\mu$m, and none peaking at 250$\,\mu$m.
  About 3\,\% are candidate lensed systems, all 12 of which have
  secure spectroscopic confirmations, placing them at redshifts $z >
  2.2$.  Only four targets are Galactic cirrus, yielding a success
  rate in our search strategy for identifying extragalactic sources
  within the \Planck\ beam of better than 98\,\%.  The galaxy
  overdensities are detected with high significance, half of the
    sample showing statistical significance above $10\,\sigma$. The
  SPIRE photometric redshifts of galaxies in overdensities suggest a
  peak at $z\simeq 2$, assuming a single common dust temperature for
  the sources of $T_{\rm d} = 35\,$K. Under this assumption, we derive
  an infrared (IR) luminosity for each SPIRE source of about
  $4\times10^{12}\,{\rm L}_{\odot}$, yielding star formation rates of
  typically $700\,{\rm M}_{\odot}\,{\rm yr}^{-1}$.  If the observed
  overdensities are actual gravitationally-bound structures, the total
  IR luminosity of all their SPIRE-detected sources peaks at
  $4\times10^{13}\,{\rm L}_{\odot}$, leading to total star formation
  rates of perhaps $7\times10^3\,{\rm M}_{\odot}\,{\rm yr}^{-1}$ per
  overdensity.  Taken together, these sources show the signatures of
  high-$z$ ($z>2$) protoclusters of intensively star-forming galaxies.
  All these observations confirm the uniqueness of our sample compared
  to reference samples and demonstrate the ability of the all-sky
  \Planck -HFI cold sources to select populations of cosmological and
  astrophysical interest for structure formation studies.}

\keywords{Galaxies: high-redshift, clusters, evolution, star
  formation -- Cosmology: observations, large-scale structure of
  Universe -- Submillimetre: galaxies -- Gravitational lensing: strong}

\maketitle

%-----------------------------------------------------------------------
%-----------------------------------------------------------------------
\section{Introduction}
\label{sect:intro}

Cosmological structure formation in the linear regime is now fairly
well constrained by observations at the largest scales
(e.g., \citealt{tegmark2004}, \citealt{cole2005}). In the non-linear
regime the situation is less clear because structure formation
becomes a complex interplay between dark matter collapse and the
hydrodynamics of baryonic cooling. In the particularly dense environs
of the most massive dark matter halos, this interplay should lead to
vigorous episodes of rapid star formation and galaxy growth, giving
rise to copious amounts of FIR (far-infrared) emission from dust
heated by young stellar populations in massive galaxies during periods
of intense star formation.

From the standpoint of galaxy evolution, studying this intense star
formation epoch in massive dark matter halos may provide a wealth of
observational constraints on the kinematics and evolutionary history
of galaxies in massive galaxy clusters, as well as the intracluster
gas. From the point of view of cosmology, clusters yield information
on non-Gaussianities of primordial fluctuations and can challenge the
$\Lambda$CDM (cold dark matter) model
\citep{brodwin2010,hutsi2010,williamson2011,harrison2012,holz2012,waizmann2012a,trindade2013},
while lensed sources act as probes of dark matter halos, and both are
probes of cosmological parameters such as $\Omega_{\rm M}$ (matter
density today divided by the critical density of the Universe) and
$\sigma_8$ (the rms fluctuation in total matter -- baryons + CDM +
massive neutrinos -- in 8 $h^{-1}$Mpc spheres today at $z = 0$)
\citep{planck2013-p11}. Furthermore, clusters of galaxies are crucial
objects that bridge astrophysics and cosmology, sometimes with some
tensions, particularly regarding the measurement of $\sigma_8$
\citep{planck2013-p11,planck2013-p15,planck2014-a01,planck2014-a15},
all of which has led to a debate between cluster phenomenology and
cosmological physics.

The extragalactic sky in the submillimetre (submm) and millimetre (mm)
regime has been of considerable scientific interest for over two
decades, with the distinct advantage that the steep rise in the redshifted
Rayleigh-Jeans tail of the modified blackbody emitted by the warm dust in
infrared galaxies largely compensates for cosmological dimming, the
``negative k-correction'' \citep{franceschini91,blain93,guiderdoni97}. As a
consequence, the flux density of galaxies depends only weakly on
redshift, opening up a particularly interesting window into the
high-redshift Universe (typically $2 < z < 6$). Constant improvements
in bolometer technology have led to impressive samples of high-$z$ galaxies
being identified with ground-based, balloon and space-borne telescopes
(e.g., \citealt{hughes98, barger98, chapman2005, lagache2005,
  patanchon2009, devlin2009, chapin2009, negrello2010, vieira2010,
  oliver2010a,eales2010}).

Nevertheless, only with the recent advent of wide-field surveys with
astrophysical and cosmological scope have the systematic
searches become efficient enough to identify the brightest of these
objects with flux densities above about 100\,mJy at 350$\,\mu$m, e.g.,
with {\it Herschel}\footnote{{\it Herschel\/} is an ESA space
  observatory with science instruments provided by European-led
  Principal Investigator consortia and with important participation
  from NASA.}  \citep{pilbratt2010,eales2010,oliver2010a}, the
South Pole Telescope \citep{vieira2010}, {\it WISE\/}
\citep{wright2010,stanford2014a}, and {\it Spitzer}
\citep{papovich2008,stanford2012}. Such sources are very rare.  For
example, the surface density of red sources brighter than 300\,mJy at
500$\,\mu$m is $10^{-2}\,{\rm deg}^{-2}$ for strongly lensed galaxies,
$3\times10^{-2}\,{\rm deg}^{-2}$ for AGN (active galactic nuclei, here
radio-loud, mostly blazars), and $10^{-1}\,{\rm deg}^{-2}$ for
late-type galaxies at moderate redshifts
\citep[e.g.,][]{negrello2007,negrello2010}. Other models predict
similar trends \citep{paciga2009, lima2010, bethermin2011a,
  hezaveh2012}. This makes even relatively shallow submm surveys
interesting for searches of high-redshift objects, as long as they
cover large parts of the sky. Studies of gravitationally lensed
galaxies at high redshifts originating from these surveys
\citep{negrello2010, combes2012, bussmann2013, herranz2013, rawle2014,
  wardlow2013} illustrate the scientific potential of such surveys for
identifying particularly interesting targets for a subsequent detailed
characterization of high-$z$ star formation through multi-wavelength
follow-up observations.

The power of wide-field surveys in detecting the rarest objects on the
submm sky is even surpassed with genuine all-sky surveys, which
systematically and exhaustively probe the brightest objects in their
wavelength domain down to their completeness limits. Here we present a
search for the rarest, most extreme high-redshift ($z\gtrsim$2)
candidates on the submillimetre sky, which was performed with the
\Planck\footnote{\Planck\ is a project of the European Space Agency --
  ESA -- with instruments provided by two scientific Consortia funded
  by ESA member states (in particular the lead countries: France and
  Italy) with contributions from NASA (USA), and telescope reflectors
  provided in a collaboration between ESA and a scientific Consortium
  led and funded by Denmark.}  all-sky survey
\citep{planck2013-p01,planck2014-a01}. The \Planck\ Catalogue of
Compact Sources (PCCS, \citealt{planck2013-p05}) has a completeness
limit of about 600\,mJy at the highest frequencies, which corresponds
to $L_{\rm FIR}\simeq 5\times10^{13}\,{\rm L}_{\odot}$ at $z=2$. With
a 5\arcmin\ beam \citep{planck2013-p03c} at the four highest
frequencies (corresponding to a physical distance of about 2.5\,Mpc at
$z=2$), we expect that sources with bona fide colours of high-$z$
galaxies in the \Planck-HFI bands are either strongly gravitationally
lensed galaxies, or the combined dust emission of multiple galaxies in
a shared vigorously star-forming environment in the high redshift
Universe. The latter case is consistent with the result that submm
galaxies or ULIRGs (Ultra Luminous Infrared Galaxies) are strongly
clustered \citep{blain2004, farrah2006, magliocchetti2007,
  austermann2009, santos2011, ivison2012, valtchanov2013, noble2013,
  clements2014}, and may include massive galaxy clusters during their
major growth phase.  It is also possible of course that the sources
are chance alignments of multiple high-redshift galaxies projected
onto the same line of sight \citep{negrello2005, negrello2007,
  negrello2010,chiang2013}, or of multiple, lower-mass galaxy groups
or clusters.

Identifying high-redshift cluster candidates directly by the
signatures of their total star formation is a very useful complement
to the diagnostics used to identify galaxy clusters so far. Most
systematic searches today focus on the primary constituents of more
evolved, lower-redshift clusters, like their populations of massive,
passively evolving galaxies ('red-sequence galaxies'), the hot
intracluster medium (either through X-ray emission or the
Sunyaev-Zeldovich effect), or the suspected progenitors of today's
brightest cluster galaxies, in particular high-redshift galaxies
\citep[for instance]{steidel2000, brand2003, kodama2007, scoville2007,
  venemans2007, papovich2008,daddi2009, brodwin2010, galametz2010, papovich2010,
  capak2011, hatch2011, kuiper2011, ivison2013a, muzzin2013,
  wylezalek2013, chiang2014, cooke2014a,
  cucciati2014a,mei2014,rettura2014a}. With the limited sensitivity and
the large beam of \Planck, in turn, we effectively select the most
intensely star-forming Mpc-scale environments in the high-redshift
Universe.

%__________________________________________________________________ 
% Figures: MAPS
%__________________________________________________________________ 
\begin{figure*}[!ht] 
   \centering 
  \includegraphics[width=1.0\textwidth]{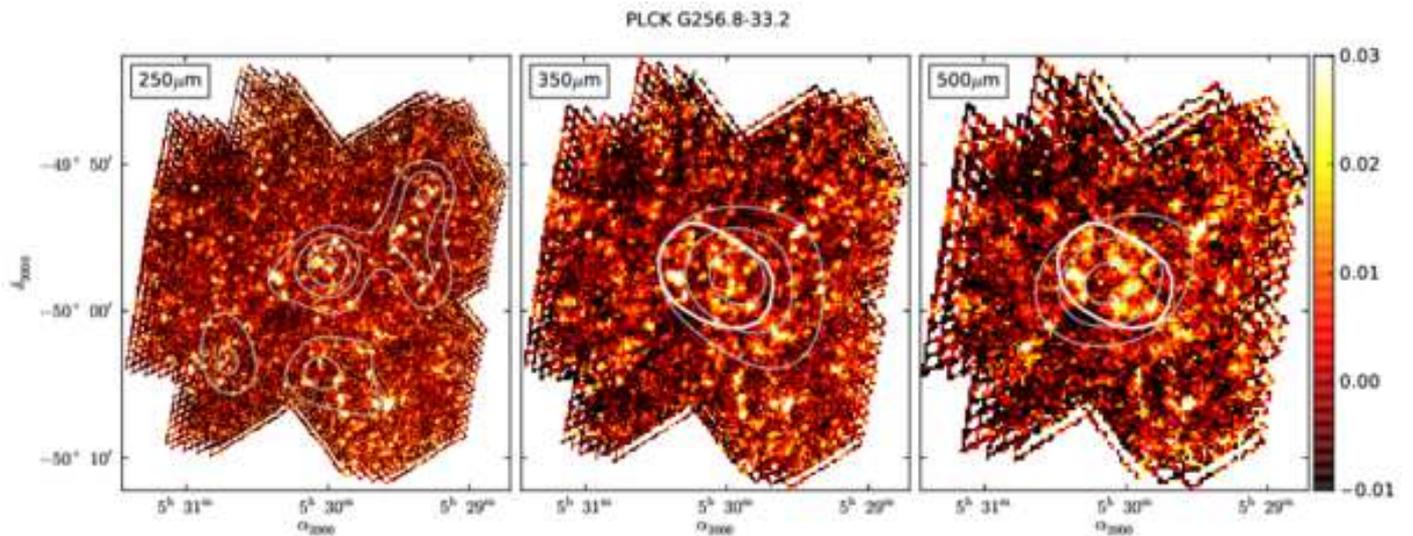}
  \caption{SPIRE maps at 250, 350, and 500$\,\mu$m of one typical
    overdensity in our sample (a \Planck\ cold source of the CIB --
    Cosmic Infrared Background).  The thick white contour at
    350$\,\mu$m shows the \Planck\ IN region at 857\,GHz, and the same
    is true for the map at 500$\,\mu$m at 545\,GHz (i.e., the 50\,\%
    contour of the \Planck\ source maximum).  Thin contours correspond
    to the overdensity of SPIRE sources at each wavelength, marked at
    2$\,\sigma$, 3$\,\sigma$, etc. (see Sect.~\ref{sect:overdensities}
    for details). SPIRE identifies a few (typically 5 to 10) sources
    inside the \Planck\ contour.  We use these contours to separate
    the IN and OUT zones. These data come from a
    $7\arcmin\times7\arcmin$ SPIRE scan (see Table~\ref{tab:otsummary}
    and Sect.~\ref{sect:inout}).}
  \label{fig:overdens_map} 
\end{figure*}

\begin{figure*}[!ht] 
   \centering 
  \includegraphics[width=1.0\textwidth]{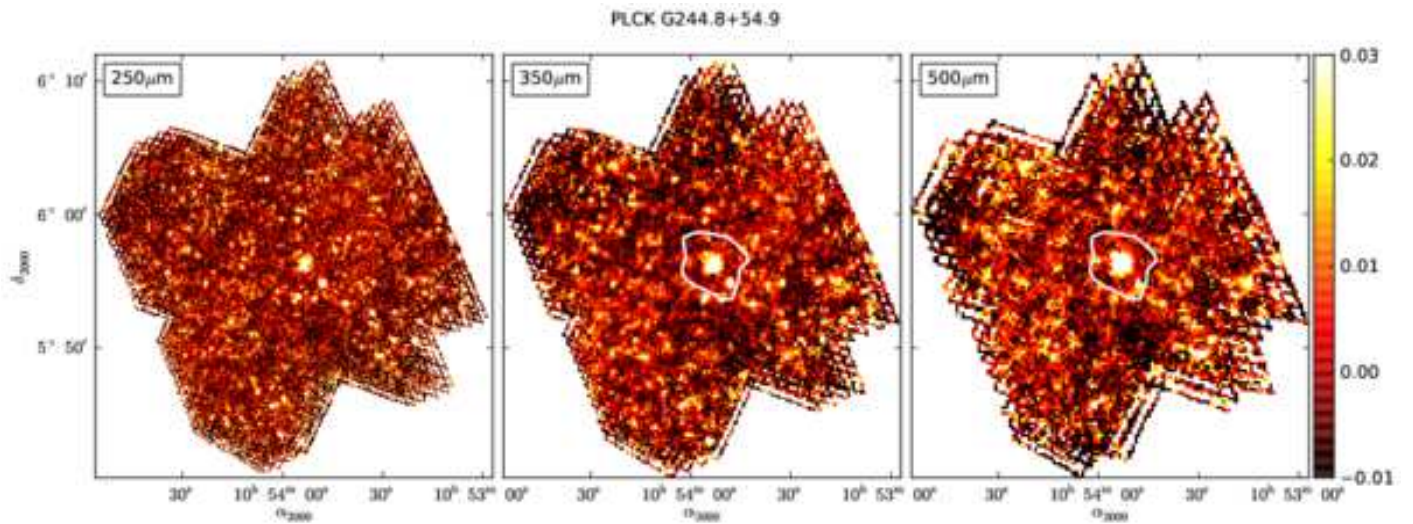}
  \caption{SPIRE maps at 250, 350, and 500$\,\mu$m for a lensed
    source.  In this case a single very bright SPIRE source is
    detected.  This source is confirmed at the Jy level and has a
    redshift of $z=3.0$, measured from multi-line spectroscopy
    acquired at IRAM using EMIR \citep{canameras2015}.  The white
    contour is the \Planck\ IN region, defined in the same way as in
    Fig.~\ref{fig:overdens_map}. These data come from a
    $7\arcmin\times7\arcmin$ SPIRE scan (see Table~\ref{tab:otsummary}
    and Sect.~\ref{sect:inout}). }
  \label{fig:lens_map} 
\end{figure*}

Using \Planck, we have selected putative high-redshift objects with
spectral energy distributions (SEDs) of warm dust, peaking between
observed frequencies of 353 and 857\,GHz. This has a net effect of
selecting sources either peaking at 545\,GHz, or to a lesser extent
sources having their infrared peak between 353 and 857\,GHz.  This
equates, in principle, to redshifted infrared galaxies at $z \sim
2$--$4$.  We refer to these as ``cold'' sources of the Cosmic Infrared
Background (or CIB), which have a red and thus potentially redshifted
SED. The CIB
\citep{puget96,hauser98,hauser2001,dole2006,planck2013-pip56} is the
integrated relic emission in the broad infrared range, typically
$8\,\mu$m to 1\,mm, where the emission reaches a maximum
\citep{dole2006}.  Physically, such objects correspond to galaxy and
AGN formation and evolutionary processes, and more generally the
history of energy production in the post-recombination Universe
\citep[e.g.,][]{kashlinsky2005,planck2011-6.6}. The CIB, as observed
in the submillimetre, is considered as a proxy for intense star
formation at redshifts $z>1$ \citep{planck2013-pip56} as well as for
the mass of structures \citep{planck2013-p13}.

The \Planck\ collaboration has also released the PCCS
(\citealt{planck2013-p05}), containing 24381 sources at 857\,GHz, with
7531 sources at Galactic latitudes $\left| b \right| > 30^{\circ}$, of
which many are of interest for extragalactic studies.  The main difference
between the cold sources of the CIB and the red sources of the PCCS
(both used in this work) can be summarized as a threshold difference:
cold CIB sources are detected in the CIB fluctuations with a combined
spectral and angular filtering method \citep{montier2010}, while PCCS
sources are detected independently in the frequency maps using an
angular filtering method with a higher threshold
\citep{planck2013-p05}.

This paper presents the observations and analysis of our extensive dedicated
{\it Herschel}-SPIRE \citep{griffin2010} follow-up of 234
\Planck\ sources (either selected from the CIB fluctuations or from
the PCCS). The paper is structured into seven sections.  In
Sect.~\ref{sect:samples}, we detail the \Planck\ parent sample and
the {\it Herschel\/} observations. Section~\ref{sect:analysis} gives a 
technical description of the algorithms used in the generation of the SPIRE
photometry and the catalogue. In Sect.~\ref{sect:statanalysis} we
use statistics to characterize the {\it Herschel\/} observations; in
particular we quantify the overdensities and the colours of the SPIRE
counterparts and propose a classification of either overdensities or
lensed candidates.  In Sect.~\ref{sect:lensed}, we discuss the properties
of the lensed source candidates, while in Sect.~\ref{sect:highz} we focus on
the overdensities and their characterization, including a stacking
analysis. Conclusions are reported in Sect.~\ref{sect:conclusion}.  The
Appendices contain information on the SPIRE catalogue generation and the
number counts, and a gallery of sample fields. We use the \Planck\
2013 cosmology \citep[table~5: \Planck+WP+highL+BAO]{planck2013-p11}
throughout the paper.

%-----------------------------------------------------------------------
%-----------------------------------------------------------------------
\section{Sample selection and observations}
\label{sect:samples}

\subsection{\Planck\ observations and selection}
\Planck\ \footnote{\Planck\ data (maps and catalogues) can be
  downloaded from the \Planck\ Legacy Archive
  \url{http://pla.esac.esa.int/pla/aio/planckProducts.html}} observed
the whole sky at frequencies between 30 and 857\,GHz
\citep{planck2013-p01}. We made two different selections to follow-up
with {\it Herschel\/}: first, using the maps and looking for cold sources of
the CIB; second, using the public catalogue of compact
sources (PCCS).

\subsubsection{Cold sources of the CIB in the \Planck\ maps}
\label{sect:coldsources}
We make use of the \Planck-HFI (High Frequency Instrument,
\citealt{planck2013-p03}) data as well as the {\it IRAS}/IRIS data
\citep{miville-deschenes2005} at 100$\,\mu$m. For this purpose we use
the cleanest 26\,\% of the sky (in the \Planck\ 857\,GHz map), defined
by a minimal cirrus contamination,
$N_{\rm{HI}}<3\times10^{20}\mathrm{cm^{-2}}$.  The detection and
selection algorithm, based on \cite{montier2010} can be summarized in
the following seven steps \cite[]{planck_collaboration2015}.

\begin{enumerate}[(i)]
\item {\it CMB cleaning}: the 143\,GHz \Planck-HFI map is
  extrapolated to the other bands according to a CMB spectrum and
  removed from the maps at other frequencies.
\item {\it Galactic cirrus cleaning}: the {\it IRAS}/IRIS
  $100\,\mu$m map, considered as a ``warm template'' of
  Galactic cirrus, is extrapolated to the {\Planck}-HFI bands, taking
  into account the local colour, and removed from the maps following
  the prescription of the {\tt CoCoCoDeT} algorithm \citep{montier2010}.
\item {\it Construction of excess maps}: the 545\,GHz excess map
  is defined as the difference between the cleaned 545\,GHz map and a
  power law interpolated between the cleaned maps at 857\,GHz and
  353\,GHz.
\item {\it Point source detection in the excess maps}: we apply a
  Mexican hat type detection algorithm \citep{gonzalez-nuevo2006} with
  a size parameter of $R=5$\arcmin\ in the 545\,GHz excess maps.
\item {\it Single frequency detection}: we also require a
  detection in each cleaned map at 857 and 353\,GHz.
\item {\it Colour-colour selection}: we apply two criteria on the
  $S_{545} / S_{857}$ (i.e., 545\,GHz to 857\,GHz) and $S_{353} /
  S_{545}$ (i.e., 353\,GHz to 545\,GHz) flux density
  ratios to select the reddest sources.
\end{enumerate}
%\end{description}

This produces a dozen hundred candidates on the cleanest 26\,\% region
of the submm sky. We note that this procedure has been set-up early on
in the \Planck\ project. The final catalog of \Planck\ high-z
candidates is being generated using a similar (but not exactly
identical) method which will be described in a forthcoming paper
\cite[]{planck_collaboration2015}. In particular, the CMB estimate
will not be the 143\,GHz HFI map, but instead the CMB derived by
component separation. The present paper focuses on the first
candidates followed up by {\it Herschel}.

\subsubsection{\Planck\ PCCS sources}
\label{sect:pccssources}
We make use of the PCCS to choose a sample of high-$z$ sources
selected by the expected peak in their thermal dust spectrum in the
rest-frame far infrared.  Our four step procedure here is based on the
work of \cite{negrello2010}.  First of all, we use the 857\,GHz
Galactic mask, keeping 52\,\% of the sky \citep{planck2012-VII}.
Secondly, we select all the sources with ${\rm S/N}>4$ at 545\,GHz and
the colours $S_{857} / S_{545} < 1.5$ (where $S_{\nu}$ is the flux
density at frequency $\nu$ in GHz) and $S_{217} < S_{353}$.  Thirdly,
we inspect each source with respect to the NASA/IPAC Extragalactic
Database (NED), {\it IRAS\/} maps \citep{neugebauer84}, and optical
maps using ALADIN. Any object identified as a local galaxy, a bright
radio source or Galactic cirrus is removed (about half of the
  objects).  Finally, we remove any PCCS source already falling in
the H-ATLAS or South Pole Telescope (SPT) survey fields.

%___________________________________________________________________________
% TABLE: summary of OT and HPASSS programmes
%___________________________________________________________________________
\begin{table}[!t]
\begingroup
\newdimen\tblskip \tblskip=5pt
\caption{SPIRE programmes following-up \Planck\ cold
          sources of the CIB.  HPASSS is composed of 124 sources
          selected from the maps (two were repeated, so 126 fields),
          together with 28 from the PCCS (four were cirrus dominated, so 24 net
          fields).  A total of
          228 sources are useable (Sect.~\ref{sect:3samples}).}
\label{tab:otsummary}
\vskip -5mm
\footnotesize
\setbox\tablebox=\vbox{
 \newdimen\digitwidth
 \setbox0=\hbox{\rm 0}
 \digitwidth=\wd0
 \catcode`*=\active
 \def*{\kern\digitwidth}
 \newdimen\signwidth
 \setbox0=\hbox{+}
 \signwidth=\wd0
 \catcode`!=\active
 \def!{\kern\signwidth}
 \halign{\tabskip=0pt\hbox to 0.75in{#\leaderfil}\tabskip=1em&
 \hfil#\hfil\tabskip=1em&
 \hfil#\hfil\tabskip=1em&
 \hfil#\hfil\tabskip 0pt\cr
\noalign{\doubleline}
\omit Programme& No.\ of fields& Map size& AOR$^{\rm a}$ time\cr
\noalign{\vskip 4pt\hrule\vskip 6pt}
OT1&    10& $15\arcmin\times15\arcmin$& 3147\,s\cr
OT2&    70&   $7\arcmin\times7\arcmin$& *838\,s\cr
HPASSS& 124 (126) (9$^{\rm b}$) + 24 (28)& $7\arcmin\times7\arcmin$& *838\,s\cr
Total& 228 (234)& & \cr
\noalign{\vskip 4pt\hrule\vskip 6pt}
}}
\endPlancktable
\tablenote {{\rm a}} Astronomical Observation Request.\par
\tablenote {{\rm b}} Nine HPASSS sources are from archival data.\par
\endgroup
\end{table}

\subsection{The \Planck\ parent sample and three subsamples}
\label{sect:3samples}

\Planck\ and {\it Herschel\/} were launched on 14 May 2009, and
started routine scientific observations a few months later
\citep{planck2011-1.1,pilbratt2010}.  Our \Planck\ parent sample has
evolved with time as a function of the delivery of internal releases
of the intensity maps (typically every 6 months). We thus built three
samples: one for each main {\it Herschel\/} Open Time call for proposals
(OT1 and OT2), and a last one for the {\it Herschel\/} 'must-do' DDT
(Director's Discretionary Time) observations (which we call HPASSS for
{\it Herschel\/} and \Planck\ All Sky Legacy Survey Snapshot). We
requested only SPIRE data, as it is better optimized than PACS for the targeted
redshift range. Our main goal was to identify the counterparts
contributing to the \Planck-HFI detections, hence the focus 
on rapid SPIRE follow-up of a maximum number of targets. Obtaining
PACS observations would have enhanced the angular resolution and
wavelength coverage, but for these short observations would not have
resulted in further constraints in most cases.

The samples are summarized in Table~\ref{tab:otsummary}, together with
the map sizes and time per Astronomical Observation Request (AOR).
The OT1 sample refers to the {\it Herschel\/} first call for Open Time
observations in July 2010 (P.I.: L. Montier). The OT2 sample refers to
the {\it Herschel\/} second call for Open Time observations in September
2011 (P.I.: H. Dole). Figs.~\ref{fig:overdens_map} and
\ref{fig:lens_map} show examples of this observation strategy.  The
{\it Herschel\/} and \Planck\ All-Sky Source Snapshot Legacy Survey
(HPASSS) was set up by the \Planck\ collaboration (P.I.: H. Dole)
in the response to 'must-do' DDT programme in June 2012. At that
time, the preliminary PCCS was internally released to the
collaboration, and we could benefit from the unique and timely
{\it  Herschel\/} and \Planck\ synergy when the \Planck\ products were
approaching their final state and when {\it Herschel\/} surveys already
demonstrated the efficiency of the SPIRE observations. We note that the
SPIRE data from 9 \Planck\ fields included in the HPASSS sample are
{\it Herschel\/} archival data: 4C24.28-1 (program: OT2\_rhuub\_2);
NGP\_v1, NGP\_h1, NGP\_h2, SGP\_sm3-h, GAMA12\_rn1, NGP\_v8, NGP\_h6
(KPOT\_seales01\_2); Lockman\_SWIRE\_offset\_1-1 (SDP\_soliver\_3); and
Spider-1  -- (OT1\_mmiville\_2).

%__________________________________________________________________ 
% Figures: PHOTOMETRIC ACCURACY
%__________________________________________________________________ 
\begin{figure*}[!ht] 
   \centering 
  \includegraphics[width=1.0\textwidth]{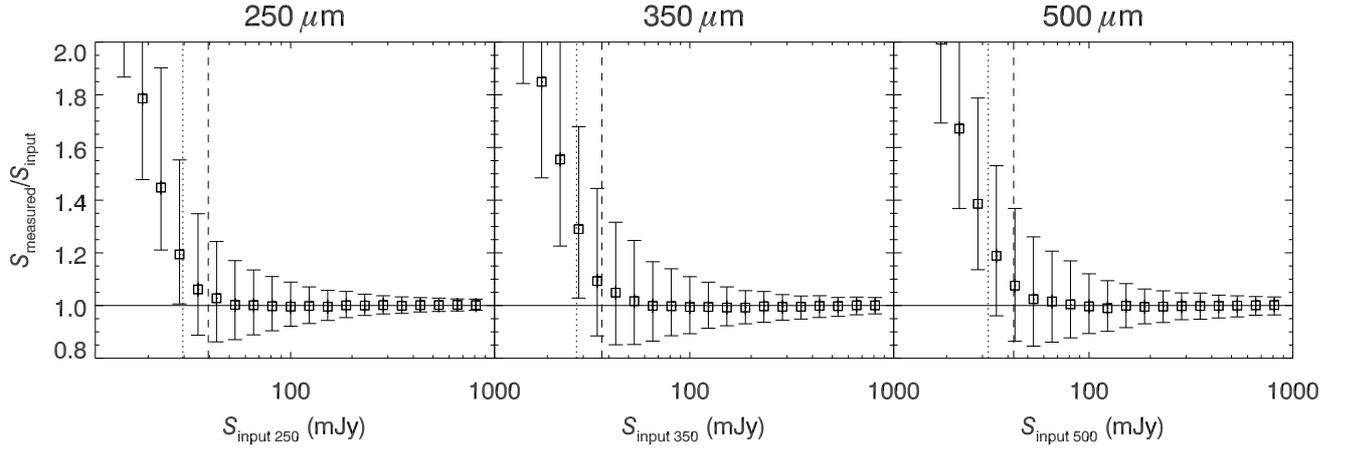}
  \caption{Photometric accuracy from Monte Carlo source injections in
    real maps at 250, 350 and 500$\,\mu$m. The $3\,\sigma$ (dotted line)
    and $4\,\sigma$ (dashed line) levels are also shown. See
    Sect.~\ref{sect:blind} for details.}
  \label{fig:photometricaccuracymc} 
\end{figure*}

%__________________________________________________________________ 
% Figures: COMPLETENESSES
%__________________________________________________________________ 
\begin{figure*}[!ht] 
   \centering 
  \includegraphics[width=1.0\textwidth]{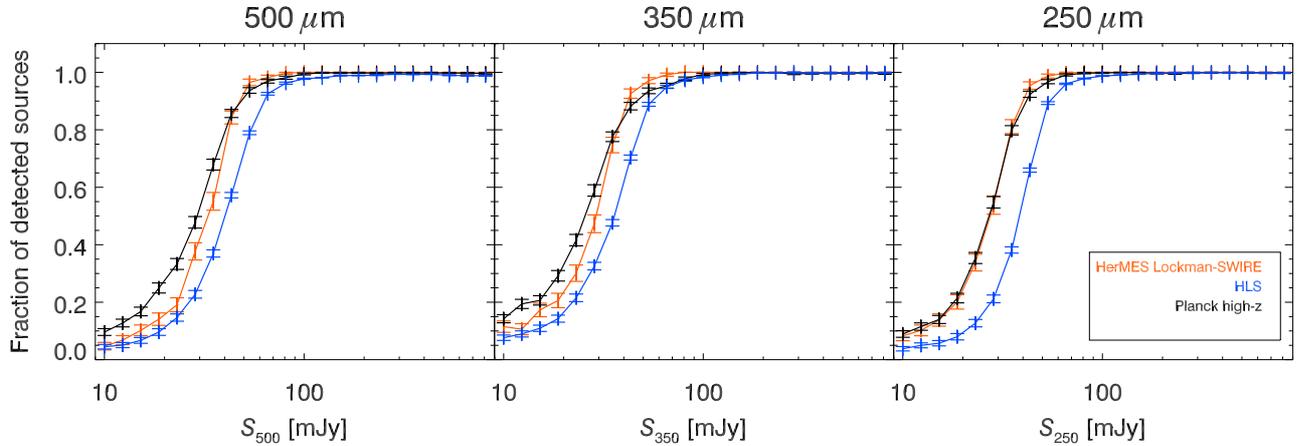}
  \caption{Completeness levels from Monte Carlo source injections in
    SPIRE maps of different data sets at 250, 350 and
    500$\,\mu$m. Plotted are our \Planck\ fields (black), HerMES
    sources (orange), and Herschel Lensing Survey (HLS) clusters
    (blue). The 50\,\% (dotted line) and 80\,\% (dashed line) completeness
    levels are also shown. See Sect.~\ref{sect:blind} for details.}
  \label{fig:completeness} 
\end{figure*}

\subsection{SPIRE data processing, total SPIRE sample and definition
  of IN and OUT \Planck\ regions}
\label{sect:inout}
The SPIRE data were processed starting with 'Level~0' using {\tt HIPE}
10.0 \citep{ott2010} and the calibration tree 'spire\_cal\_10\_1'
using the Destriper module\footnote{Scan map destriper details:
 % \href{https://nhscsci.ipac.caltech.edu/sc/index.php/Spire/PhotScanMapDestriper}{https://nhscsci.ipac.caltech.edu}}
  https://nhscsci.ipac.caltech.edu/sc/index.php/Spire/PhotScanMapDestriper}{https://nhscsci.ipac.caltech.edu}
as mapmaker (baselines are removed thanks to an optimum fit between
all timelines) for most observations. For 33 AORs, we had also to
remove some particularly noisy detectors (PSWB5, PSWF8, and PSWE9
affected for Operational Days 1304 and 1305) from the Level~1
timelines. In that case, we processed the data with the naive scan
mapper using a median baseline removal (the destriper module worked
with all bolometers in Level~1).  Turnarounds have been taken
into account in the processed data. The useable sky surface area is
thus extended, and goes beyond the nominal $7\arcmin\times7\arcmin$ or
$15\arcmin\times15\arcmin$ as specified in the AORs
(Table~\ref{tab:otsummary}). Since we are close to the confusion
limit, we included a check to confirm that the non-uniform coverage
imposed by including the edges does not change the source detection
statistics.

We have a total of 234 SPIRE targets.  Of these, two fields were repeated
observations, and have been used to check the robustness of our
detections. Four fields from the PCCS appear as cirrus structures:
diffuse, extended submillimetre emission without noticeable point
sources. These fields were removed from the sample. This means that
four out of 234 fields (1.7\,\%) were contaminated by Galactic
cirrus. The success rate of avoiding Galactic cirrus features is thus
larger than 98\,\%, thanks to our careful selection on the \Planck\ maps
of the cleanest 35\,\% of the sky.

Our final sample thus contains 228 fields (i.e., 234 minus two
repeated fields minus four cirrus-dominated fields). They are composed
of (See Table~\ref{tab:otsummary}): 10 sources from OT1; 70 sources
from OT2; 124 objects from HPASSS CIB; and 24 sources from HPASSS
PCCS.

Each SPIRE field of a \Planck\ target is then separated into two
zones: IN and OUT of the \Planck\ source at 545\,GHz
(the frequency where our selection brings the best S/N ratio).  The IN
region boundary is defined as the 50\,\% \Planck\ intensity contour,
i.e., the isocontour corresponding to 50\,\% of the peak \Planck\
flux, and encompasses the \Planck-HFI beam. The OUT region is
defined to be outside this region (see the thick white contours
in Figs.~\ref{fig:overdens_map} and \ref{fig:lens_map}).

%___________________________________________________________________________
% TABLE: 1 sigma measured noise
%___________________________________________________________________________
\begin{table}[!t]
\begingroup
\newdimen\tblskip \tblskip=5pt
\caption{SPIRE 1$\,\sigma$ total noise (instrument + confusion) levels
measured in various fields, in mJy. Our \Planck\ fields are denoted
'\Planck\ high-$z$'.}
\label{tab:1sigma}
\vskip -5mm
\footnotesize
\setbox\tablebox=\vbox{
 \newdimen\digitwidth
 \setbox0=\hbox{\rm 0}
 \digitwidth=\wd0
 \catcode`*=\active
 \def*{\kern\digitwidth}
 \newdimen\signwidth
 \setbox0=\hbox{+}
 \signwidth=\wd0
 \catcode`!=\active
 \def!{\kern\signwidth}
 \halign{\tabskip=0pt\hbox to 1.75in{#\leaderfil}\tabskip=1em&
 \hfil#\hfil\tabskip=1em&
 \hfil#\hfil\tabskip=1em&
 \hfil#\hfil\tabskip 0pt\cr
\noalign{\doubleline}
\omit& \multispan3\hfil 1$\,\sigma$ noise level\hfil\cr
\noalign{\vskip -5pt}
\omit& \multispan3\hrulefill\cr
\omit Field& 250$\,\mu$m& 350$\,\mu$m& 500$\,\mu$m\cr
\noalign{\vskip 4pt\hrule\vskip 6pt}
   \Planck\ high-$z$& *9.9& *9.3& 10.7\cr
HerMES Lockman-SWIRE& 10.1& 10.5& 12.0\cr
                 HLS& 14.1& 12.6& 14.2\cr
\noalign{\vskip 4pt\hrule\vskip 6pt}
}}
\endPlancktable
\endgroup
\end{table}

\subsection{Ancillary SPIRE data sets} 

For a first characterization of our sample, and to infer whether it is
different from other samples of distant galaxies observed with SPIRE,
we will compare with the data obtained in other SPIRE
programmes. Since we suspect that most of our targets contain
overdensities like proto-clusters of galaxies, it will be useful to contrast it
with samples of galaxy clusters at lower redshift, as well as against blank
fields.  Our two comparison samples are as follows.

\begin{enumerate}
\item The HerMES\footnote{\url{http://hedam.oamp.fr/HerMES/}}
\citep{oliver2010a} and H-ATLAS\footnote{\url{http://www.h-atlas.org/}}
\citep{eales2010} public data as reference fields. In particular, we
will be using the 'level~5' Lockman HerMES field, which has a similar depth
to our SPIRE observations.
\item The SPIRE snapshot programme of local or massive galaxy clusters
\citep{egami2010} including: the ``Herschel Lensing Survey'' (HLS);
the ``SPIRE Snapshot Survey of Massive Galaxy Clusters'' (OT1); and
the ``SPIRE Snapshot Survey II, Using SPT/CODEX Massive Clusters as
Powerful Gravitational Lenses'' (OT2)\footnote{{\it Herschel} public
data can be downloaded from the Herschel Science Archive:
\url{http://herschel.esac.esa.int/Science_Archive.shtml}}.
\end{enumerate}

Finally, we use the OT2 data from ``The highest redshift strongly
lensed dusty star-forming galaxies'' (P.I. J. Vieira) programme to
carry out technical checks, since those AORs are very similar to ours.

%-----------------------------------------------------------------------
%-----------------------------------------------------------------------
\section{SPIRE Photometry}
\label{sect:analysis}

\subsection{Photometric analysis}

We first measure the noise in each map at each wavelength (channel) by
fitting a Gaussian function to the histogram of the maps and deriving
the standard deviation, $\sigma_{\rm channel}$, a mixture of confusion
noise \citep{dole2003,nguyen2010} and detector noise. The values are
reported in Table~\ref{tab:1sigma}.  Data are sky subtracted and at
mean 0, as done in HIPE 10. All pixels are used to plot the noise
histogram. Histogram have a Gaussian shape, with a bright pixel
tail. The bright pixel tail contribution (larger than 3 sigma) to the
histogram is nearly 1\% in number for each field at 350um. Then we
extract the sources using a blind method and by a band-merging
procedure described in the following sections.

%__________________________________________________________________ 
% Figures: TOTAL NUMBER COUNTS
%__________________________________________________________________ 
\begin{figure*}[!ht] 
   \centering 
  \includegraphics[width=1.0\textwidth]{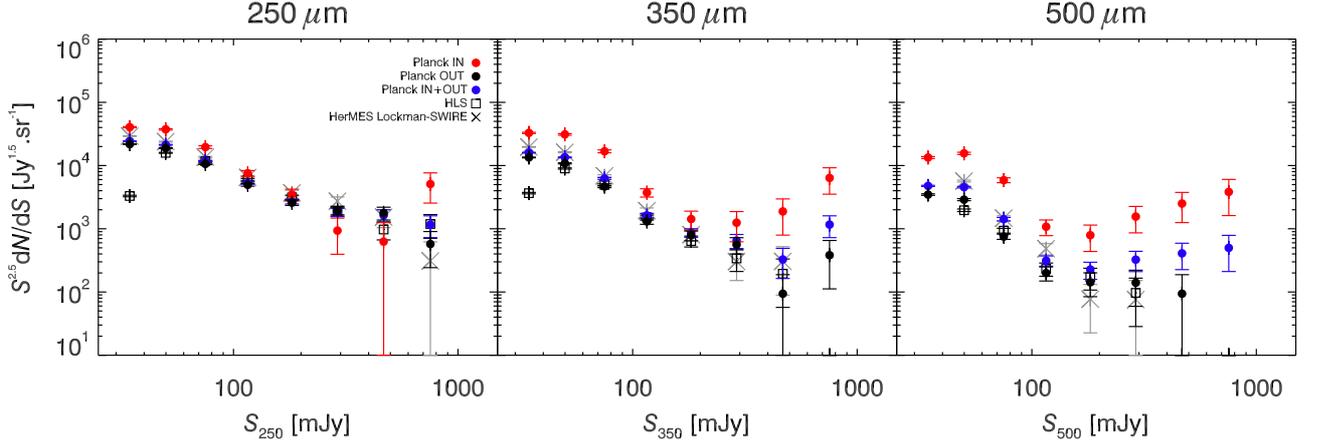}
  \caption{Differential Euclidean-normalized number counts,
    $S^{2.5}dN/dS$, for various data sets (not corrected for
    flux-boosting or incompleteness) used to measure the relative
    behaviour. Symbols show: \Planck\ IN region (red circles, 228
    fields); \Planck\ OUT regions (black circles); \Planck\ whole
    field, i.e., IN+OUT (blue circles); sample of 535 $z<1$
    clusters of the HLS (open square); and HerMES Lockman SWIRE blank
    field (crosses).  Our SPIRE sources corresponding to the \Planck\
    IN regions clearly show an excess in number density at 350 and
    500$\,\mu$m, illustrating their red colour, and a potentially
    higher redshift than average. See Sect.~\ref{sect:counts} for
    details.}
  \label{fig:totalcounts} 
\end{figure*}

\subsection{Blind catalogues}
\label{sect:blind}
We detect blindly, i.e., independently at each frequency, the
sources using {\tt StarFinder} \citep{diolaiti2000}. We use Gaussian
point spread functions (PSFs) with FWHM of 18.15\arcsec\ for 250$\,\mu$m,
25.15\arcsec\ for 350$\,\mu$m, and 36.3\arcsec\ for
500$\,\mu$m\footnote{for nominal SPIRE map pixels of 6, 10, and
14\arcsec, respectively.}.  These individual band catalogues are used for
checking photometric accuracy and completeness, and to produce number count
estimates. The catalogues are then band-merged in order to quantify
the colours of the sources.

We employ Monte Carlo simulations to check our photometry at each
wavelength: injection of sources in the data, and blind extraction, in
order to measure the photometric accuracy
(Fig.~\ref{fig:photometricaccuracymc}). As expected, the photometric
accuracy is of the order of 10\,\% at flux densities larger than a few
tens of mJy, and decreases towards smaller flux densities closer to
our noise level (dominated by confusion) at around 30\,mJy.  The
completeness levels are also measured, and reported in
Fig.~\ref{fig:completeness} and Table~\ref{tab:80completeness}.

\subsection{Band-merged catalogues}
\label{sect:bandmerged}
We use the 350$\,\mu$m map-based catalogue from {\tt StarFinder} as
the input catalogue.  This wavelength has two advantages: first, its
angular resolution meets our need to identify the sources; and second,
it is consistent with the \Planck\ colour selection, which avoids the
many low-$z$ galaxies peaking at 250$\,\mu$m. As will be shown in
Sect.~\ref{sect:counts}, our sample has an excess in number at
350$\,\mu$m compared to reference samples drawn from the HerMES
Lockman SWIRE level 5 field or the low-redshift HLS cluster fields, in
agreement with the existence of an overdensity of SPIRE sources in the
\Planck\ beam.

Our band-merging procedure has three steps. First of all, we optimize
the measurement of the source position using the 250$\,\mu$m channel,
where available.  Secondly, we measure a preliminary flux density on
each source, which will serve as a prior to avoid unrealistic flux
measurements while deblending. Thirdly, we perform
spatially-simultaneous PSF-fitting and deblending at the best measured
positions with the newly determined prior flux densities as inputs for
{\tt FastPhot} \citep{bethermin2010a}. The details are described in
Appendix~\ref{sect:bandmergingappendix}.  This method provides better
matched statistics (more than 90\,\% identifications) than the blind
extraction method performed independently at each wavelength
(Sect.~\ref{sect:blind}, Table~\ref{tab:ratio_undetect} vs.
Table~\ref{tab:matches250_500}). We finally have a total of about 7100
SPIRE sources, of which about 2200 are located in the \Planck\ IN
regions (giving an average of about 10 SPIRE sources per \Planck\ IN
field).

%__________________________________________________________________ 
% TABLE: 80% COMPLETENESS
%__________________________________________________________________ 
\begin{table}[!t]
\begingroup
\newdimen\tblskip \tblskip=5pt
\caption{80\,\% completeness levels obtained from Monte Carlo source
injection in the SPIRE maps for different fields, in mJy.  Our \Planck\ fields
are denoted ``\Planck\ high-$z$''.}
\label{tab:80completeness}
\vskip -5mm
\footnotesize
\setbox\tablebox=\vbox{
 \newdimen\digitwidth
 \setbox0=\hbox{\rm 0}
 \digitwidth=\wd0
 \catcode`*=\active
 \def*{\kern\digitwidth}
 \newdimen\signwidth
 \setbox0=\hbox{+}
 \signwidth=\wd0
 \catcode`!=\active
 \def!{\kern\signwidth}
 \halign{\tabskip=0pt\hbox to 1.75in{#\leaderfil}\tabskip=1em&
 \hfil#\hfil\tabskip=1em&
 \hfil#\hfil\tabskip=1em&
 \hfil#\hfil\tabskip 0pt\cr
\noalign{\doubleline}
\omit& \multispan3\hfil 80\,\% Completeness\hfil\cr
\noalign{\vskip -5pt}
\omit& \multispan3\hrulefill\cr
\omit Field& 250$\,\mu$m& 350$\,\mu$m& 500$\,\mu$m\cr
\noalign{\vskip 4pt\hrule\vskip 6pt}
   \Planck\ high-$z$& 35.2& 37.0& 40.7\cr
HerMES Lockman-SWIRE& 35.0& 38.4& 42.7\cr
                 HLS& 49.3& 48.5& 54.3\cr
\noalign{\vskip 4pt\hrule\vskip 6pt}
}}
\endPlancktable
\endgroup
\end{table}

\section{Statistical Analysis of the Sample}
\label{sect:statanalysis}

\subsection{Number counts: significant excess of red sources}
\label{sect:counts}

We compute the differential Euclidean-normalized number counts
$S^{2.5}dN/dS$, with $S$ being the flux density at wavelength
$\lambda$, and $N$ the number of sources per steradian. The counts are
not corrected here for incompleteness, or for flux boosting (Eddington
bias), since we are interested in the relative behaviour between the
samples. The detected sources used here are extracted using the blind
technique (Sect.~\ref{sect:blind}). We cut the samples at $4\,\sigma$
for these counts ($\sigma$ values in Table~\ref{tab:1sigma}). We used
five different data sets to estimate the number counts:

\begin{itemize}
\item[$\bullet$] \Planck\ IN+OUT, our 228 \Planck\ entire fields,
  each covering about $20\arcmin\times20\arcmin$;
\item[$\bullet$] \Planck\ IN, our 228 \Planck\ fields, using only the
 central parts corresponding to the \Planck\ 50\,\% contour level
 (determined separately for each source), called the IN region;
\item[$\bullet$] \Planck\ OUT, our 228 \Planck\ fields, using only the part
  exterior to the \Planck\ 50\,\% level, called the OUT region;
\item[$\bullet$] HerMES Lockman SWIRE, HerMES level 5 field in Lockman
  \citep{oliver2010a}, covering $18.2\,{\rm deg}^2$;
\item[$\bullet$] HLS, 535 cluster-fields of \cite{egami2010}, five from
  the Herschel Lens Survey KPOT, 282 from OT1, and 248 from OT2,
  hereafter refered to as HLS.
\end{itemize}

The number counts of all these data sets are plotted in
Fig.~\ref{fig:totalcounts} and reported in
Appendix~\ref{sect:countstables} in
Tables~\ref{tab:number_counts_250}, \ref{tab:number_counts_350}, and
\ref{tab:number_counts_500}. We derive from the total number count the
following results.

\begin{enumerate}[(i)]

\item The counts measured in the entire \Planck\ fields (blue
  circles, meaning IN + OUT), HerMES Lockman, and HLS clusters, are
  generally compatible with each other at 250$\,\mu$m, except for a few
  points at large flux density, above 300\,mJy, where the numbers are
  small. This means that, on average, those fields do not show any
  strong deviations between them.

\item The observed counts (uncorrected for incompleteness) show the
  characteristic shape of the Eddington bias: a cutoff below about
  40\,mJy, an excess around 50\,mJy, and a behaviour compatible with
  the models at higher flux densities. It is beyond the scope of this
  paper to re-derive unbiased number counts
  \citep[e.g.,][]{glenn2010,oliver2010a,clements2010a}, as we only
  focus on the relative trends.

%__________________________________________________________________ 
% Figure Histo 350um OverDensities and Significance
%__________________________________________________________________ 
\begin{figure*}[!ht] 
   \centering 
  \includegraphics[width=0.49\textwidth]{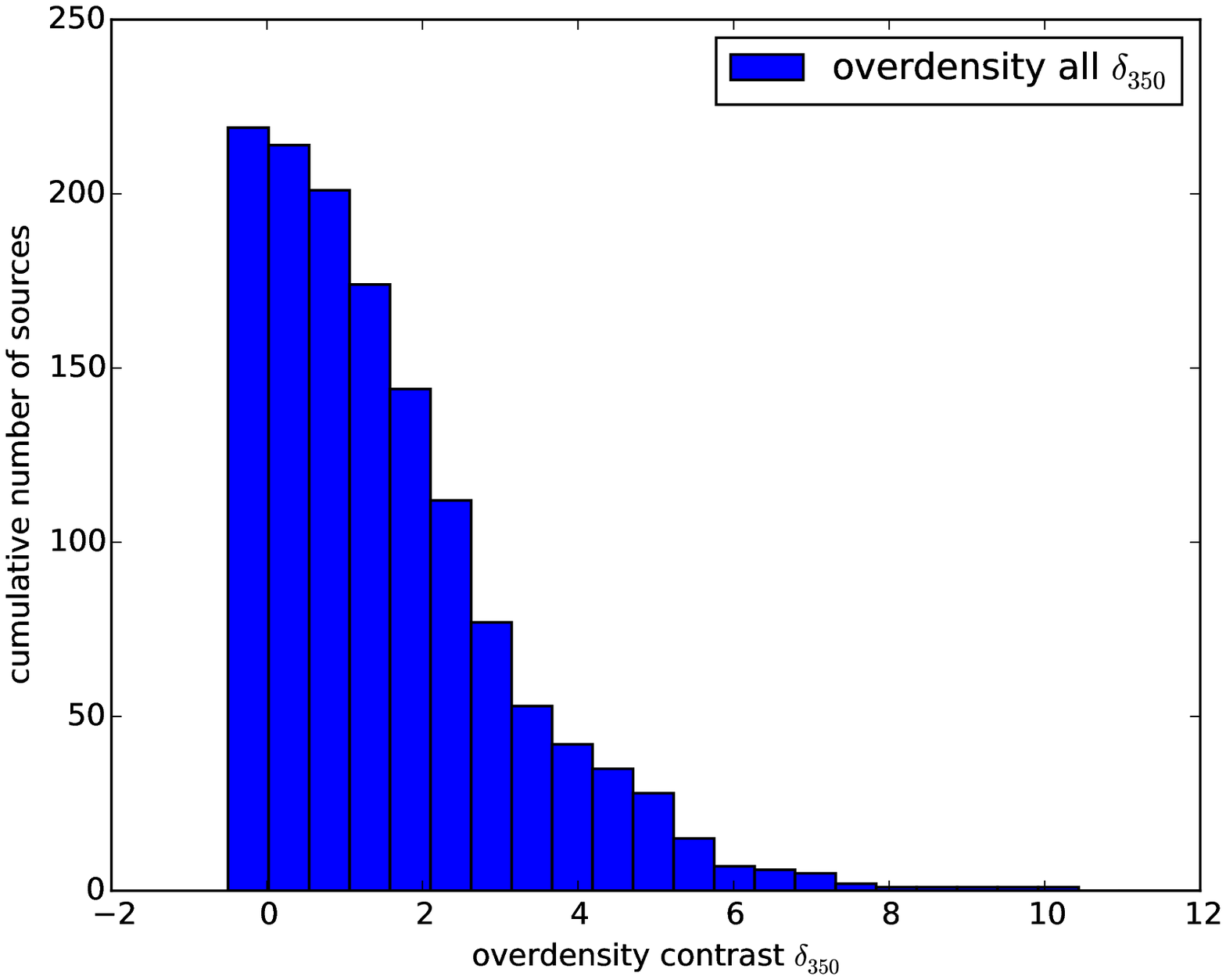}
  \includegraphics[width=0.49\textwidth]{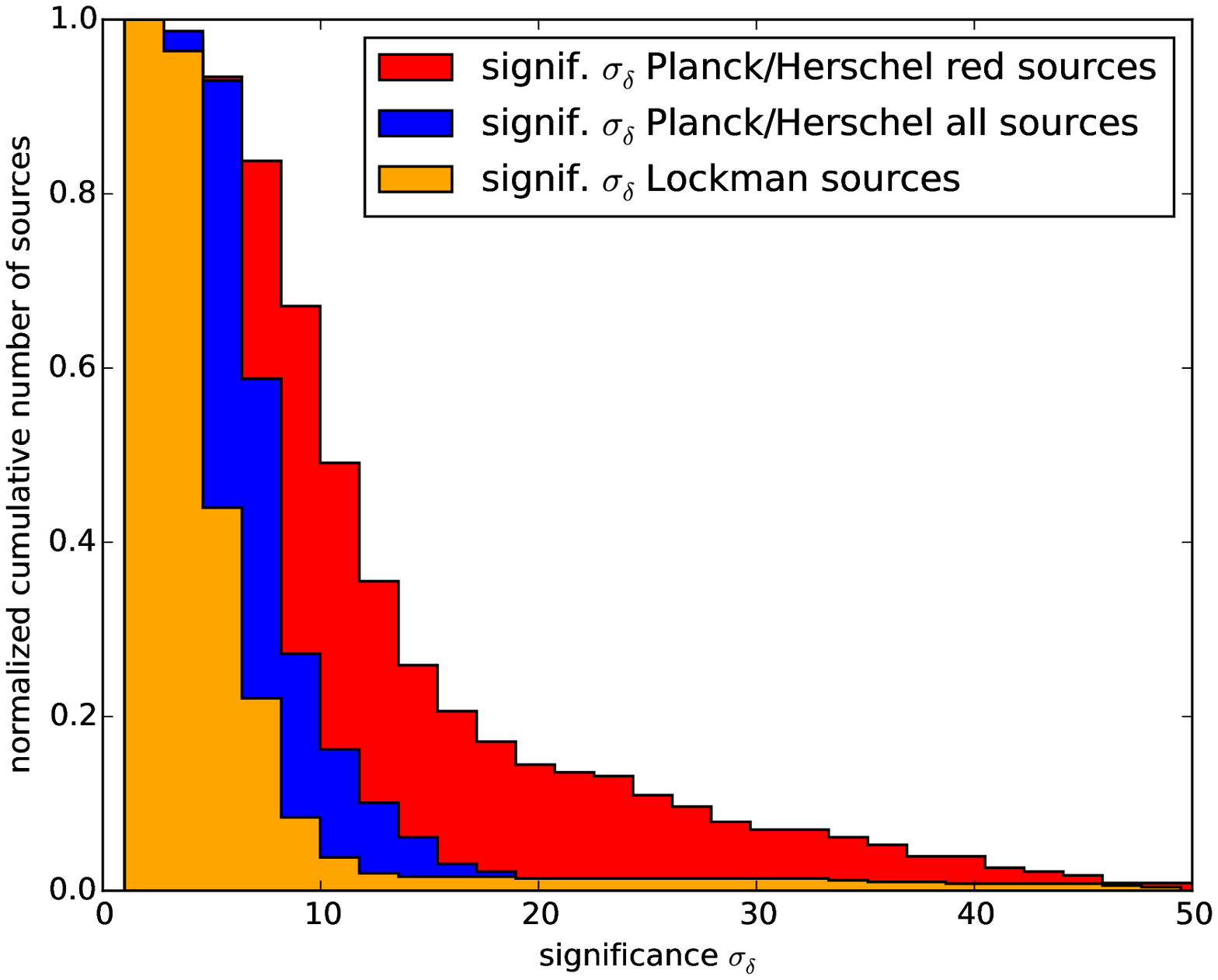}

  \caption{{\it Left}: Cumulative histogram of the overdensity
    contrast $\delta_{350}$ (blue) of each \Planck\ field (based on
    350$\,\mu$m SPIRE sources). Overdensities are fairly large, with
    59 fields having $\delta_{350} > 3$.  {\it Right}: Cumulative
    (normalized) statistical significance in $\sigma$ derived from the
    density maps. Blue represents all our SPIRE sources, red
    represents only redder SPIRE sources, defined by $S_{350}/S_{250}
    > 0.7$ and $S_{500}/S_{350} > 0.6$, and orange 500 random fields
    in Lockman. Most of our fields have a significance greater than
    $4\,\sigma$, and the significance is higher still for the redder
    sources. See Sect.~\ref{sect:overdensities} for details.}
\label{fig:overdensity350}
\end{figure*}

%__________________________________________________________________ 
% Figure colour counts
%__________________________________________________________________ 
\begin{figure*}[!ht] 
   \centering 
   \includegraphics[width=0.95\textwidth]{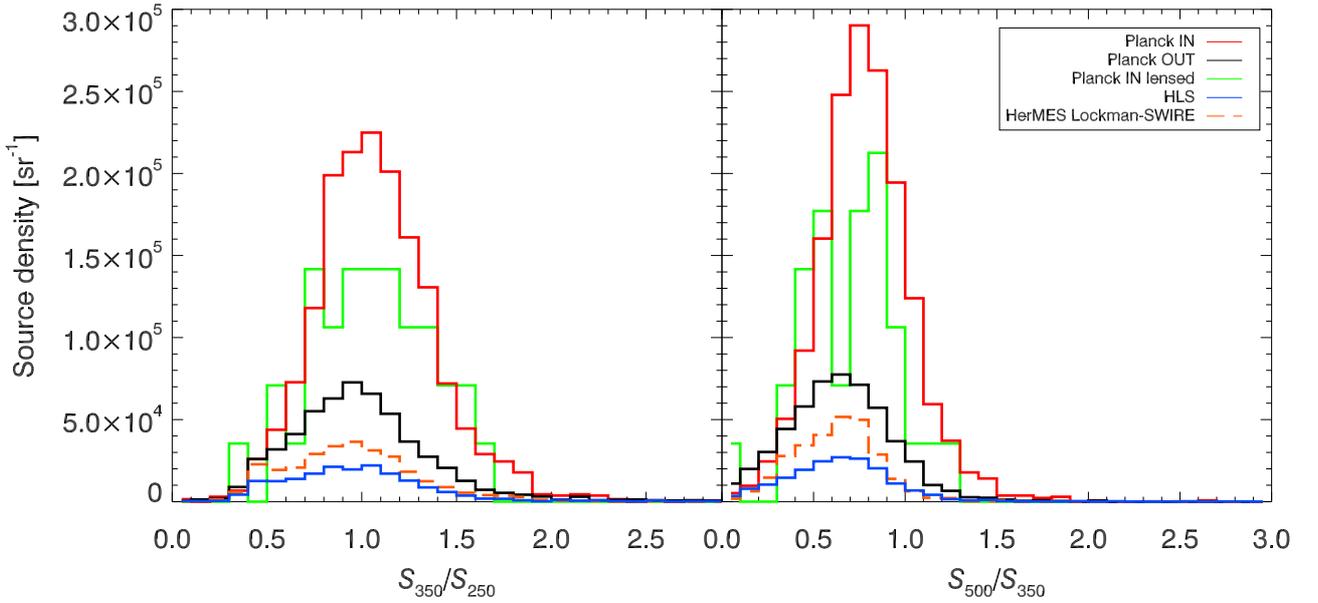}
   \caption{Colour counts: source surface density as a function of the
     SPIRE colour, $S_{350}/S_{250}$ (left), and
     $S_{500}/S_{350}$ (right). Histograms are: red solid line, \Planck\
     IN; black line, \Planck\ OUT; green, \Planck\ IN lensed fields only; blue
     line, $z<1$ HLS clusters; and orange dashes, Lockman SWIRE. The
     \Planck\ IN sources (total and/or lensed sources) show a much
     higher surface density than other samples, owing mainly to our all-sky
     search strategy.  See Sect.~\ref{sect:color} for details.}
  \label{fig:colorcount}
\end{figure*}

%__________________________________________________________________ 
% Figure c-c- a la Amblard: T et z
%__________________________________________________________________ 
\begin{figure*}[!ht] 
   \centering 
   \includegraphics[width=0.95\textwidth]{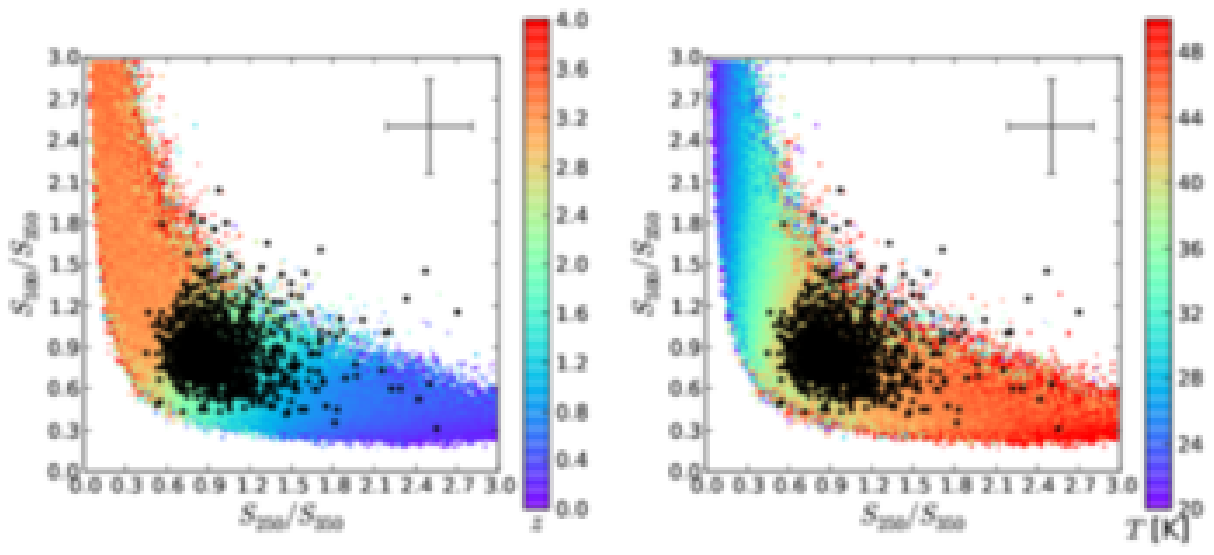}
   \caption{{\it Herschel}-SPIRE colour-colour diagram of IN sources
     (black dots) on top of redshifted blackbodies.  {\it Left}:
     colour codes redshift. {\it Right}: colour codes dust temperature
     in kelvin.  Although there is a degeneracy between redshift and
     temperature, the colours suggest a range of $z \sim 1.5$--3 for
     the SPIRE sources corresponding to the \Planck\ IN regions.  The
     upper right symbol gives the typical (median) error bars of the
     measured points. See Sect.~\ref{sect:color} for details.}
   \label{fig:c-c-amblard} 
\end{figure*}

\item The IN counts at large flux densities ($S>300\,$mJy)
  show a systematic excess at 350 and 500$\,\mu$m over the counts in
  the wide blank field (Lockman) as well as cluster fields. At
  300\,mJy, the overdensity factors are, respectively, 4.1 and 3.7 at
  350$\,\mu$m, and about 20 and 16 at 500$\,\mu$m.  This chromatic
  excess (i.e., larger excess at longer wavelengths) is
  consistent with the hypothesis of the presence of a population of
  high-redshift lensed candidates at $z=2$--4. We will show in
  Sect.~\ref{sect:lensed} that this is indeed the case.

\item The IN counts at 250$\,\mu$m are a bit higher (for
  $S_{250} < 100\,$mJy) than in Lockman, and are lower at larger flux
  densities. This means that the SPIRE counterparts of the \Planck\
  sources have, on average, counts that deviate little from blank
  fields at this wavelength.

\item The IN counts at 350$\,\mu$m for $S_{350} \simeq 50\,$mJy are
  a factor between 1.9 and 3.4 higher than in Lockman and in the
  cluster fields. This means that, on average, the SPIRE counterparts
  of the \Planck\ sources have a significant excess of 350$\,\mu$m
  sources compared to wide blank fields or $z<1$ HLS cluster
  fields. This is expected, given our \Planck\ selection criteria.

\item The IN counts at 500$\,\mu$m for $S_{500} \simeq 50\,$mJy
  are a factor of 2.7--8 higher than in the Lockman and cluster
  fields. As in item (v) above, this means that, on average, the SPIRE
  images of the \Planck\ source targets show a significant excess of
  500$\,\mu$m sources compared to wide blank field or $z<1$ HLS cluster
  fields.
 
\item The OUT counts are compatible with the wide blank
  extragalactic fields, as well as the cluster fields. Our OUT
  zones can thus also be used as a proxy for the same statistics in
  blank fields.
\end{enumerate}

As a conclusion, the SPIRE observations of the \Planck\ fields reveal
``red'' sources. The SPIRE images exhibit a significant excess of 350
and 500$\,\mu$m sources in number density compared with wide blank
fields (HerMES Lockman SWIRE of the same depth) or fields targeting
$z<1$ galaxy clusters (HLS). This significant excess should be
expected given the \Planck\ colour selection, and is now demonstrated
with secure SPIRE detections. It is therefore clear that there is no
significant contamination by cirrus confusion in our \Planck\ sample,
and that where there is a \Planck\ high-$z$ candidate, {\it Herschel\/}
detects galaxies.

\subsection{Overdensities}
\label{sect:overdensities}
We compute the dimensionless overdensity contrast $\delta_{\lambda}$
of our fields at wavelength $\lambda$ via
\begin{equation}
\delta_{\lambda} = \frac{\rho_{{\rm IN}} - \rho_{\rm OUT}}{\rho_{\rm OUT}},
\end{equation}
where $\rho_{{\rm IN}}$ is the surface density of SPIRE sources in the
\Planck\ IN region, and $\rho_{\rm OUT}$ is the mean surface density
of SPIRE sources computed in the \Planck\ OUT region, at SPIRE
wavelength $\lambda$.  We have already shown (Sect.~\ref{sect:counts}
and Fig.~\ref{fig:totalcounts}) that the OUT region has a density
equivalent to that of blind surveys, and is thus a good estimate of
$\bar{\rho}$, the mean surface density.  To reduce the Poisson noise, we
use the counts from {\it all\/} the OUT regions.

The overdensity contrasts $\delta_{\lambda}$ extend up to 10, 12, and
50 at 250, 350, and 500$\,\mu$m, respectively, with a median
overdensity $\delta_{\lambda}$ of $\delta_{250} = 0.9$, $\delta_{350}
= 2.1$, and $\delta_{500} = 5.0$.  This means that our \Planck\ IN
regions have an excess of SPIRE sources. Indeed, there are 50 fields
with $\delta_{500} > 10$, and 129 with $\delta_{500} > 4$ (with
significance levels always higher than $4\,\sigma_{\delta}$, see
below). At 350$\,\mu$m, there are 19 fields with $\delta_{350} > 5$,
37 fields with $\delta_{350} > 4$ and 59 fields with $\delta_{350} >
3$, as shown in Fig.~\ref{fig:overdensity350} (left panel, in
blue). In Appendix \ref{sect:overdensities_using_akde}, we also use
the densities measured with AKDE to estimate the overdensities.

How significant are these overdensity contrasts?  To quantify this we
compute the mean density field using the AKDE algorithm (adaptative
kernel density estimator, see \citealt{valtchanov2013} and also
\citealt{pisani1996,ferdosi2011}).  The principle is to generate a
two-dimensional density field based on the positions of the sources from
a catalogue, filtered (smoothed) according to the source surface
density. From this smoothed field, we compute the standard deviation
and hence we derive the significance $\sigma_{\delta}$ of the
overdensity. We also run 1000 Monte Carlo runs to get a better
estimate of the scatter on $\sigma_{\delta}$ (by creating AKDE density
maps using random source positions -- but otherwise using the real
catalogs of each field -- and measuring the RMS over those 228000
realizations). All our fields show overdensities larger than
$1.8\,\sigma_{\delta}$, with a median of $7\,\sigma_{\delta}$
(Fig.~\ref{fig:overdensity350} right panel).  The typical number of
sources in an overdensity is around 10, but with a fairly wide
scatter.

We can then choose to select the reddest sources, this being a
possible signature of a higher redshift or a colder dust
temperature. We define the SPIRE red sources with the following cuts
in colour based on the distributions shown in
Fig.~\ref{fig:colorcount}: $S_{350}/S_{250} > 0.7$; and
$S_{500}/S_{350} > 0.6$.  When selecting only the red sources, the
overdensity significance increases: The mean is now
$12\,\sigma_{\delta}$ and the median $9\,\sigma_{\delta}$ .  We have
50\,\% of the sample at $10\,\sigma_{\delta}$ or more (when selecting
the red sources), which is more than a factor of 3 larger than when
selecting all SPIRE sources.  23\% of the sample is above
$15\,\sigma_{\delta}$ (Fig.~\ref{fig:overdensity350} right panel),
i.e., 51 fields.

%__________________________________________________________________ 
% Figure ID code  of Identification class
%__________________________________________________________________ 
\begin{figure}[!ht] 
   \centering 
  \includegraphics[width=0.48\textwidth]{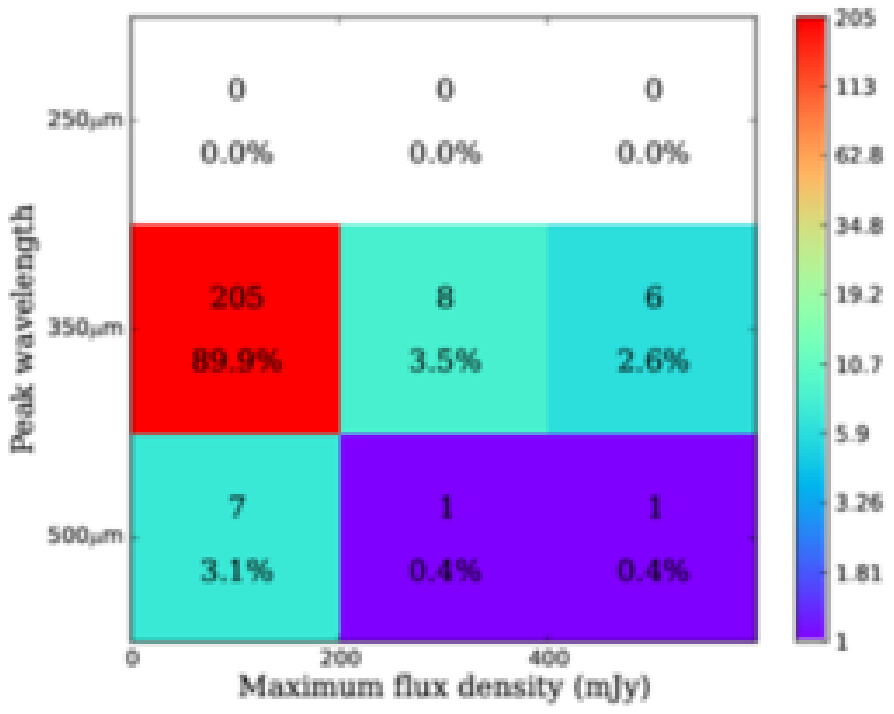} 
  \caption{Classification of 228 \Planck\ fields with {\it
      Herschel}-SPIRE. The $x$-axis represents the maximum flux of the
    SPIRE source within the IN region: lensed source candidates are
    selected if they fall above 400\,mJy (i.e., the right column,
    seven sources).  The $y$-axis represents the wavelength at which the
    brightest SPIRE source peaks in the IN region. Overdensities are
    thus selected in the lower left four cells. The colour represents
    the number of fields in each cell (as shown in the colour
    bar). the numbers in the cells are the number of fields and percentage in
    each cell. Our sample is thus dominated by overdensities peaking
    at 350$\,\mu$m. See Sect.~\ref{sect:classification} for details.}
   \label{fig:pie_chart} 
\end{figure}

These high significance levels can be contrasted with the mean
$\delta_{500} \simeq 0.25$ obtained by \cite{rigby2014} at the
locations of 26 known protoclusters around very powerful
radiogalaxies, drawn from the list of 178 radiogalaxies at $z>2$ of
\cite{miley2008}. They can also be compared to the few similar
examples in \cite{clements2014}, with at most 4.7$\,\sigma_{\delta}$
at 350$\,\mu$m. We show in Fig.~\ref{fig:overdensity350} right panel,
in orange, the cumulative normalized significance $\sigma_{\delta}$ of
500 random positions in the Lockman field; Using this test sample
illustrates the high significance of the overdensities of our sample.

Examples of some overdensities are shown in Appendix~\ref{sect:gallery},
where we present a gallery of representative SPIRE data.

\subsection{Colours of the sources}
\label{sect:color}
Using the band-merged catalogue (Sect.~\ref{sect:bandmerged}), we can
derive the colours of the sources, i.e., the ratios of the
observed flux densities $S_{500}/S_{350}$ vs.\ $S_{250}/S_{350}$. In
this colour-colour space, redder sources will have higher
$S_{500}/S_{350}$ ratios and lower $S_{250}/S_{350}$ ratios, and will
tend to lie in the lower left part of the diagram.

We provide in Fig.~\ref{fig:colorcount} the source surface density
histograms for the two SPIRE colours: $S_{350}/S_{250}$ and
$S_{500}/S_{350}$ (red, \Planck\ IN; black, \Planck\ OUT; green, only
the lensed fields; blue, HLS; and dashes, HerMES). The IN sources show
three times larger surface densites in $S_{350}/S_{250}$, and four
times larger surface densities in $S_{500}/S_{350}$ than OUT and
Lockman sources. The IN source distribution peaks at much higher
surface density than any other sample, suggesting that our sample is
dominated by red and overdense SPIRE sources.

%__________________________________________________________________ 
% Figure G256 Spitzer
%__________________________________________________________________ 
\begin{figure}[!ht] 
   \centering 
  \includegraphics[width=0.41\textwidth]{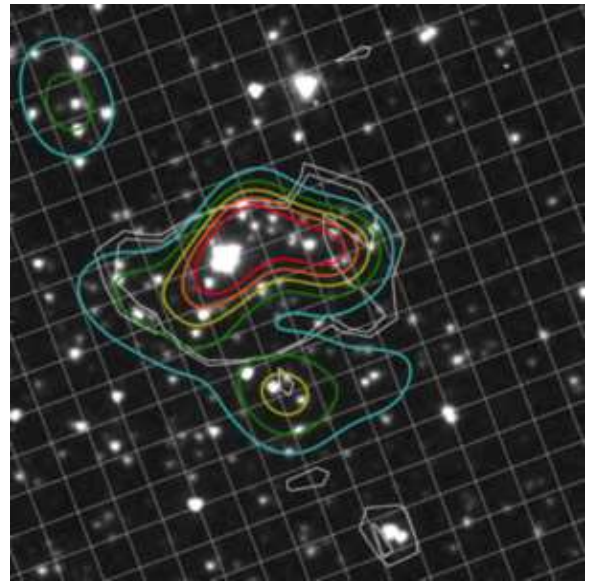}
  \caption{A high-$z$ cluster candidate observed by \Planck, {\it
      Herschel}, and {\it Spitzer}-IRAC (image covering $0.5\arcmin
    \times 0.3\arcmin$). We show the IRAC channel 1 (3.6$\,\mu$m)
    image, with SPIRE 350$\,\mu$m white contours overlaid. Colour
    contours represent statistical significance of the local
    overdensity, from light blue to red: 3, 4, 5, 6, and
    7$\,\sigma_{\delta}$. See Sect.~\ref{sect:validover} for details.}
   \label{fig:g256} 
\end{figure}

Following the approach of \cite{amblard2010}, we generate the SED of
$10^6$ modified blackbodies with 10\,\% Gaussian noise and explore
three parameters: the blackbody temperature ($T$) in the range 10--60\,K;
the emissivity ($\beta$) in the range 0--2; and the redshift
for $z=0$--5.  For each set of parameters we convert the fixed,
observed SPIRE wavelengths into rest frame wavelength (at redshift
$z$, thus varying with $z$) using: $\lambda_{\rm rest}=\lambda_{{\rm
    SPIRE}}/(1+z)$. We then calculate the flux at each wavelength
using $\lambda_{\rm rest}$ and compute the colours $S_{250}/S_{350}$
and $S_{500}/S_{350}$ for each set of parameters.
Fig.~\ref{fig:c-c-amblard} shows those realizations, colour-coded in
redshift (deep purple for $z=0$, up to red for $z=4$). Our SPIRE IN
sources are shown in black. They predominantly fall in the redshift
region corresponding to $z \sim 1.5-3$. We note, however, that the
redshift-temperature degeneracy is present in this suggestive result
\citep{amblard2010,pope2010,greve2012}, and will be discussed in
Sect.~\ref{sect:dustsfr}.

\subsection{Classification of the sources}
\label{sect:classification}

We classify the {\it Herschel\/} fields in an automated way and for
preliminary analysis into two main categories: (1) {\it
  overdensities}, appearing as clustered sources, and (2) {\it lensed
  candidates}, appearing as a bright single source as a counterpart of
the \Planck\ source.  To make this classification, we measure within
the IN \Planck\ region the flux density and colour of the brightest
SPIRE source.

We sort the fields using two criteria: (1) the SPIRE channel at which
the source has its maximum flux density; and (2) its flux density at
350$\,\mu$m. We will use this two-parameter space to classify the
sources. This classification provides the first set of information about
the sources, although a definitive classification will require
follow-up data to confirm the nature of the sources.

In this space, the lensed candidates will populate the ``red'' and
``bright'' areas (typically 350 or 500$\,\mu$m peakers, and $S_{\nu} >
400\,$mJy).  This does not necessarily mean that all the sources in
this area will be strongly gravitationally lensed (although this is
confirmed by further follow-up, see following sections).

The overdensity candidates will be located in a different area of this
two-parameter space than the lensed candidates. They will tend to populate
the fainter end (typically $S_{\nu} < 200\,$mJy) and will have red
colours (typically 350 or 500$\,\mu$m peakers).

%__________________________________________________________________ 
% Figure G106.8 SANTOS z=1.58 CLUSTER
%__________________________________________________________________ 
\begin{figure*}[!ht] 
   \centering 
  \includegraphics[width=0.40\textwidth]{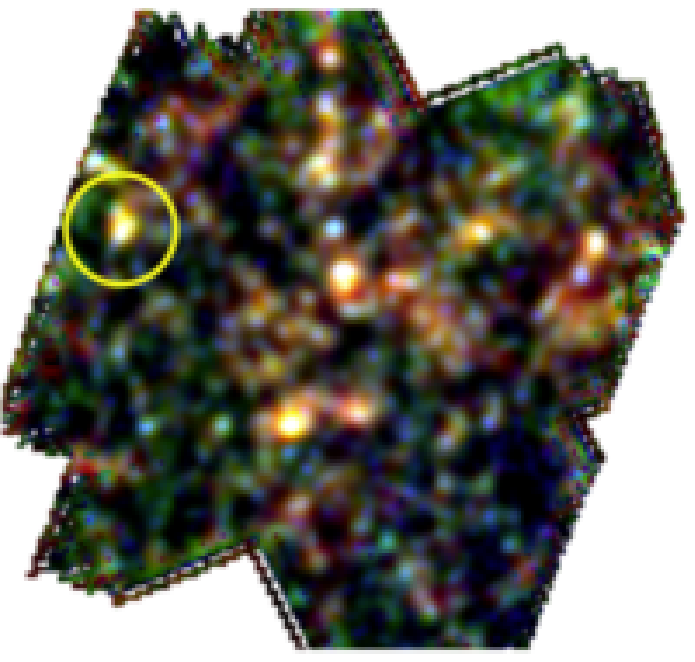} 
  \includegraphics[width=0.47\textwidth]{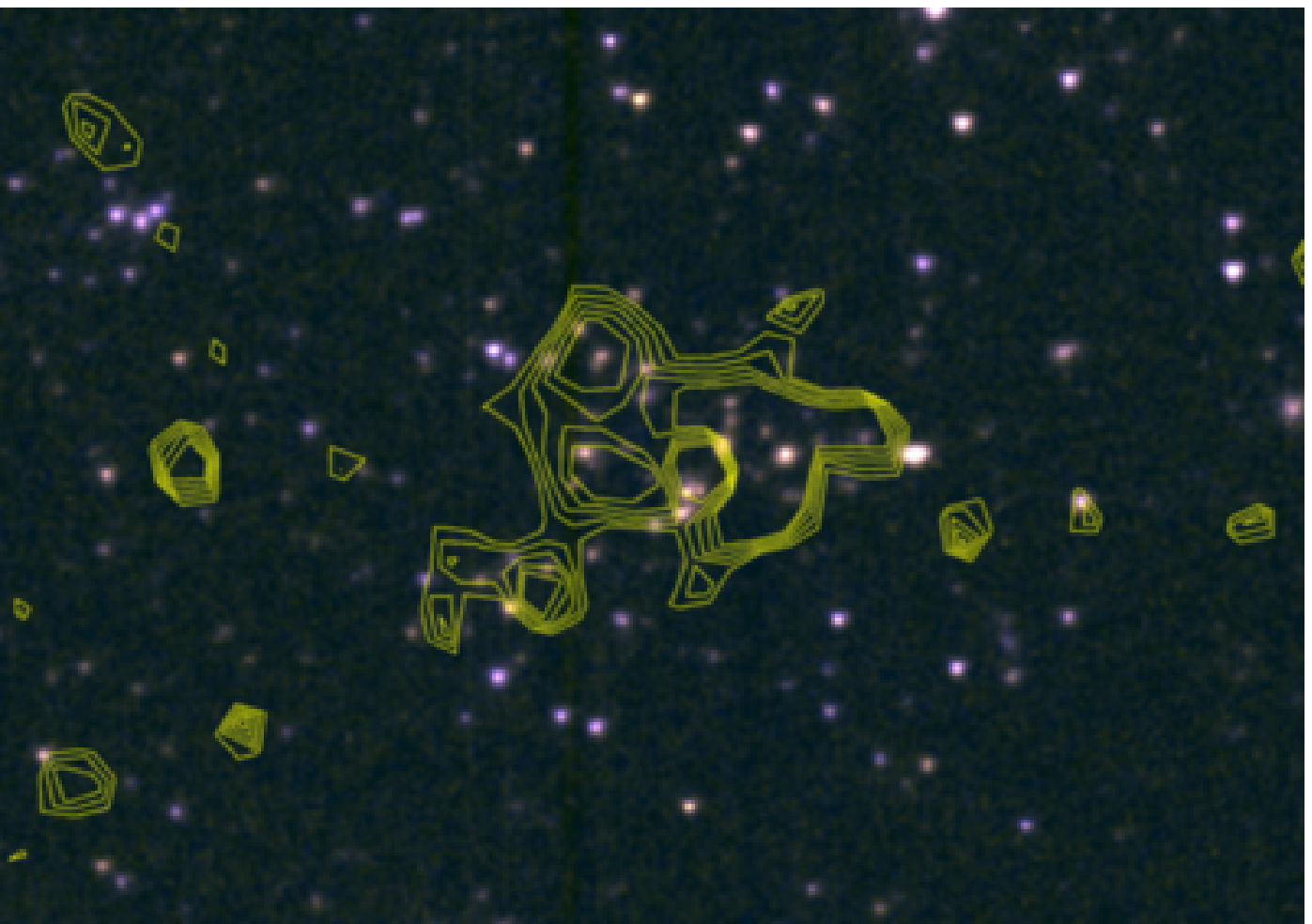} 
  \caption{{\it Left}: SPIRE 3-colour image of one \Planck\ high-$z$
    overdensity candidate (located in the centre), in an image
    covering about $20\arcmin\times20\arcmin$. We note the bluer
    source on the left (circled), which corresponds to the galaxy
    cluster shown on the right.  {\it Right}: a confirmed $z=1.58$ galaxy
    cluster XMMU J0044.0-2033 \citep{santos2011} which lies close to
    our \Planck\ high-$z$ source. The background image is a two colour
    composite of {\it Spitzer}-IRAC channel~1 (3.6$\,\mu$m) and channel
    2 (4.5$\,\mu$m), covering about $3\arcmin\times2\arcmin$. The contours
    are {\it Herschel}-SPIRE 350$\,\mu$m. See Sect.~\ref{sect:validover}
    for details.}
   \label{fig:santos} 
\end{figure*} 

We do not expect many 250$\,\mu$m peakers because of the \Planck\ colour
selection, for which there is a bias towards redder colours and thus
potentially higher-redshift sources.

Fig.~\ref{fig:pie_chart} summarizes the classes of SPIRE
identifications of the \Planck\ high-$z$ (meaning $z>1.5$) candidates,
based on those criteria. We show that the vast majority of our sample
is composed of overdensities of 350$\,\mu$m peakers. The number of
lensed candidates is rather small in comparison, amounting to seven
sources based on this criterion. All are confirmed to be strongly
gravitationally lensed galaxies \citep{canameras2015}. We will show
later that visual inspection led us to discover a few more lensed
sources.

The dashed lines in Fig.~\ref{fig:colorcount} show the colour vs.\
surface density of the lensed candidate population, whose red colours
are confirmed.

Finally, we derived number counts (as in Fig.~\ref{fig:totalcounts})
but now by separating the overdensities from the lensed sources
that dominate the counts at large flux densities, typically $S >
250\,$mJy (shown in Appendix~\ref{sect:countspertype} in
Fig. ~\ref{fig:totalcounts_lens_over}).  This separation suggests that
the excess at large flux densities in the number counts of our sample
is due to the presence of bright lensed sources, compared to reference
samples (HLS, HerMES).

%-----------------------------------------------------------------------
%-----------------------------------------------------------------------
\section{Strongly gravitationally lensed source candidates}
\label{sect:lensed}

\subsection{Validation of existing lensed sources}

\cite{negrello2007} and \cite{bethermin2012} predicted that a small,
but significant fraction of very bright high-redshift ($z>2$)
submillimetre galaxies are strongly gravitationally lensed, dusty
starbursts in the early Universe. \cite{negrello2010} presented
observational evidence of these predictions. In addition, \cite{fu2012}
confirmed the nature of the source H-ATLAS J114637.9-001132 as a
strongly gravitationally lensed galaxy at $z=3.3$; this source was
part of the first release of the \Planck\ ERCSC Catalogue
\citep{planck2011-1.10,planck2011-1.10sup}, and fell fortuitously into
the H-ATLAS survey field \citep[see also][]{herranz2013}.
Another independent confirmation that
\Planck\ sources can be strongly gravitationally lensed came
from the source HLS~J091828.6$+$514223, confirmed as a bright, $z=5.2$
gravitationally lensed galaxy behind the massive intermediate-redshift
galaxy cluster Abell~773 \citep{combes2012}, as part of the
Herschel Lensing Survey \citep{egami2010}. This source was
independently found in our survey and is part of the {\it Herschel}/SPIRE
OT2 selection.  It is excluded from our sample
to satisfy the condition of non-redundancy.

\subsection{Previously unknown gravitationally lensed sources}

In the absence of extensive follow-up, it is challenging to
distinguish between single or multiple strongly gravitationally lensed
galaxies behind the same foreground structure and overdensities of
intrinsically bright submillimetre galaxies, for all but the brightest
gravitationally lensed sources. Moreover, at flux densities of about
$S_{350}=250\,$mJy and above, isolated SPIRE point sources may turn
out to be associations of multiple FIR galaxies when observed at
higher spatial resolution \citep{ivison2013a}.

%__________________________________________________________________ 
% Figure stack SPIRE
%__________________________________________________________________ 
\begin{figure*}[!ht] 
   \centering 
  \includegraphics[width=0.70\textwidth]{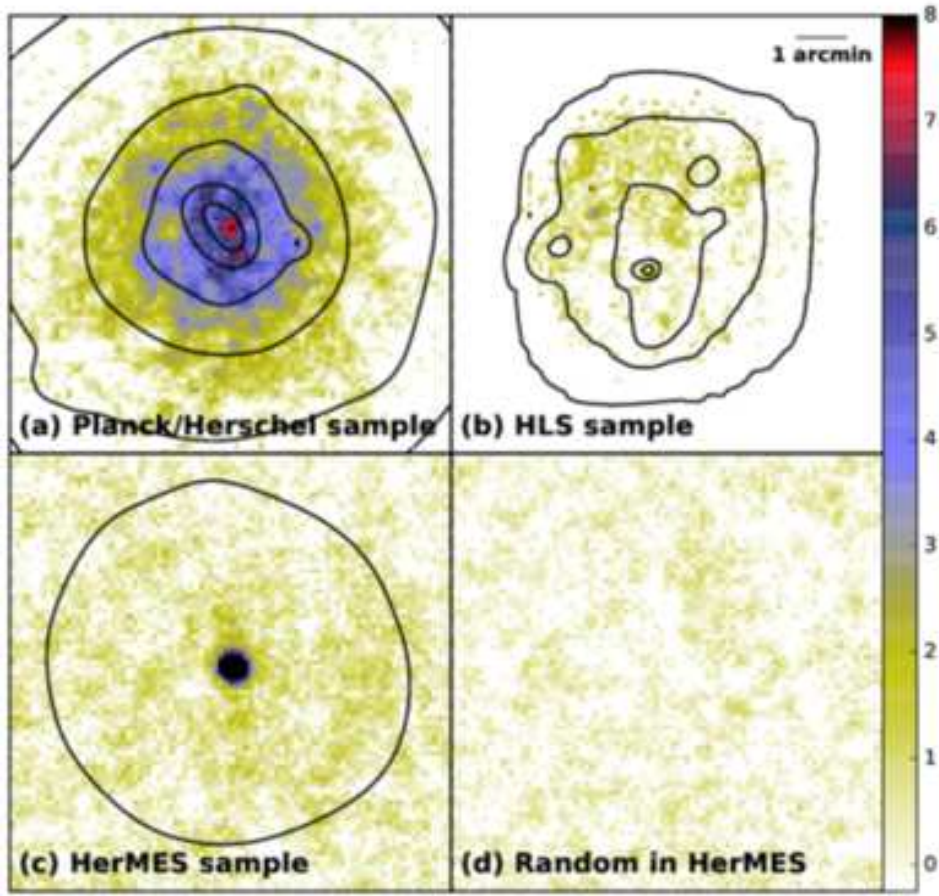}
  \caption{Stacks of SPIRE 350$\,\mu$m data ($8.7\arcmin \times
    8.7\arcmin$): (a) 220 \Planck\ {\it Herschel\/} fields; (b) 278
    HLS clusters; (c) 500 sources in the HerMES Lockman field, peaking
    at 350$\,\mu$m; and (d) 500 random positions in the HerMES Lockman
    field. Black contours show the density of red sources (using
    AKDE). Our \Planck\ fields clearly exhibit a high significance of
    extended (a few arcminutes) submm emission, due to the presence of
    many red point sources, not seen in the HLS or HerMES sources. See
    Sect.~\ref{sect:stacking} for details.}
   \label{fig:stack_spire} 
\end{figure*}

As a first step towards identifying the gravitational lensed
candidates in our sample (Sect.~\ref{sect:classification}), we
therefore focused on those targets where SPIRE shows only a single,
very bright source, with the typical FIR colours of high-redshift
($z>2$) galaxies associated within one \Planck\ beam. We thus
identified seven isolated SPIRE point sources, as described in
Sect.~\ref{sect:classification}, plus five others at slightly fainter
flux densities.  All have peak flux densities at 350$\,\mu$m,
including 11 with $S_{350}=300$--1120\,mJy \citep{canameras2015}. Six
of these galaxies were taken from the PCCS, the remaining six
originate from the sample of \cite{planck_collaboration2015}.
Although the initial selection of lensed candidates was carried out by
eye upon the reception of the SPIRE imaging, seven of these targets
were also identified with our automatic classification
(Sect.~\ref{sect:classification}). We used the IRAM 30-m telescope to
obtain firm spectroscopic redshifts via a blind CO line survey with
the wide-band receiver EMIR. We identify 2$-$6 lines per source, which
confirms they are at redshifts $z=2.2$--3.6. Interferometry obtained
with the IRAM Plateau de Bure interferometer and the Submillimeter
Array, as well as several empirical calibrations based on FIR
luminosity and dust temperature \citep[following][]{harris2012}, and
CO line luminosity and line width \citep[following][]{greve2012},
demonstrate that these are indeed strongly gravitationally lensed
galaxies, amongst the brightest on the submm sky
\citep{canameras2015}.
  
%-----------------------------------------------------------------------
%-----------------------------------------------------------------------
\section{Candidate high-$z$ overdensities}
\label{sect:highz}

Without rejecting the possibility that the observed overdensities of
red SPIRE sources could be chance alignments of structures giving
coherent colours \citep[e.g.,][]{chiang2013}, we can ponder the nature
of those overdensities. Could they be high-redshift intensively
star-forming galaxy proto-clusters? Indeed, recent studies, such as
\cite{gobat2011}, \cite{santos2011,santos2013,santos2014}, and
\cite{clements2014}
confirm the presence of high redshift ($z>1.5$) galaxy clusters
emitting enough energy in the submillimetre to be detected. Those
previously detected clusters are, however, in a different state of
evolution than our candidates. Many of the confirmed clusters exhibit
X-ray emission, suggesting already mature and massive clusters. The
following sections investigate different aspects of our sample, which
is uniquely selected by strong submm emission over the whole sky,
enabling us to unveil a rare population.

\subsection{First confirmations}
\label{sect:validover}

It is beyond the scope of this paper to summarize all the follow-up
observations conducted so far, but we give here three highlights,
which will be discussed in more detail in subsequent papers.

Our sample contains a source from {\it Herschel\/} OT1 that was followed
up early as a pilot programme. Using CFHT and {\it Spitzer\/} imaging
data, and VLT and Keck spectroscopy, a structure was identified at
$z\sim1.7-2.0$ (\textcolor{blue}{Flores-Cacho et
  al., 2014, in prep.}). By fitting a modified blackbody and fixing
redshifts at respectively 1.7 and 2.0, we find an average dust
temperature for the sources of, respectively, 27K and 35K.

%__________________________________________________________________ 
% Histo z for various TDust
%__________________________________________________________________ 
\begin{figure}[!ht] 
   \centering 
  \includegraphics[width=0.49\textwidth]{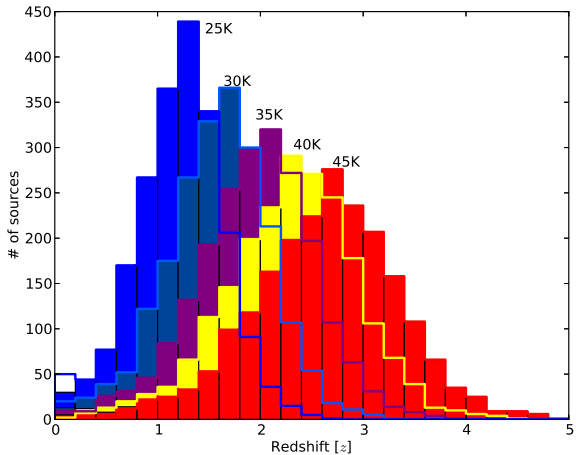}
  \caption{SPIRE photometric redshift distribution of the roughly 2200
    SPIRE sources in the IN regions, as a function of the fixed
    assumption for dust temperature: $T_{\rm d}$ = 25, 30, 35, 40, and
    45\,K (from left to right). See Sect.~\ref{sect:dustsfr} for
    details.}
   \label{fig:histoztdust} 
\end{figure}

Another field has been imaged by {\it Spitzer\/} as part of our GO-9
programme -- see Fig.~\ref{fig:g256} for the IRAC 3.6$\,\mu$m image
with the SPIRE 350$\,\mu$m contour. The colour ratios of IRAC flux
densities 3.6$\,\mu$m/4.5$\,\mu$m of the sources detected in the
overdensity exhibit red colours, indicating their probable
high-redshift nature \citep{papovich2008}.  The overdensity of IRAC
3.6$\,\mu$m sources has a statistical significance of
7$\,\sigma_{\delta}$ (\textcolor{blue}{Martinache et al., 2014, in
  prep.}). While not yet confirmed as a proto-cluster or cluster
through spectroscopy, this structure, seen with {\it Spitzer\/} at
near-infrared wavelengths, supports the hypothesis of strongly
clustered highly star-forming infrared sources.

Finally, we have also found one SPIRE field with overlapping data from
the literature. The observations from \cite{santos2011} confirm the
presence of a $z=1.58$ cluster using optical/NIR spectroscopy.  This
particular high-$z$ cluster has also been detected in X-rays, leading
to a total mass estimate of about 3--$5\times10^{14}\,{\rm M}_{\odot}$. It is
located a few arcminutes from one of our targets
(Fig.~\ref{fig:santos} right), so is not formally an identification of
our \Planck\ source, but rather a coincidence. While not connected to
our target source here, this cluster can nevertheless bring useful
information. The SPIRE colours of this cluster are bluer than our
sources. Since our target shows redder SPIRE colours, this suggests it
has a higher redshift or a cooler dust temperature.

\subsection{Large clustering of our sample revealed with stacking}
\label{sect:stacking}
We can investigate if the observed overdensities have different
clustering properties than two test samples: (1) the HLS massive
$z<1$ clusters; and (2) 350$\,\mu$m peaker sources in the HerMES
Lockman field brighter than 50\,mJy. To do so, we use a stacking
analysis \citep[see, e.g.,][]{montier2005,dole2006,braglia2011}, here
applying the method of \cite{bethermin2010}. We stack the following SPIRE
350$\,\mu$m data: (a) the 220 overdensities of our sample; (b) 278 HLS
clusters; (c) 500 bright 350$\,\mu$m peaker sources in the Lockman
field; and (d) 500 random positions in the HerMES Lockman field as a
null test. The stacks are presented in Fig.~\ref{fig:stack_spire}.

We can see that the \Planck\ fields show clear and significant
extended (a few arcminutes) emission due to the clustering of bright
submm sources. This kind of extended emission of clustered submm
sources is not observed in the HLS clusters, nor around 350$\,\mu$m
peaking HerMES sources. The random stack validates the absence of a
systematic effect in stacking
\citep{dole2006,bethermin2010,viero2013,planck2013-p13}. This proves
that our sources are of a different nature than mature HLS clusters or
average HerMES submillimetre sources.

We overplot in Fig.~\ref{fig:stack_spire} the contours in density (see
Sect.~\ref{sect:overdensities}) of the SPIRE red sources (defined in
Sect.~\ref{sect:overdensities}). The stacks indicate that the \Planck\
sample exhibits strong overdensities of red SPIRE sources.  By
contrast, the HLS sample shows a smooth and weak density of red
sources (except for the presence of some point sources -- the
background lensed galaxies). As expected, the HerMES sample shows a
stack consistent with the SPIRE PSF, while the random sample is
consistent with noise.

\subsection{Dust temperatures, photometric redshifts,
  luminosities, and star formation rates}
\label{sect:dustsfr}

With the hypothesis that our overdensities are actually
gravitationally bound structures, i.e., clusters of
star-forming, dusty galaxies, the overdensity colours
(Fig.~\ref{fig:colorcount}) suggest a peak redshift around $z=1.5$--3
(Fig.~\ref{fig:c-c-amblard}).  Using a modified blackbody fit (with
$\beta=1.5$) and fixing the dust temperatures to $T_{\rm d}=25$, 30,
35, 40, and 45\,K, we obtain the redshift distributions shown in
Fig.~\ref{fig:histoztdust} for the roughly 2200 sources found in the
IN regions for these overdensities. For $T_{\rm d}=35\,$K, the
distribution peaks at $z=2$. For higher dust temperatures, the peak
of the distribution is shifted towards higher redshifts, since dust
temperature and redshift are degenerate for a modified blackbody,
where $T_{\rm d}/(1+z)=$constant.

\begin{figure}[!ht] 
   \centering 
  \includegraphics[width=0.49\textwidth]{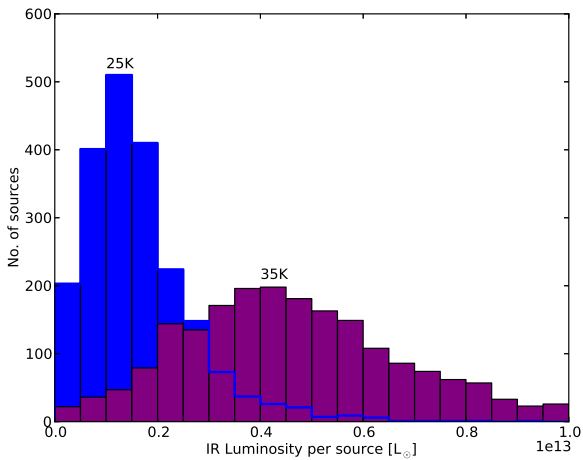}
  \caption{Infrared luminosities ($L_{\rm IR}$) of the approximately 2200
    SPIRE sources in the IN regions, as a function of the fixed dust
    temperature: 35\,K (purple, right histogram), and also shown for
    illustration 25\,K (blue, left histogram). See
    Sect.~\ref{sect:dustsfr} for details.}
   \label{fig:histolir} 
\end{figure}

\begin{figure}[!ht] 
   \centering 
  \includegraphics[width=0.49\textwidth]{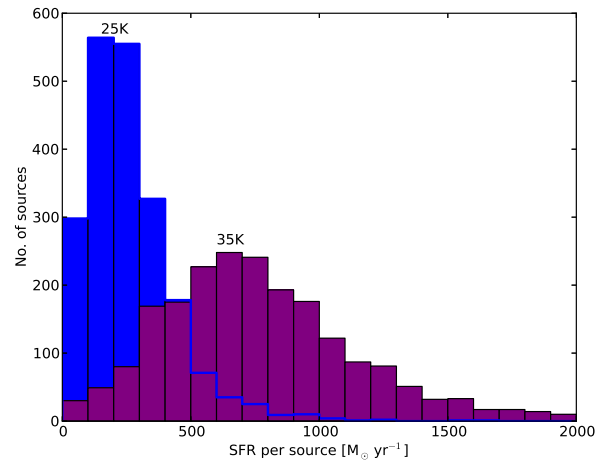}
  \caption{Star formation rate (SFR) for individual SPIRE sources in
    the overdensity sample (i.e., in the \Planck\ IN region). For
    $T_{\rm d}=35\,$K, a dust temperature favoured by the literature,
    (purple, right histogram) the distribution peaks at
    $700\,{\rm M}_{\odot}\,{\rm yr}^{-1}$.  We also show for illustration
    $T_{\rm d}=25\,$K (blue left histogram peaking at
    $200\,{\rm M}_{\odot}\,{\rm yr}^{-1}$).  See Sect.~\ref{sect:dustsfr}
    for details.}
   \label{fig:histosfr} 
\end{figure}

Many studies
\citep[e.g.,][]{magdis2010,elbaz2011,greve2012,magdis2012,symeonidis2013,weiss2013,magnelli2014}
suggest that dust temperatures of $z \sim 2$ sources are typically of
the order of 35\,K, consistent with the few measurements in hand (see
Sect.~\ref{sect:validover}). With the conservative assumption that
$T_{\rm d}=35$\,K for all sources \citep[e.g.,][]{greve2012}, we can
derive the IR luminosity for each SPIRE source (by integrating between
8$\,\mu$m and 1\,mm), and find that it has a broad distribution
peaking at $4\times10^{12}\,{\rm L}_{\odot}$
(Fig.~\ref{fig:histolir}). Assuming that the conventionally assumed
relationship between IR luminosity and star formation
\citep{kennicutt98,bell2003} holds, and that AGN do not dominate (as
was also found for the objects previously discussed by
\citealt{santos2014} and \citealt{clements2014}), this would translate
into a peak SFR of $700\,{\rm M}_{\odot}\,{\rm yr}^{-1}$ per SPIRE
source (Fig.~\ref{fig:histosfr}). If we were to use colder dust
temperatures, the peak SFR would be at about $200\,{\rm
  M}_{\odot}\,{\rm yr}^{-1}$ per SPIRE source
(Fig.~\ref{fig:histosfr}). If we were to use warmer dust temperatures,
the SFR would increase because the implied redshift would be higher
and still compatible with other observations, e.g., the compilation of
submillimetre galaxies in \cite{greve2012}, where $L_{\rm IR}$ is
measured in the range $3\times10^{12}$--$10^{14}\,{\rm L}_{\odot}$ and
$T_{\rm d}$ in the range 30--100\,K, for redshifts above 2.

\begin{figure}[!ht] 
   \centering 
  \includegraphics[width=0.49\textwidth]{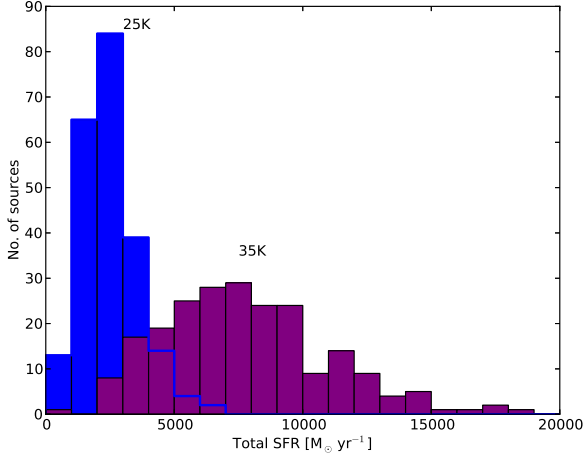}
  \caption{Total star formation rate (SFR) per \Planck\ IN region,
    i.e., for the overdensity sample, assuming each overdensity
    is an actual high-$z$ cluster of dusty galaxies, located at a
    redshift given by the SPIRE photometric redshift, and using a
    modified blackbody fit with a fixed dust temperature of
    $T_{\rm d}=35\,$K (purple, right histogram peaking at
    $7\times10^3\,{\rm M}_{\odot}\,{\rm yr}^{-1}$),
    this temperature being favoured by the
    literature. We also show the total SFR for $T_{\rm d}=25\,$K,
    giving a peak at $2\times10^3\,{\rm M}_{\odot}\,{\rm yr}^{-1}$. The
    structures could be intensively star-forming clusters. See
    Sects.~\ref{sect:dustsfr} and \ref{sect:massiveclusters} for
    details.}
   \label{fig:histosfrtot} 
\end{figure}

We can also estimate the IR luminosities and SFRs of the
overdensities, i.e., potential clusters of dusty galaxies at
high-$z$, rather than the single SPIRE sources that we used
above. With typically 10 SPIRE sources in the IN region, we sum up the
IR luminosities of SPIRE sources within the IN region (under the same
assumption of $T_{\rm d}=35\,$K) to obtain a peak IR luminosity of
$4\times10^{13}\,{\rm L}_{\odot}$.  This translates to a peak SFR of
$7\times10^3\,{\rm M}_{\odot}\,{\rm yr}^{-1}$ per structure, as seen in
Fig.~\ref{fig:histosfrtot}.

Our estimates of $L_{\rm IR}$ are within the range of FIR luminosities
expected for massive, vigorously star-forming high-$z$ structures
\citep[e.g.,][]{brodwin2013}, which are perhaps protoclusters in their
intense star formation phase, and are consistent with the four bound
structures found in \cite{clements2014}.  These $z>2$ intensively star
forming proto-clusters are also expected in some models, e.g., the
proto-spheroids of \cite{cai2013}, or the halos harbouring intense
star formation discussed in \cite{bethermin2013}.

%__________________________________________________________________ 
% Histo IR Luminosity for T=25 and 35
%__________________________________________________________________ 

\subsection{Massive galaxy clusters in formation?}
\label{sect:massiveclusters}

It is of course tempting to postulate that all (or at least a subset) of
our sources include massive galaxy clusters in the process of
formation. However, demonstrating this conclusively with the present
{\it Herschel\/} photometry alone is not possible. We have therefore
undertaken a comprehensive multi-wavelength photometric and
spectroscopic follow-up campaign to directly constrain the nature of our
sources and in particular, to investigate whether they
are good candidates for being the progenitors of today's massive galaxy
clusters seen during their most rapid phase of baryon cooling and star
formation.

Even without such an explicit analysis of the astrophysical nature of
our candidates, we can already state that: (1) our estimated
luminosities are consistent with the expected star formation
properties of massive galaxy clusters, as obtained from stellar
archaeology in galaxies in nearby clusters and out to redshifts
$z\sim1$; and (2) our total number of overdensities are consistent
with the expected numbers of massive high-$z$ galaxy clusters, as
obtained from models of dark matter halo structure formation.

For point (1), a broad consensus has now been developed whereby most of the
stellar mass in massive cluster galaxies was already in place by
$z\simeq1$. The most stringent observational constraint is perhaps the
tight red sequence of cluster galaxies in colour-magnitude diagrams of
massive galaxy clusters in the optical and near-infrared
\citep[e.g.,][]{renzini2006}, which suggests that massive cluster
galaxies formed most of their stellar populations within a short timescale
of $\le1\,$Gyr, i.e., within one cluster dynamical time,
with little star formation thereafter. Rest-frame optical studies of
distant clusters suggest that the bright end of the red sequence was
already in place by $z\simeq1$ \citep[][]{rudnick2009}, with a likely
onset amongst the most massive galaxies by $\simeq2$, or even before
\citep[][]{kodama2007}.

%__________________________________________________________________ 
% Histo SFR
%__________________________________________________________________ 

The shortness of the star formation period in nearby clusters allows
us to derive an order-of-magnitude estimate of the total star
formation rates during this period, as implied by the fossil
constraints, and the resulting FIR fluxes. To obtain upper limits on
the stellar mass formed during this epoch, we use stellar mass
estimates obtained for 93 massive X-ray-selected clusters with
$M_{500}=(10^{14}$--$2\times10^{15})\,{\rm M}_{\odot}$ at $z=0$--0.6
by \citet{lin2012} based on {\it WISE\/} $3.4\,\mu$m
photometry. $M_{500}$ is the total mass within the central cluster
regions where the mass surface density exceeds the cosmological value
by at least a factor of 500. \citeauthor{lin2012} find that in this
mass range, stellar mass scales with $M_{500}$ as
$(M_\ast/10^{12}\,{\rm M}_\odot) = (1.8\pm0.1)\ (M_{500}/10^{14}\,{\rm
  M}_\odot)^{0.71\pm 0.04}$.  For clusters with
$M_{500}=1\times10^{14}\,{\rm M}_{\odot}$ (or $2\times10^{15}\,{\rm
  M}_{\odot}$, their highest mass), this corresponds to
$M_{\ast}=2\times10^{12}\,{\rm M}_{\odot}$ (or $1.5\times10^{13}\,{\rm
  M}_{\odot}$). We do not expect that including the intracluster light
would increase these estimates by more than a few tens of percent
\citep[e.g.,][]{gonzalez2007}, which is a minor part of the
uncertainty in our rough order-of-magnitude estimate.

These mass and timescale estimates suggest total
star formation rates in the progenitors of massive galaxy clusters at
lower redshifts of a few $\times 10^3$ and up to about
$2\times10^4\,{\rm M}_{\odot}\,{\rm yr}^{-1}$.
Using the conversion of \citet{kennicutt98}
between star formation rate and infrared luminosity
(Sect.~\ref{sect:dustsfr}),
this corresponds to $L_{\rm IR}\simeq (10^{13}$--$10^{14})\,{\rm L}_{\odot}$.
This is well within the range of the global star formation rates and
$L_{\rm FIR}$ values in our sample of overdensities.

%__________________________________________________________________ 
% Histo SFR TOTAL
%__________________________________________________________________ 

For point (2), the absence of such objects, we use the
\cite{tinker2008} halo model to compute the expected surface density
of dark matter halos. We expect between 8500 and $1 \times 10ˆ7$ dark
matter halos at $z>2$ having masses $M_{\rm tot}>10^{14}\,{\rm
  M}_{\odot}$ and $10^{13}\,{\rm M}_{\odot}$, repectively, over 35\,\%
of the sky. With the assumption that the cluster galaxies have formed
most of their stellar populations within a short timescale (see
above), we estimate that only a fraction of those halos will be
observationally caught during their intense star formation phase, when
they are infrared- and submm-bright. Assuming that this phase happened
mainly between $z=5$ and $z=1.5$--2.0, we find that the upper limit on
the formation period of each individual cluster, about 1\,Gyr, is 3--4
times shorter than the cosmic time elapsed over this epoch,
2--3\,Gyr. We would therefore expect to find a few thousand dark
matter halos on the submm sky at any given observing epoch. This
order-of-magnitude estimate is above with our finding of about 200
overdensities with \Planck\ in this study, and with a few hundred to be
detected in the full \Planck\ data set by
\cite{planck_collaboration2015}, but not more than an order of
magnitude, illustrating the need to understand the detailed processes
of star formation in the most massive halos \citep{bethermin2013}.

%-----------------------------------------------------------------------
%-----------------------------------------------------------------------
\section{Conclusions}
\label{sect:conclusion}

Our \Planck\ sample based on a colour selection of cold sources of the
CIB is overwhelmingly dominated in number by significant galaxy
overdensities peaking at 350$\,\mu$m, and a minority of rare, bright
$z>1.5$ strongly gravitationally lensed sources, among which are the
brightest ever detected in the submillimetre. This confirms the
efficiency of \Planck\ to select extreme and rare submillimetre
sources over the whole sky, as well as the need for higher angular
resolution imaging using {\it Herschel\/} (and current/planned
ground-based submm/mm observatories) to identify and study them.

From the analysis of {\it Herschel\/} observations of the sample of 228
\Planck\ cold sources of the CIB, we draw several conclusions.

\begin{itemize}

\item Less than 2\,\% of the fields are Galactic cirrus
structures. When a clean and controlled selection is performed on
\Planck, there is thus no reason to expect a large cirrus
contamination when using a conservative Galactic mask.

\item With about 93\,\% of the overdensities peaking at 350$\,\mu$m,
and 3.5\,\% peaking at 500$\,\mu$m, our sample unambiguously selects sky
areas with the largest concentrations of red SPIRE galaxies, consisting of
highly significant overdensities.

\item Some of the overdensities are confirmed high-$z$
structures, e.g., a source at $z \simeq 1.7$--2.0
(\textcolor{blue}{Flores-Cacho et al. 2014, in prep.}).

\item The significance in density contrast of the overdensities (e.g.,
  half of the fields above 10$\,\sigma_{\delta}$ when selecting red
  SPIRE sources) is higher than any other sample targeting
  proto-clusters or high-$z$ clusters in the submillimetre, confirming
  the relevance of the strategy of using \Planck\ data on the cleanest
  parts of the sky to uncover high-$z$ candidates.

\item The SPIRE sources in the overdensity fields have a peak
redshift of $z = 2$ or 1.3, if we fix the dust temperature at
$T_{\rm d}=35\,$K or 25\,K, respectively.

\item With the assumption of $T_{\rm d} =35\,$K, each SPIRE
source has an average IR luminosity of $4\times10^{12}\,{\rm L}_{\odot}$,
leading to star formation rates for each source peaking at
$700\,{\rm M}_{\odot}\,{\rm yr}^{-1}$. If confirmed, these exceptional
structures harbouring vigorous star formation could be proto-clusters
in their starburst phase.

\item Assuming the SPIRE sources are located in the same
large-scale overdensity, we derive a total IR luminosity of
$4\times10^{13}\,{\rm L}_{\odot}$, leading to total star formation rates of
$7\times10^3\,{\rm M}_{\odot}\,{\rm yr}^{-1}$, and with around 10 detected
sources per structure.

\item About 3\,\% of our sample is composed of intensely
  gravitationally lensed galaxies. This sample is unique, as it
  targets the brightest observed such sources, typically above
  400\,mJy at 350$\,\mu$m and reaching up to the jansky level. They
  are all spectroscopically confirmed to lie at redshifts $z =
  2.2$--3.6 \citep{canameras2015,nesvadba2015}.

\item The novelty and efficiency of our new sample is that it
provides about 50 times more fields for a \Planck-{\it Herschel\/}
co-analysis than in existing {\it Herschel\/} surveys searching for
serendipitous \Planck\ sources \citep[e.g.,][]{clements2014}, for only
a fraction of the {\it Herschel\/} observing time.

\item Our new sample exhibits high density contrasts with a high
significance: for the red sources, 30\,\% of our sample (about 70
fields) shows significance levels higher than 4.5$\,\sigma_{\delta}$, and
15\,\% of our sample (34 fields) are higher than 7$\,\sigma_{\delta}$.

\end{itemize}

The presence of the overdensities (and the hypothesis that these could
be actual forming, bound structures) is consistent with other
findings \citep{santos2014,clements2014,noble2013,wylezalek2013a}.
Our \Planck\ extraction and confirmation with {\it Herschel\/} provides
an efficient selection over the whole sky, biased towards star-forming
sources, giving a few hundred candidates (Fig.~\ref{fig:histosfr}).
This selection nicely complements the structures found at $z \sim
2$ already detected by different means, e.g., with the X-ray signature
of hot IGM gas, with stellar mass overdensities, Ly$\alpha$ emission, or in
association with radiogalaxies
\cite[]{pentericci1997,pentericci2000,brodwin2005,miley2006,nesvadba2006,venemans2007,doherty2010,papovich2010,brodwin2010,brodwin2011,hatch2011,gobat2011,santos2011,stanford2012,santos2013,santos2014,brodwin2013,galametz2013,rigby2014,wylezalek2013,chiang2014,cooke2014a,cucciati2014a}. However,
it is hard to estimate the fraction that may be comprised of random
alignments of unrelated clumps \citep[e.g.,][]{chiang2013}, since the
fluctuation field due to clustering is highly non-Gaussian.  Ancillary
data (particularly redshifts) are needed to confirm the associations,
and an ambitious follow-up programme is underway.

To further characterize the overdensities discovered by \Planck\ and
confirmed by {\it Herschel}, we also started to stack them in many
ancillary data sets. The first data set is the \Planck\ lensing map
\citep{planck2013-p12}. The goal is to detect a signature of the
presence of large gravitational potential, as has been demonstrated already by
correlating the CIB fluctuations with the lensing map
\citep{planck2013-p13}.  The second data set is to use the nine
frequency maps and the diffuse tSZ map obtained by MILCA
\citep{hurier2013}. The goal here is to detect a signature of the
thermal Sunyaev-Zeldovich (tSZ) effect \citep{sunyaev1969,sunyaev1972}
and thus the presence of hot gas. These analyses will be published in
a subsequent paper.

The \Planck\ data thus provide unique and powerful samples to
uncover rare populations, exploitable for extragalactic studies, as was
shown with our successful follow-up with {\it Herschel\/} of 228
\Planck\ high-$z$ candidates. There are hundreds more \Planck\
high-$z$ candidates (the full catalogue is being finalized and will be
published in \citealt{planck_collaboration2015}) to be studied and
characterized.

%-----------------------------------------------------------------------
%-----------------------------------------------------------------------
\begin{acknowledgements}
  The development of \Planck\ has been supported by: ESA; CNES and
  CNRS/INSU-IN2P3-INP (France); ASI, CNR, and INAF (Italy); NASA and
  DoE (USA); STFC and UKSA (UK); CSIC, MICINN, JA, and RES (Spain);
  Tekes, AoF, and CSC (Finland); DLR and MPG (Germany); CSA (Canada);
  DTU Space (Denmark); SER/SSO (Switzerland); RCN (Norway); SFI
  (Ireland); FCT/MCTES (Portugal); and PRACE (EU). A description of
  the \Planck\ Collaboration and a list of its members, including the
  technical or scientific activities in which they have been involved,
  can be found at
 % \href{http://www.sciops.esa.int/index.php?project=planck&page=Planck_Collaboration}{http://www.sciops.esa.int/index.php?project=planck}.
  http://www.sciops.esa.int/index.php?project=planck.

  The {\it Herschel\/} spacecraft was designed, built, tested, and launched
  under a contract to ESA managed by the {\it Herschel}/\Planck\ Project
  team by an industrial consortium under the overall responsibility of
  the prime contractor Thales Alenia Space (Cannes), and including
  Astrium (Friedrichshafen) responsible for the payload module and for
  system testing at spacecraft level, Thales Alenia Space (Turin)
  responsible for the service module, and Astrium (Toulouse)
  responsible for the telescope, with in excess of a hundred
  subcontractors.

  This work is based in part on observations made with the {\it
    Spitzer Space Telescope}, which is operated by the Jet Propulsion
  Laboratory, California Institute of Technology under a contract with
  NASA. Support for this work was provided by NASA through an award
  issued by JPL/Caltech.  Based in part on observations obtained with
  MegaPrime/MegaCam, a joint project of CFHT and CEA/DAPNIA, at the
  Canada-France-Hawaii Telescope (CFHT) which is operated by the
  National Research Council (NRC) of Canada, the Institute National
  des Sciences de l'Univers of the Centre National de la Recherche
  Scientifique of France, and the University of Hawaii. Based in part
  on observations obtained with WIRCam, a joint project of CFHT,
  Taiwan, Korea, Canada, France, and the Canada-France-Hawaii
  Telescope (CFHT) which is operated by the National Research Council
  (NRC) of Canada, the Institute National des Sciences de l'Univers of
  the Centre National de la Recherche Scientifique of France, and the
  University of Hawaii.  Based in part on observations carried out
  with the IRAM 30-m Telescope. IRAM is supported by INSU/CNRS
  (France), MPG (Germany) and IGN (Spain). Based in part on
  observations carried out with the IRAM Plateau de Bure
  Interferometer. IRAM is supported by INSU/CNRS (France), MPG
  (Germany) and IGN (Spain).  Based in part on observations made at
  JCMT with SCUBA-2. The James Clerk Maxwell Telescope is operated by
  the Joint Astronomy Centre on behalf of the Science and Technology
  Facilities Council of the United Kingdom, the Netherlands
  Organisation for Scientific Research, and the National Research
  Council of Canada.  This research has made use of the SIMBAD
  database, operated at CDS, Strasbourg, France.  This research has
  made use of the NASA/IPAC Extragalactic Database (NED) which is
  operated by the Jet Propulsion Laboratory, California Institute of
  Technology, under contract with the National Aeronautics and Space
  Administration.  We acknowledge the support from the CNES, the PNCG
  (Programme National de Cosmologie et Galaxies), ANR HUGE
  (ANR-09-BLAN-0224-HUGE) and ANR MULTIVERSE (ANR-11-BS56-015). We
  also acknowledge the support from R\'egion Ile-de-France with
  DIM-ACAV.  We acknowledge the Integrated Data \& Operation Center
  (IDOC) at Institut d'Astrophysique Spatiale and Observatoire des
  Sciences de l'Univers de l'Universit\'e Paris Sud (OSUPS). Support
  for IDOC is provided by CNRS and CNES.  We acknowledge final support
  from ASI/INAF agreement I/072/09/0 and PRIN-INAF 2012 project
  ``Looking into the dust-obscured phase of galaxy formation through
  cosmic zoom lenses in the Herschel Astrophysical Large Area
  Survey.''  We acknowledges financial support from the Spanish CSIC
  for a JAE-DOC fellowship, cofunded by the European Social Fund and
  from the Ministerio de Economia y Competitividad, project
  AYA2012-39475-C02-01.  This research made use of {\tt matplotlib}
  \cite{hunter2007}, and of {\tt APLpy}, an open-source plotting
  package for Python hosted at http://aplpy.github.com.  We thank
  E. Egami, B. Cl\'ement, E. Daddi, H.J. McCracken and A. Boucaud and
  for fruitful discussions and helpful advice.

\end{acknowledgements}

%-----------------------------------------------------------------------
%-----------------------------------------------------------------------
\bibliographystyle{aa}

% Automatic References w/ Gabi elsewhere
%\bibliography{all2planck.bib}
%%%%%%\bibliography{planck_herschel_highz_pip102rev,Planck_bib.bib}

%Insert bbl file at the end

%-----------------------------------------------------------------------
%-----------------------------------------------------------------------

%-----------------------------------------------------------------------
%-----------------------------------------------------------------------

\appendix

\section{SPIRE band-merging procedure}
\label{sect:bandmergingappendix}

%__________________________________________________________________ 
% TABLE: percentage of matches BASIC
%__________________________________________________________________ 
\begin{table}[!ht]
\begingroup
\newdimen\tblskip \tblskip=5pt
\caption{Percentages of SPIRE sources matched at 250 and 500$\,\mu$m
         from the 350$\,\mu$m input SPIRE catalogue, and the frequency of
         matches (i.e., number of sources found per 350$\,\mu$m source). Here,
         IN refers to the \Planck\ source region, and OUT refers to the zone
         outside the \Planck\ 50\,\% contour (see Sect.~\ref{sect:inout}).
         The rows are: no match (top); one match; two matches; and
         three matches (bottom).  More than 60\,\%
         of the SPIRE 350$\,\mu$m sources corresponding to the \Planck\ sources
         (i.e., the IN regions) have counterparts at 250 and 500$\,\mu$m.}
\label{tab:matches250_500}
\vskip -5mm
\footnotesize
\setbox\tablebox=\vbox{
 \newdimen\digitwidth
 \setbox0=\hbox{\rm 0}
 \digitwidth=\wd0
 \catcode`*=\active
 \def*{\kern\digitwidth}
 \newdimen\signwidth
 \setbox0=\hbox{+}
 \signwidth=\wd0
 \catcode`!=\active
 \def!{\kern\signwidth}
 \newdimen\pointwidth
 \setbox0=\hbox{\rm .}
 \signwidth=\wd0
 \catcode`?=\active
 \def?{\kern\pointwidth}
 \halign{\tabskip=0pt\hbox to 1.0in{#\leaderfil}\tabskip=1em&
 \hfil#\hfil\tabskip=1em&
 \hfil#\hfil\tabskip=1em&
 \hfil#\hfil\tabskip=1em&
 \hfil#\hfil\tabskip 0pt\cr
\noalign{\doubleline}
\omit& \multispan2\hfil 250$\,\mu$m\hfil& \multispan2\hfil 500$\,\mu$m\hfil\cr
\noalign{\vskip -5pt}
\omit& \multispan4\hrulefill\cr
\omit& IN& OUT& IN& OUT\cr
\noalign{\vskip 4pt\hrule\vskip 6pt}
No match& 10.9& 16.7& 38.8& 67.0\cr
 1 match& 82.0& 77.0& 60.7& 32.9\cr
 2 matches& *6.9& *6.0& *0.5& 0.1\cr
 3 matches& *0.1& *0.2& *0?*&   0?*\cr
\noalign{\vskip 4pt\hrule\vskip 6pt}
}}
\endPlancktable
\endgroup
\end{table}

%__________________________________________________________________ 
% TABLE: percentage of matches BANDMERGED
%__________________________________________________________________ 
\begin{table}[!ht]
\begingroup
\newdimen\tblskip \tblskip=5pt
\caption{After the SPIRE band-merging process, the percentage of SPIRE sources
         that do not have any flux density measurement (i.e., are
         blended or not detected or have multiple matches in
         Table~\ref{tab:matches250_500}).  More than about 90\,\%
         of the \Planck\ IN sources have counterparts at all three SPIRE
         wavelengths.}
\label{tab:ratio_undetect}
\vskip -5mm
\footnotesize
\setbox\tablebox=\vbox{
 \newdimen\digitwidth
 \setbox0=\hbox{\rm 0}
 \digitwidth=\wd0
 \catcode`*=\active
 \def*{\kern\digitwidth}
 \newdimen\signwidth
 \setbox0=\hbox{+}
 \signwidth=\wd0
 \catcode`!=\active
 \def!{\kern\signwidth}
 \newdimen\pointwidth
 \setbox0=\hbox{\rm .}
 \signwidth=\wd0
 \catcode`?=\active
 \def?{\kern\pointwidth}
 \halign{\tabskip=0pt\hbox to 1.0in{#\leaderfil}\tabskip=1em&
 \hfil#\hfil\tabskip=1em&
 \hfil#\hfil\tabskip=1em&
 \hfil#\hfil\tabskip=1em&
 \hfil#\hfil\tabskip=1em&
 \hfil#\hfil\tabskip=1em&
 \hfil#\hfil\tabskip 0pt\cr
\noalign{\doubleline}
\omit& \multispan2\hfil 250$\,\mu$m\hfil& \multispan2\hfil 350$\,\mu$m\hfil&
 \multispan2\hfil 500$\,\mu$m\hfil\cr
\noalign{\vskip -5pt}
\omit& \multispan6\hrulefill\cr
\omit& IN& OUT& IN& OUT& IN& OUT\cr
\noalign{\vskip 4pt\hrule\vskip 6pt}
No detections& 0.2\,\%& 0.6\,\%& 0.2\,\%& 0.6\,\%& *0.5\,\%& *3.6\,\%\cr
${\rm S/N}<3$& 3.2\,\%& 7.6\,\%& 0.9\,\%& 3.2\,\%& 10.6\,\%& 39.4\,\%\cr
\noalign{\vskip 4pt\hrule\vskip 6pt}
}}
\endPlancktable
\endgroup
\end{table}

We describe here the details of the band-merging procedure,
highlighted in Sect.~\ref{sect:bandmerged}.

{\it First step\/}: positional optimization of blind catalogues using shorter
wavelength data.
\begin{itemize}
\item Apply a $4\,\sigma$ cut to the 350$\,\mu$m blind catalogue obtained
  with {\tt StarFinder} (Sect.~\ref{sect:blind}). This cut is imposed
  to obtain a high purity of detections.
\item Match by position the 350$\,\mu$m sources to the 250$\,\mu$m
  catalogue, within a radius of one 250$\,\mu$m FWHM. 
\end{itemize}
The following depends on whether there are 0, 1, 2, or more matches:
\begin{itemize}
\item if zero sources are found at 250$\,\mu$m, this is a non-detection at
250$\,\mu$m;
\item if one source is found, we take this source to be the 250$\,\mu$m
counterpart;
\item if two sources are found, we replace the position of the 350$\,\mu$m
unique source by the two positions of the 250$\,\mu$m sources (the
350$\,\mu$m flux density will be measured later) and take the two
250$\,\mu$m flux densities;
\item we apply the same procedure for triply or more matched sources:
  we keep the positions of the 250~$\mu$m sources as priors (only one
  case, outside the \Planck\ beam.
\end{itemize}
We repeat this operation at 500$\,\mu$m using one 350$\,\mu$m FWHM
as the search radius. The statistics of those basic positional matches
are reported in Table~\ref{tab:matches250_500}. We have more than 60\,\%
unique matches in the IN regions, i.e., for the \Planck\
sources. This simple method is, however, biasing the sample towards
strong and unique detections at the three SPIRE wavelengths. The two
following steps address this issue.

{\it Second step\/}: prior flux density determination.
\begin{itemize}
\item At 350$\,\mu$m: take the {\tt StarFinder} flux densities
  (Sect.~\ref{sect:blind}) for single-matched source with 250$\,\mu$m;
  assign $S_{350}=0$ if there are two or more matched sources at
  250$\,\mu$m. 
\item At 250$\,\mu$m: take the flux density for single-, double- or
  triple- matched sources with the 350$\,\mu$m source; assign
  $S_{250}=0$ if there is no match with 350$\,\mu$m at this
  step.
\item At 500$\,\mu$m: take the {\tt StarFinder} flux densities
  (Sect.~\ref{sect:blind}) for single-matched source with the
  350$\,\mu$m source; Assign $S_{500}=0$ if there are 0, 2
  or more matched sources at 350$\,\mu$m.
\end{itemize}

{\it Third step\/}: deblending and photometry using {\tt FastPhot}.
\begin{itemize}
\item Use the 250$\,\mu$m positions obtained in the first step as prior
  positions.
\item Use flux densities at each SPIRE wavelength obtained in the
  second step.
\item Perform simultaneous PSF-fitting and deblending;
\item Assign the measured flux densities that were previously missing.
\end{itemize}

Table~\ref{tab:ratio_undetect} presents some statistics following the
band-merging process.

%-----------------------------------------------------------------------
%-----------------------------------------------------------------------
\section{SPIRE number counts tables}
\label{sect:countstables}

The measured number counts shown in Fig.~\ref{fig:totalcounts}, are
provided in tables~\ref{tab:number_counts_250},
\ref{tab:number_counts_350}, and \ref{tab:number_counts_500}. We note
that those counts are not corrected for incompleteness or for flux
boosting, since here we are interested only in the relative
distribution of the samples (and the quantification of the excess
of one sample to another).

%__________________________________________________________________ 
% TABLES: counts
%__________________________________________________________________ 

%
% Table NUMBER COUNTS
%---------------------------------------------------------------
\begin{table*}[tmb]                 % table* is a two-column table.  Drop the * for one column.
\begingroup
\newdimen\tblskip \tblskip=5pt
\caption{Differential Euclidean-normalized number counts
  $S^{2.5}\frac{dN}{dS}$ at 250~$\mu$m, not corrected for
  incompleteness or flux boosting. They are given here only for relative
  distributions.}                          % Caption goes here.
\label{tab:number_counts_250}                            % Label goes here.
\nointerlineskip
\vskip -3mm
\footnotesize
\setbox\tablebox=\vbox{
  \newdimen\digitwidth 
  \setbox0=\hbox{\rm 0} 
   \digitwidth=\wd0 
   \catcode`*=\active 
   \def*{\kern\digitwidth}
   \newdimen\signwidth 
   \setbox0=\hbox{+} 
   \signwidth=\wd0 
   \catcode`!=\active 
   \def!{\kern\signwidth}
\halign{ # & # & # & # & # & # & #  \cr                        % Template goes here.
\noalign{\doubleline}
                        % Table headings go here.
    $\left < S_{250} \right >$  &  $S_{250}$ range  & IN \Planck  & OUT \Planck  & Total \Planck & Lockman & HLS    \cr
    $[$mJy$]$ &   $[$mJy$]$   &   [Jy$^{1.5}$\,sr$^{-1}$] & [Jy$^{1.5}$\,sr$^{-1}$] & [Jy$^{1.5}$\,sr$^{-1}$] & [Jy$^{1.5}$\,sr$^{-1}$] & [Jy$^{1.5}$\,sr$^{-1}$]  \cr
\noalign{\vskip 3pt\hrule\vskip 5pt}
                        % Body of table goes here.
     11. &      11.-     12. &  --  &  --  &  --  &  --  &  --  \cr 
     12. &      12.-     13. &  --  &  --  &  --  &  --  &      20. $\pm$       7. \cr 
     13. &      13.-     14. &  --  &  --  &  --  &  --  &     103. $\pm$      14. \cr 
     16. &      14.-     17. &  --  &  --  &  --  &  --  &     271. $\pm$      21. \cr 
     19. &      17.-     21. &     235. $\pm$      50. &     476. $\pm$      28. &     445. $\pm$      25. &  --  &     447. $\pm$      27. \cr 
     25. &      21.-     29. &    9276. $\pm$     343. &    6075. $\pm$     108. &    6493. $\pm$     104. &    4828. $\pm$     122. &     750. $\pm$      38. \cr 
     34. &      29.-     40. &   40486. $\pm$     838. &   21789. $\pm$     238. &   24227. $\pm$     234. &   29237. $\pm$     351. &    3299. $\pm$      94. \cr 
     50. &      40.-     59. &   37392. $\pm$    1001. &   18877. $\pm$     275. &   21291. $\pm$     273. &   23910. $\pm$     395. &   15891. $\pm$     257. \cr 
     75. &      59.-     90. &   19583. $\pm$     944. &   10611. $\pm$     269. &   11781. $\pm$     264. &   13593. $\pm$     388. &   11697. $\pm$     288. \cr 
    116. &      90.-    141. &    7545. $\pm$     791. &    4960. $\pm$     248. &    5297. $\pm$     239. &    6494. $\pm$     362. &    6011. $\pm$     278. \cr 
    182. &     141.-    223. &    3482. $\pm$     742. &    2587. $\pm$     248. &    2704. $\pm$     236. &    3696. $\pm$     377. &    3127. $\pm$     277. \cr 
    290. &     223.-    357. &     935. $\pm$     540. &    1963. $\pm$     303. &    1829. $\pm$     273. &    2730. $\pm$     455. &    1891. $\pm$     303. \cr 
    467. &     357.-    576. &     626. $\pm$     626. &    1785. $\pm$     409. &    1634. $\pm$     365. &    1524. $\pm$     482. &     975. $\pm$     308. \cr 
    753. &     576.-    931. &    5100. $\pm$    2550. &     573. $\pm$     331. &    1164. $\pm$     440. &     310. $\pm$     310. &    1190. $\pm$     486. \cr 
\noalign{\vskip 5pt\hrule\vskip 3pt}}}
%\endPlancktable                    % ends one-column \halign
\endPlancktablewide                 % ends two-column \halign
%\tablenote a Footnote a.\par
%\tablenote b Footnote b.\par
\endgroup
\end{table*}                        % table* is a two-column table.  Drop the * for one column.
%---------------------------------------------------------------
% written by make_tables_counts_herschel.pro on Tue Feb  4 14:22:52 2014

%
% Table NUMBER COUNTS
%---------------------------------------------------------------
\begin{table*}[tmb]                 % table* is a two-column table.  Drop the * for one column.
\begingroup
\newdimen\tblskip \tblskip=5pt
\caption{Differential Euclidean-normalized number counts $S^{2.5}\frac{dN}{dS}$ at 350~$\mu$m, not corrected for
  incompleteness or flux boosting. They are given here only  for relative
  distributions.}                          % Caption goes here.
\label{tab:number_counts_350}                            % Label goes here.
\nointerlineskip
\vskip -3mm
\footnotesize
\setbox\tablebox=\vbox{
  \newdimen\digitwidth 
  \setbox0=\hbox{\rm 0} 
   \digitwidth=\wd0 
   \catcode`*=\active 
   \def*{\kern\digitwidth}
   \newdimen\signwidth 
   \setbox0=\hbox{+} 
   \signwidth=\wd0 
   \catcode`!=\active 
   \def!{\kern\signwidth}
\halign{ # & # & # & # & # & # & #  \cr                        % Template goes here.
\noalign{\doubleline}
                        % Table headings go here.
    $\left < S_{350} \right >$  &  $S_{350}$ range  & IN \Planck  & OUT \Planck  & Total \Planck & Lockman & HLS    \cr
    $[$mJy$]$ &   $[$mJy$]$   &   [Jy$^{1.5}$\,sr$^{-1}$] & [Jy$^{1.5}$\,sr$^{-1}$] & [Jy$^{1.5}$\,sr$^{-1}$] & [Jy$^{1.5}$\,sr$^{-1}$] & [Jy$^{1.5}$\,sr$^{-1}$]  \cr
\noalign{\vskip 3pt\hrule\vskip 5pt}
                        % Body of table goes here.
     11. &      11.-     12. &  --  &  --  &  --  &  --  &  --  \cr 
     12. &      12.-     13. &  --  &  --  &  --  &  --  &  --  \cr 
     13. &      13.-     14. &  --  &  --  &  --  &  --  &       5. $\pm$       3. \cr 
     16. &      14.-     17. &  --  &  --  &  --  &  --  &      39. $\pm$       8. \cr 
     19. &      17.-     21. &     107. $\pm$      34. &     162. $\pm$      16. &     155. $\pm$      15. &  --  &     148. $\pm$      16. \cr 
     25. &      21.-     29. &    7931. $\pm$     317. &    4114. $\pm$      88. &    4611. $\pm$      87. &     648. $\pm$      45. &     415. $\pm$      29. \cr 
     34. &      29.-     40. &   32868. $\pm$     755. &   13424. $\pm$     187. &   15959. $\pm$     190. &   19667. $\pm$     288. &    3661. $\pm$      99. \cr 
     50. &      40.-     59. &   31093. $\pm$     913. &   10831. $\pm$     209. &   13473. $\pm$     217. &   16222. $\pm$     325. &    8986. $\pm$     194. \cr 
     75. &      59.-     90. &   16805. $\pm$     875. &    4650. $\pm$     178. &    6235. $\pm$     192. &    6880. $\pm$     276. &    4938. $\pm$     187. \cr 
    116. &      90.-    141. &    3731. $\pm$     556. &    1305. $\pm$     127. &    1622. $\pm$     132. &    1936. $\pm$     198. &    1509. $\pm$     140. \cr 
    182. &     141.-    223. &    1425. $\pm$     475. &     783. $\pm$     136. &     867. $\pm$     134. &     809. $\pm$     176. &     640. $\pm$     126. \cr 
    290. &     223.-    357. &    1247. $\pm$     623. &     561. $\pm$     162. &     650. $\pm$     163. &     303. $\pm$     152. &     339. $\pm$     128. \cr 
    467. &     357.-    576. &    1879. $\pm$    1085. &      94. $\pm$      94. &     327. $\pm$     163. &     305. $\pm$     216. &     195. $\pm$     138. \cr 
    753. &     576.-    931. &    6374. $\pm$    2851. &     382. $\pm$     270. &    1164. $\pm$     440. &  --  &  --  \cr 
\noalign{\vskip 5pt\hrule\vskip 3pt}}}
%\endPlancktable                    % ends one-column \halign
\endPlancktablewide                 % ends two-column \halign
%\tablenote a Footnote a.\par
%\tablenote b Footnote b.\par
\endgroup
\end{table*}                        % table* is a two-column table.  Drop the * for one column.
%---------------------------------------------------------------
% written by make_tables_counts_herschel.pro on Tue Feb  4 14:22:52 2014

%
% Table NUMBER COUNTS
%---------------------------------------------------------------
\begin{table*}[tmb]                 % table* is a two-column table.  Drop the * for one column.
\begingroup
\newdimen\tblskip \tblskip=5pt
\caption{Differential Euclidean-normalized number counts $S^{2.5}\frac{dN}{dS}$ at 500~$\mu$m, not corrected for
  incompleteness or flux boosting. They are given here only for relative
  distributions.}                          % Caption goes here.
\label{tab:number_counts_500}                            % Label goes here.
\nointerlineskip
\vskip -3mm
\footnotesize
\setbox\tablebox=\vbox{
  \newdimen\digitwidth 
  \setbox0=\hbox{\rm 0} 
   \digitwidth=\wd0 
   \catcode`*=\active 
   \def*{\kern\digitwidth}
   \newdimen\signwidth 
   \setbox0=\hbox{+} 
   \signwidth=\wd0 
   \catcode`!=\active 
   \def!{\kern\signwidth}
\halign{ # & # & # & # & # & # & #  \cr                        % Template goes here.
\noalign{\doubleline}
                        % Table headings go here.
    $\left < S_{500} \right >$  &  $S_{500}$ range  & IN \Planck  & OUT \Planck  & Total \Planck & Lockman & HLS    \cr
    $[$mJy$]$ &   $[$mJy$]$   &   [Jy$^{1.5}$\,sr$^{-1}$] & [Jy$^{1.5}$\,sr$^{-1}$] & [Jy$^{1.5}$\,sr$^{-1}$] & [Jy$^{1.5}$\,sr$^{-1}$] & [Jy$^{1.5}$\,sr$^{-1}$]  \cr
\noalign{\vskip 3pt\hrule\vskip 5pt}
                        % Body of table goes here.
     11. &      11.-     12. &  --  &  --  &  --  &  --  &  --  \cr 
     12. &      12.-     13. &  --  &  --  &  --  &  --  &  --  \cr 
     13. &      13.-     14. &  --  &  --  &  --  &  --  &  --  \cr 
     16. &      14.-     17. &  --  &  --  &  --  &  --  &       3. $\pm$       2. \cr 
     19. &      17.-     21. &  --  &       5. $\pm$       3. &       4. $\pm$       2. &  --  &      42. $\pm$       8. \cr 
     25. &      21.-     29. &     685. $\pm$      93. &     346. $\pm$      26. &     390. $\pm$      25. &  --  &     138. $\pm$      17. \cr 
     34. &      29.-     40. &   13397. $\pm$     482. &    3474. $\pm$      95. &    4768. $\pm$     104. &    4614. $\pm$     140. &     437. $\pm$      34. \cr 
     50. &      40.-     59. &   15600. $\pm$     647. &    2890. $\pm$     108. &    4547. $\pm$     126. &    5692. $\pm$     193. &    1964. $\pm$      90. \cr 
     75. &      59.-     90. &    5875. $\pm$     517. &     751. $\pm$      72. &    1419. $\pm$      92. &    1473. $\pm$     128. &     935. $\pm$      81. \cr 
    116. &      90.-    141. &    1078. $\pm$     299. &     199. $\pm$      50. &     314. $\pm$      58. &     484. $\pm$      99. &     232. $\pm$      55. \cr 
    182. &     141.-    223. &     791. $\pm$     354. &     142. $\pm$      58. &     227. $\pm$      68. &      77. $\pm$      54. &     172. $\pm$      65. \cr 
    290. &     223.-    357. &    1559. $\pm$     697. &     140. $\pm$      81. &     325. $\pm$     115. &      76. $\pm$      76. &      97. $\pm$      69. \cr 
    467. &     357.-    576. &    2506. $\pm$    1253. &      94. $\pm$      94. &     408. $\pm$     183. &  --  &  --  \cr 
    753. &     576.-    931. &    3825. $\pm$    2208. &  --  &     499. $\pm$     288. &  --  &  --  \cr 
\noalign{\vskip 5pt\hrule\vskip 3pt}}}
%\endPlancktable                    % ends one-column \halign
\endPlancktablewide                 % ends two-column \halign
%\tablenote a Footnote a.\par
%\tablenote b Footnote b.\par
\endgroup
\end{table*}                        % table* is a two-column table.  Drop the * for one column.
%---------------------------------------------------------------
% written by make_tables_counts_herschel.pro on Tue Feb  4 14:22:52 2014

%-----------------------------------------------------------------------
%-----------------------------------------------------------------------
\section{Number counts by type: overdensity or lensed sources}
\label{sect:countspertype}

\begin{figure*}[!ht] 
   \centering 
  \includegraphics[width=0.95\textwidth]{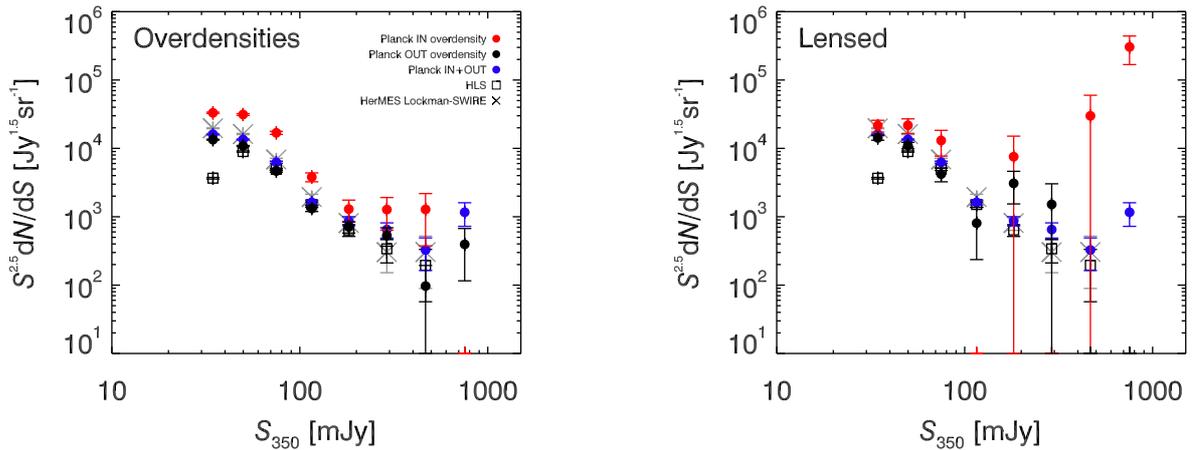}
  \caption{Differential Euclidean-normalized number counts,
    $S^{2.5}dN/dS$, for various data sets at 350$\,\mu$m, to illustrate
    the difference between the overdensity fields (left) and the
    lensed fields (right). This is the same as
    Fig.~\ref{fig:totalcounts}, except that we have split the fields
    into only overdensity fields and only lensed fields.
    Lensed sources thus dominate the statistics at
    large flux densities, while overdensities dominate at smaller flux
    densities. See Sect~\ref{sect:classification} for details.}
  \label{fig:totalcounts_lens_over} 
\end{figure*}

As discussed in Sect.~\ref{sect:classification},
Fig.~\ref{fig:totalcounts_lens_over} represents the counts for
overdensity fields (left) and lensed fields (right). This shows that
the bright part of the number counts is dominated by single bright
objects, i.e., the lensed sources. However, these lensed counts cannot be
used for statistical studies of the lensed sources because we have
only used the surface area of the \Planck\ IN region.

%__________________________________________________________________ 
% Figures: TOTAL NUMBER COUNTS: over (left) and lensed (right)
%__________________________________________________________________ 

%-----------------------------------------------------------------------
%-----------------------------------------------------------------------
\section{Overdensities using AKDE}
\label{sect:overdensities_using_akde}

\begin{figure}[!ht] 
   \centering 
  \includegraphics[width=0.5\textwidth]{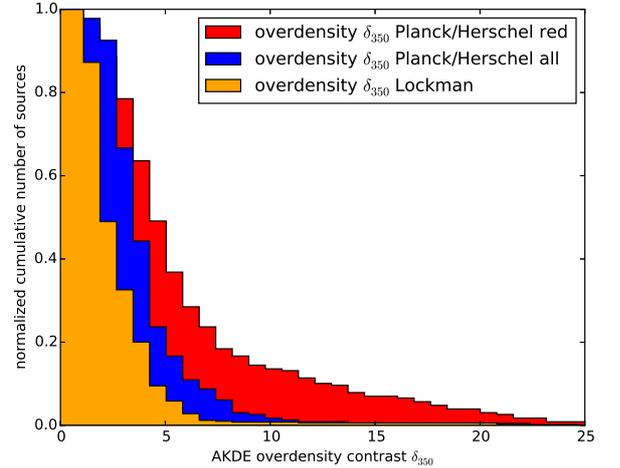}
  \caption{Cumulative histogram of the normalized overdensity contrast
    $\delta_{350}$ using the AKDE density estimator. Blue represents
    all our SPIRE sources, red represents only redder SPIRE sources,
    defined by $S_{350}/S_{250} > 0.7$ and $S_{500}/S_{350} > 0.6$,
    and orange 500 random fields in Lockman. Most of our fields show
    overdensities larger than 5. See Sect.~\ref{sect:overdensities}
    for details.}
  \label{fig:overdensities_akde} 
\end{figure}

We can also use another estimator for the density contrast
$\delta_{350}$ by taking the measured source density from AKDE. If we
compute the density minus the median density divided by the median
density, we obtain the distribution shown in
Fig.~\ref{fig:overdensities_akde}.

%-----------------------------------------------------------------------
%-----------------------------------------------------------------------
\section{Gallery of selected sources}
\label{sect:gallery}

%__________________________________________________________________ 
% Figures: GALLERY
%__________________________________________________________________ 
\begin{figure*}[!ht]
   \centering 
  \includegraphics[width=0.42\textwidth]{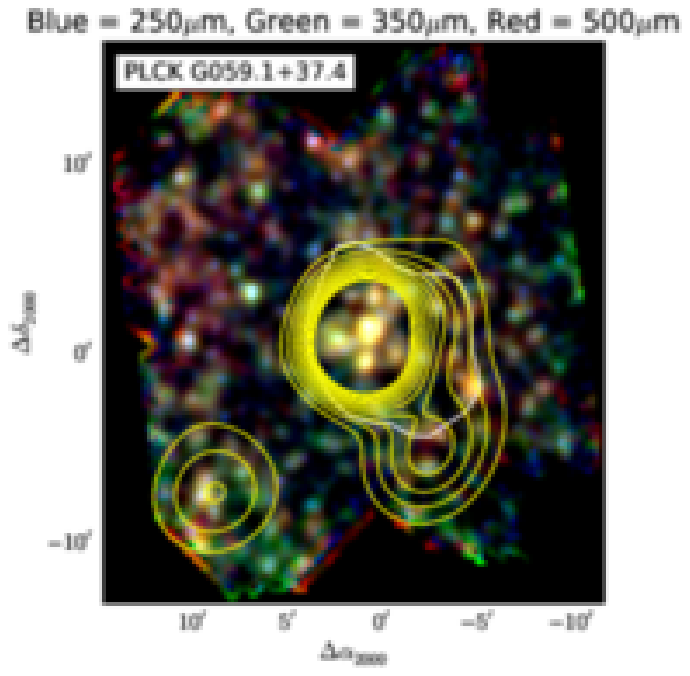}
  \includegraphics[width=0.42\textwidth]{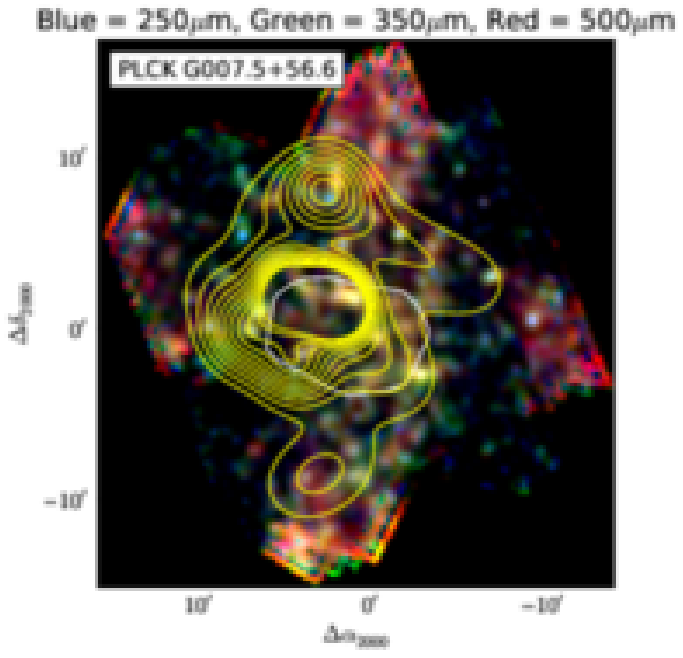}
  \includegraphics[width=0.42\textwidth]{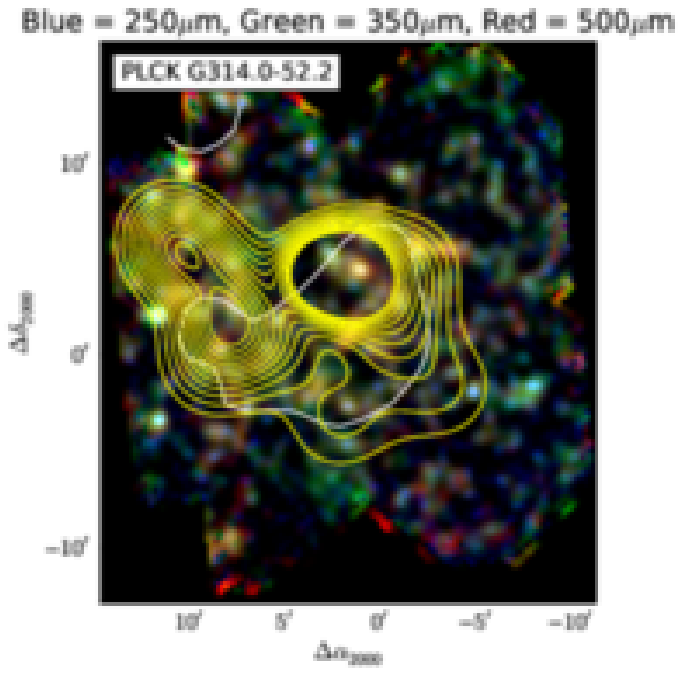}
  \includegraphics[width=0.42\textwidth]{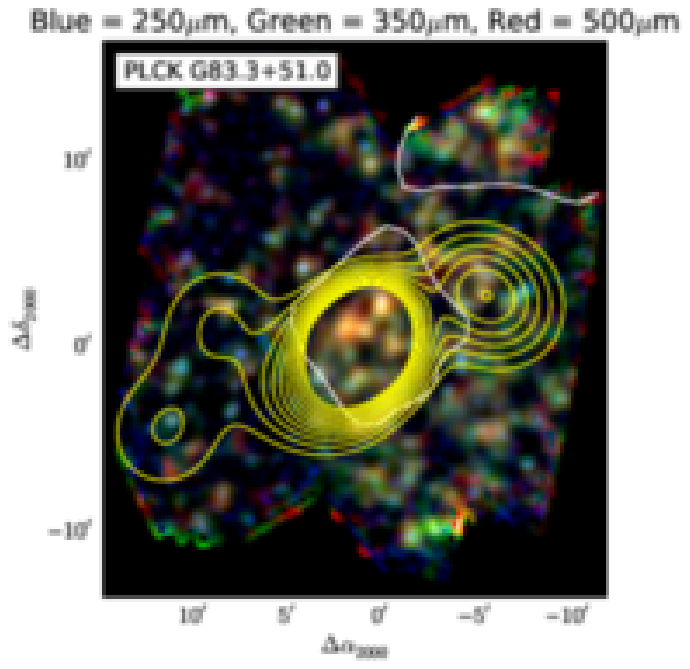}
  \caption{Representative \Planck\ targets, showing 3-colour SPIRE
    images: blue, 250$\,\mu$m; green, 350$\,\mu$m; and red, 500$\,\mu$m. The
    white contours show the \Planck\ IN region, while the yellow
    contours are the significance of the overdensity of 350$\,\mu$m sources,
    plotted 2$\,\sigma$, 3$\,\sigma$, 4$\,\sigma$, etc.}
 \label{fig:gallery} 
\end{figure*} 

\begin{figure*}[!ht]
   \centering 
  \includegraphics[width=0.44\textwidth]{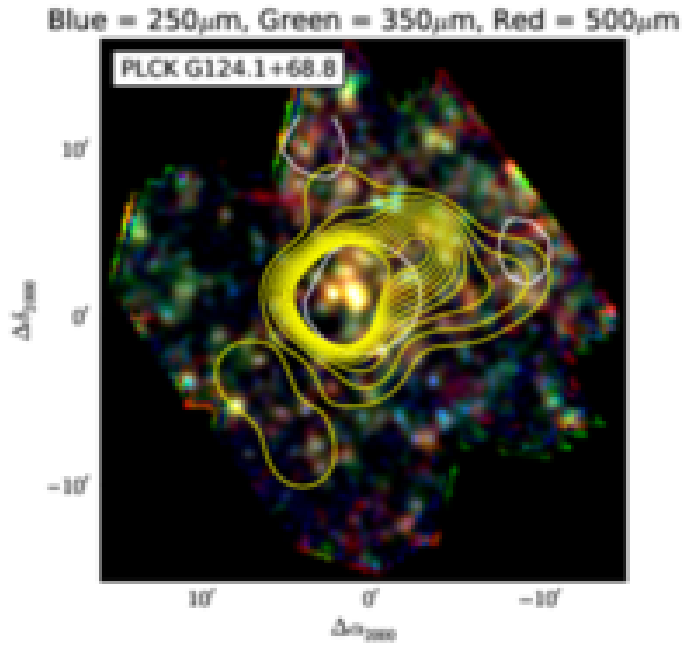}
  \label{fig:gallery2} 
  \caption{Representative \Planck\ targets. Same as Fig.~\ref{fig:gallery}.}
\end{figure*} 

\begin{figure*}[!ht]
   \centering 
  \includegraphics[width=0.44\textwidth]{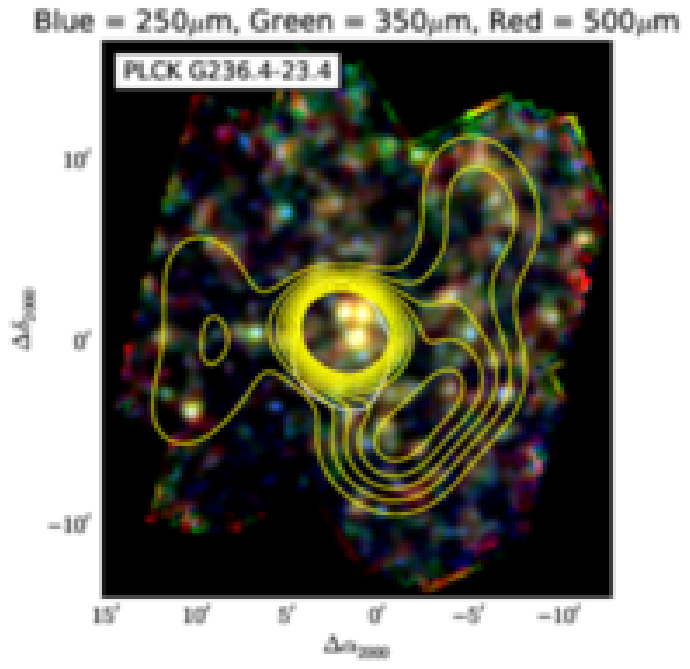}
  \label{fig:gallery3} 
  \caption{Representative \Planck\ targets. Same as Fig.~\ref{fig:gallery}.}
\end{figure*}

\begin{figure*}[!ht]
   \centering 
  \includegraphics[width=0.44\textwidth]{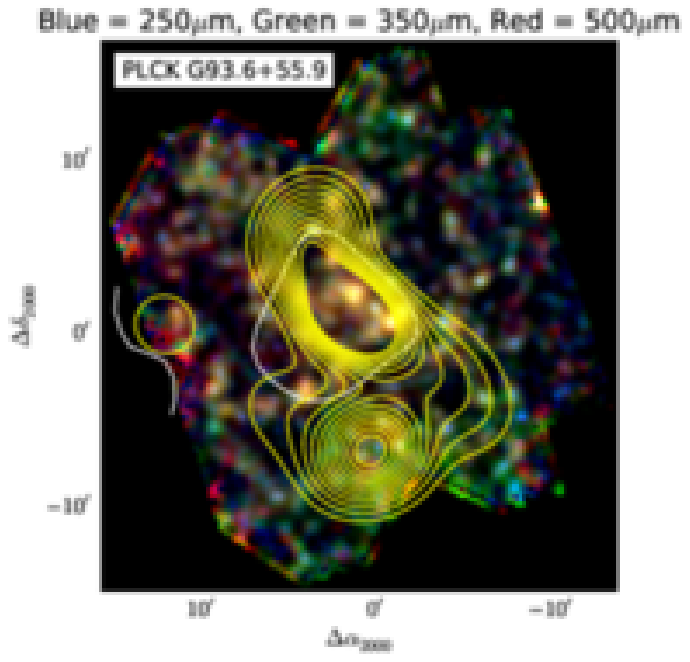}
  \label{fig:gallery4} 
  \caption{Representative \Planck\ targets. Same caption as Fig.~\ref{fig:gallery}.}
\end{figure*}

\begin{figure*}[!ht]
   \centering 
  \includegraphics[width=0.44\textwidth]{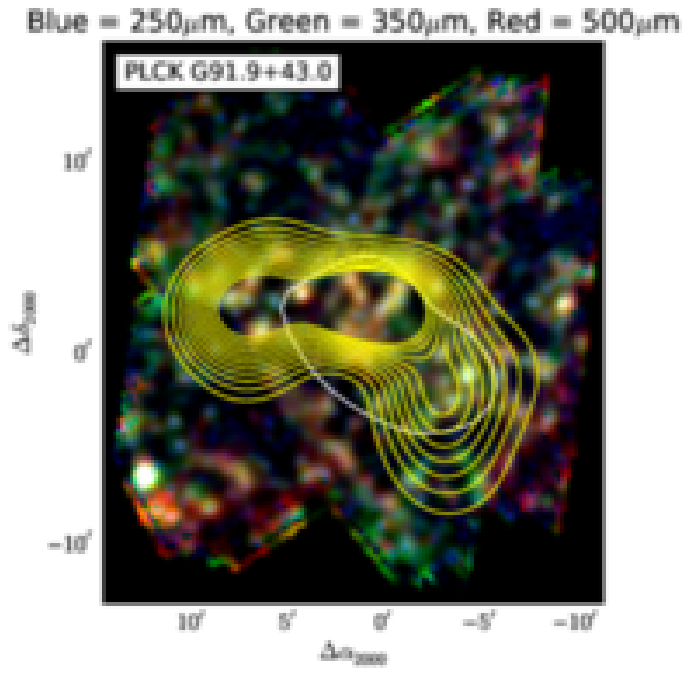}
  \label{fig:gallery5} 
  \caption{Representative \Planck\ targets. Same caption as Fig.~\ref{fig:gallery}.}
\end{figure*}

\begin{figure*}[!ht]
   \centering 
  \includegraphics[width=0.44\textwidth]{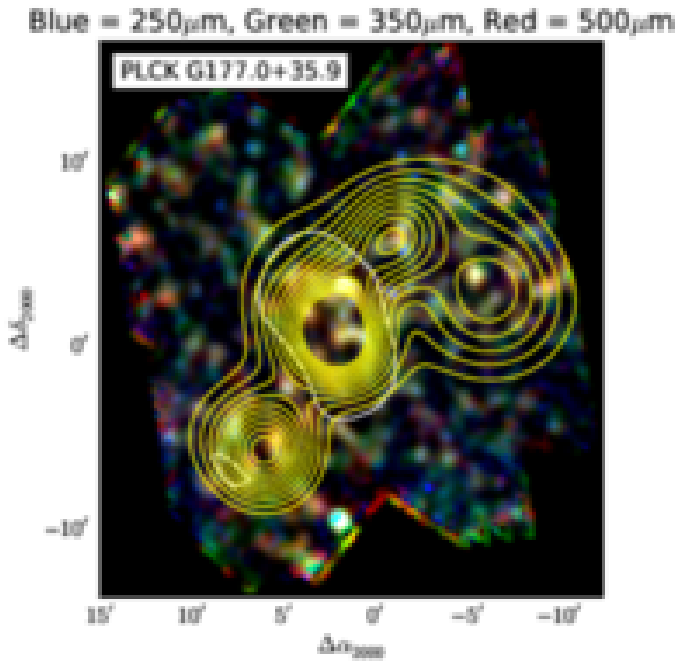}
  \label{fig:gallery6} 
  \caption{Representative \Planck\ targets. Same caption as Fig.~\ref{fig:gallery}.}
\end{figure*} 

\begin{figure*}[!ht]
   \centering 
  \includegraphics[width=0.44\textwidth]{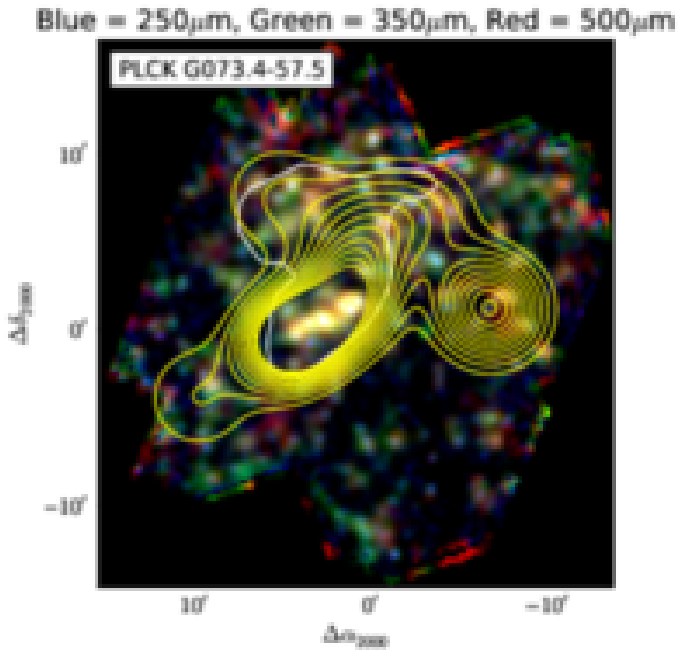}
  \label{fig:gallery7} 
  \caption{Representative \Planck\ targets. Same caption as Fig.~\ref{fig:gallery}.}
\end{figure*}

\begin{figure*}[!ht]
   \centering 
   \includegraphics[width=0.95\textheight, angle=90]{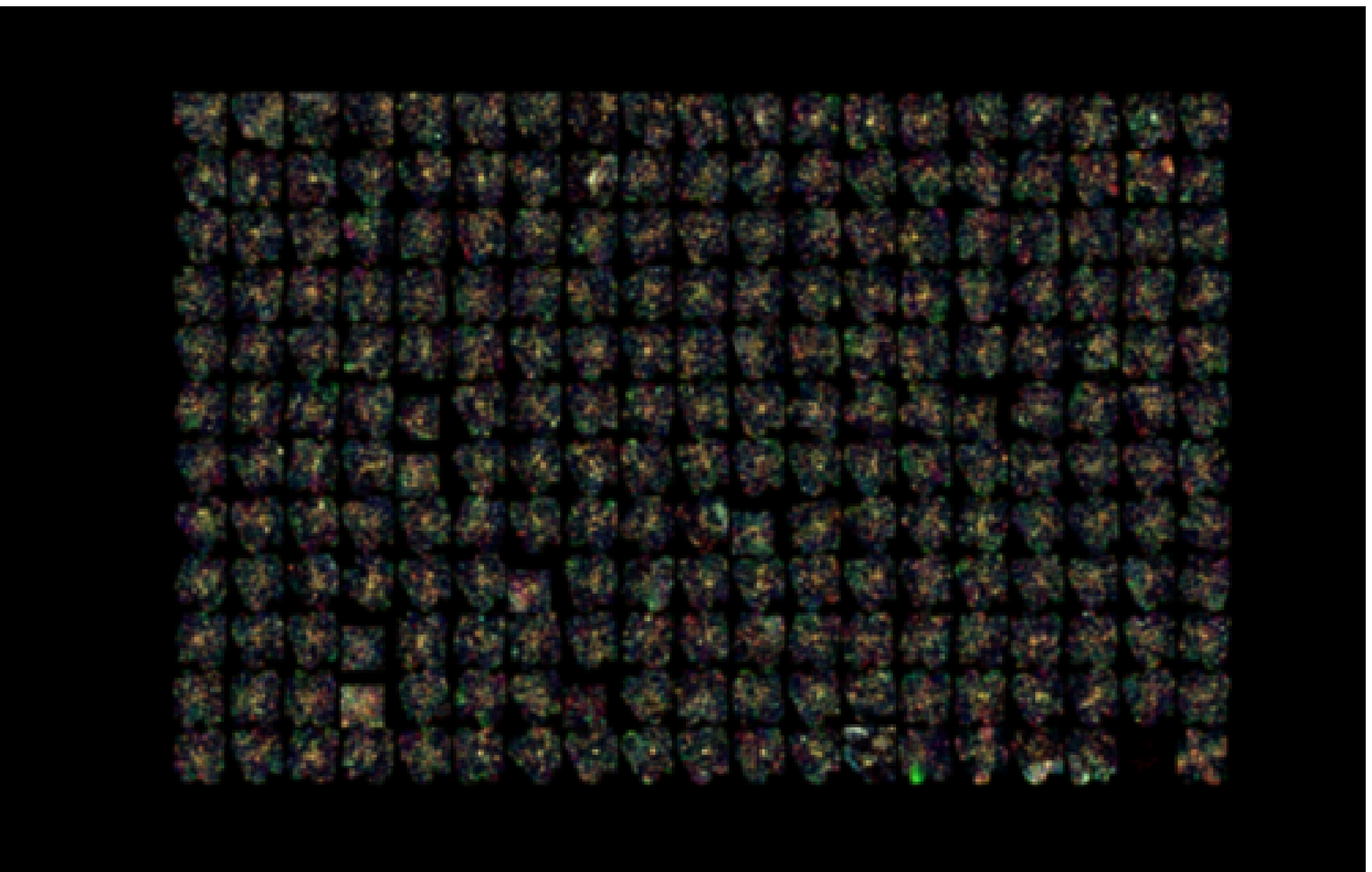}
   \label{fig:galleryall} 
   \caption{Mosaic showing the 228 SPIRE fields.}
\end{figure*}

We provide 3-colour images for a few sources drawn from the 228 of our
sample. Each plot shows:
\begin{itemize}
\item a 3-colour image of the SPIRE data, with 250$\,\mu$m in blue,
  350$\,\mu$m in green, and 350$\,\mu$m in red (using the python {\tt APLpy}
  module: \verb|show_rgb|);
\item the \Planck\ contour (IN region, i.e., 50\,\% of the peak) as
  a bold white line;
\item the significance of the SPIRE overdensity contrast (starting at
  $2\,\sigma$, and then incrementing by $1\,\sigma$) as yellow solid
  lines.
\end{itemize} 

Figure~\ref{fig:galleryall} summarizes the 228 SPIRE sample. 

\raggedright

\end{document}